\documentclass[11pt, a4paper]{article}
\pdfoutput=1
\usepackage{jheppub,amsmath,amssymb,slashed,url,bm,textgreek,upgreek}
\usepackage{jheppub}  
\usepackage{tikz,lipsum,lmodern}
\usepackage{slashed}
\usepackage[most]{tcolorbox}
\usepackage{amssymb} 
\usepackage{amsmath}
\usepackage{mathtools}
\usepackage{amsfonts}    
\usepackage{dsfont}
\usepackage{pdfpages}
\usepackage{verbatim}
\usepackage{tensor}
\usepackage{mathrsfs}
\usepackage{textgreek} 
\usepackage[mathscr]{euscript}
\usepackage[normalem]{ulem}
\usepackage{tikz}
\usepackage{makecell}
\usepackage[verbose]{placeins}
\usepackage{subcaption}
\usetikzlibrary{3d, arrows.meta, decorations.pathreplacing, decorations.markings,calc,shapes.misc,decorations.pathmorphing,patterns.meta, math}
\usepackage{pgfplots}
\usepackage[T1]{fontenc}
\pgfplotsset{compat=1.18}

\newcommand{\beq}{\begin{equation}}
\newcommand{\eeq}{\end{equation}}
\newcommand{\nn}{\nonumber\\} 
\newcommand{\bea}{\begin{eqnarray}}
\newcommand{\ea}{\end{eqnarray}}
\newcommand{\barr}{\begin{array}}
\newcommand{\earr}{\end{array}}

\def\be{\begin{equation}}
\def\ee{\end{equation}}
\def\ba#1\ea{\begin{align}#1\end{align}}
\def\bg#1\eg{\begin{gather}#1\end{gather}}
\def\bm#1\em{\begin{multline}#1\end{multline}}
\def\bmd#1\emd{\begin{multlined}#1\end{multlined}}

\def\d{\mathrm{d}}
 \def\i{\mathrm{i}}

\def\vp{\chi}

\def\wg{\wedge}

\def\({\left(}
\def\){\right)}
\def\[{\left[}
\def\]{\right]}
\def\<{\langle}
\def\>{\rangle}

\def\bea{\begin{eqnarray}}
\def\eea{\end{eqnarray}}

\newcommand{\tr}{\operatorname{tr}}

\newcommand{\zb}{{\bar z}}

\def\nn{\nonumber}

\begin{document}

\global\long\def\aad{(a\tilde{a}+a^{\dagger}\tilde{a}^{\dagger})}%

\global\long\def\ad{{\rm ad}}%

\global\long\def\bij{\langle ij\rangle}%

\global\long\def\df{\coloneqq}%

\global\long\def\bs{b_{\alpha}^{*}}%

\global\long\def\bra{\langle}%

\global\long\def\dd{\mathrm{d}}%

\global\long\def\dg{{\rm {\rm \dot{\gamma}}}}%

\global\long\def\ddt{\frac{\mathrm{d}^{2}}{\mathrm{d}t^{2}}}%

\global\long\def\ddg{\nabla_{\dot{\gamma}}}%

\global\long\def\del{\mathcal{\delta}}%

\global\long\def\Del{\Delta}%

\global\long\def\dtau{\frac{\dd^{2}}{\dd\tau^{2}}}%

\global\long\def\ul{U(\Lambda)}%

\global\long\def\udl{U^{\dagger}(\Lambda)}%

\global\long\def\dl{D(\Lambda)}%

\global\long\def\da{\dagger}%

\global\long\def\id{{\rm id}}%

\global\long\def\ml{\mathcal{L}}%

\global\long\def\mm{\mathcal{\mathcal{M}}}%

\global\long\def\mf{\mathcal{\mathcal{F}}}%

\global\long\def\ket{\rangle}%

\global\long\def\kpp{k^{\prime}}%

\global\long\def\lr{\leftrightarrow}%

\global\long\def\lf{\leftrightarrow}%

\global\long\def\ma{\mathcal{A}}%

\global\long\def\mb{\mathcal{B}}%

\global\long\def\md{\mathcal{D}}%

\global\long\def\mbr{\mathbb{R}}%

\global\long\def\mbz{\mathbb{Z}}%

\global\long\def\mh{\mathcal{\mathcal{H}}}%

\global\long\def\mi{\mathcal{\mathcal{I}}}%

\global\long\def\ms{\mathcal{\mathcal{\mathcal{S}}}}%

\global\long\def\mg{\mathcal{\mathcal{G}}}%

\global\long\def\mfa{\mathcal{\mathfrak{a}}}%

\global\long\def\mfb{\mathcal{\mathfrak{b}}}%

\global\long\def\mfb{\mathcal{\mathfrak{b}}}%

\global\long\def\mfg{\mathcal{\mathfrak{g}}}%

\global\long\def\mj{\mathcal{\mathcal{J}}}%

\global\long\def\mk{\mathcal{K}}%

\global\long\def\mmp{\mathcal{\mathcal{P}}}%

\global\long\def\mn{\mathcal{\mathcal{\mathcal{N}}}}%

\global\long\def\mq{\mathcal{\mathcal{Q}}}%

\global\long\def\mo{\mathcal{O}}%

\global\long\def\qq{\mathcal{\mathcal{\mathcal{\quad}}}}%

\global\long\def\ww{\wedge}%

\global\long\def\ka{\kappa}%

\global\long\def\nn{\nabla}%

\global\long\def\nb{\overline{\nabla}}%

\global\long\def\pathint{\langle x_{f},t_{f}|x_{i},t_{i}\rangle}%

\global\long\def\ppp{p^{\prime}}%

\global\long\def\qpp{q^{\prime}}%

\global\long\def\we{\wedge}%

\global\long\def\pp{\prime}%

\global\long\def\sq{\square}%

\global\long\def\vp{\varphi}%

\global\long\def\ti{\widetilde{}}%

\global\long\def\wg{\widetilde{g}}%

\global\long\def\te{\theta}%

\global\long\def\tr{{\rm Tr}}%

\global\long\def\ta{{\rm \widetilde{\alpha}}}%

\global\long\def\sh{{\rm {\rm sh}}}%

\global\long\def\ch{{\rm ch}}%

\global\long\def\Si{{\rm {\rm \Sigma}}}%

\global\long\def\sch{{\rm {\rm Sch}}}%

\global\long\def\vol{{\rm {\rm {\rm Vol}}}}%

\global\long\def\reg{{\rm {\rm reg}}}%

\global\long\def\zb{{\rm {\rm |0(\beta)\ket}}}%

\newcommand{\pro}{Pr\'oszy\'nski~}

\title{Collapse to extremality: The third law of black hole thermodynamics in quantum gravity
\vspace{-0.3cm}}

\author{Gustavo J. Turiaci and Chih-Hung Wu}
\affiliation{Department of Physics, University of Washington, Seattle, WA 98195, USA}
\emailAdd{turiaci@uw.edu, chwu29@uw.edu}

\abstract{We present a two-step construction to violate the classical third law of black hole thermodynamics for charged black holes in four dimensions. A near-extremal black hole with an $\mathrm{AdS}_2\times X$ throat is first prepared and then driven to extremality in finite time by a charged shell. This construction suggests that the second step is universal for
near-extremal black holes, and can be used
to construct violations of the third law for Kerr black holes. The same description allows us to study the process in quantum
gravity. The classical collapse condition selects a BF-violating matter
sector, for which we construct a holographic dictionary and derive the
quantum transition amplitude. Quantum effects replace the classical
extremal endpoint by a probability distribution that vanishes at
extremality and peaks at positive energy. This obstruction is consistent with the absence of non-supersymmetric
extremal black hole states in quantum gravity. We show that driving the peak close enough to extremality for quantum
gravity to become important requires a preparation time that scales
with the black hole entropy.
}

\maketitle

\pagenumbering{roman}

\newpage
\clearpage
\pagenumbering{arabic}
\setcounter{page}{1}
\section{Introduction}

The third law of black hole mechanics was first formulated in \cite{Bardeen:1973gs} as part of the classical laws governing black hole dynamics. Its \emph{unattainability} formulation states
that an extremal black hole with exactly zero surface gravity cannot be formed by collapse in finite time. A common intuition is that the weak energy condition (WEC) should prevent such finite-time extremalization. The precise classical arguments, however, require more than the WEC alone. They also impose boundedness and regularity
conditions on the infalling matter and assumptions about the evolution of
the horizon \cite{Sullivan:1980,Israel:1986gqz}. Our starting point is the recent realization that the stronger statement based on the WEC alone is false. In particular, the classical argument~\cite{Israel:1986gqz}
implicitly assumes a connected evolution of the outermost apparent
horizon, an assumption that need not hold even for regular collapse
\cite{Kehle:2022}. Kehle and Unger
constructed smooth collapse satisfying the dominant energy condition,
and hence the WEC, that forms an exactly extremal black hole in finite
time~\cite{Kehle:2022}. At the same time, the WEC remains a
strong restriction and prevents extremalization in broad classes of charged
collapse \cite{Fairoos:2017lnm, Bahmani:2024qkr}. This points to an important distinction between impossibility and
genericity, with extremal formation naturally associated with a finely
tuned threshold in the space of initial data~\cite{Kehle:2024vyt,East:2025nfb,Angelopoulos:2026bez, Mittal:2026ryg}.

Charged shells provided the earliest explicit tests of the third law. Their dynamics and global structure, including finite-time collapse to extremality and the associated turning-point conditions, were studied in \cite{Boulware1973,Proszynski:1978,FarrugiaHajicek1979,
Lake:1979,Proszynski:1983}. A charged thin shell can satisfy the WEC in the distributional sense while remaining localized on a timelike hypersurface. Because the shell is singular, the finite jump in the exterior ADM mass and charge does not by itself contradict formulations of the third law that assume regular continuum matter. These examples
also clarify that the newly formed extremal horizon need not arise by
continuously deforming the original subextremal horizon. Early generalizations to continuous charged matter already exhibited
closely related dynamics, beginning with Ori's charged null fluid bounce
and later extensions of this construction
\cite{Ori:1991,Chatterjee:2015cyv,Bick:2026naa}.

Recent work has uncovered that violations of the third law are not restricted to distributional matter. Smooth finite-time formation of extremal Reissner-Nordstr\"om (RN) black holes has been established using characteristic gluing, extended to maximally symmetric asymptotics, and studied numerically \cite{Kehle:2022,Marin:2024rxt,Gadioux:2025unn, Lee:2026uql}. In the characteristic gluing construction, the apparent horizon can jump discontinuously despite the smoothness of the matter and spacetime, allowing extremality to be reached in finite time. The evolution can also pass through a brief superextremal phase before
reaching extremality, without producing a naked singularity in the
physical spacetime. This behavior already appears in the charged null
fluid picture of \cite{Ori:1991} and is realized explicitly in the smooth
Einstein-Maxwell-Vlasov construction of \cite{Kehle:2024vyt}. Characteristic gluing has also been developed in vacuum gravity \cite{Kehle:2023eni}, where a finite-time violation of the third law has recently been constructed \cite{Crump:2026kgu}. Conditional versions of the third law remain valid when the matter sector obeys suitable local mass-charge bounds or supersymmetric positivity conditions \cite{Reall:2024njy,McSharry:2025iuz}. 

Quantum
effects give the classical third law statement its thermodynamic interpretation. Hawking
radiation~\cite{Hawking:1974sw} identifies the black hole temperature as
$T=\hbar\kappa/(2\pi)$, so the impossibility of reaching
$\kappa=0$ in a finite process is reminiscent of the Nernst
unattainability formulation of the third law in ordinary thermodynamics. Recent developments in quantum gravity have also sharpened our understanding
of a distinct Nernst statement concerning the entropy as the temperature approaches zero~\cite{Wald:1997qp}. In particular, quantum gravity corrections
substantially modify the low-temperature entropy and spectrum of
near-extremal black holes~\cite{Iliesiu:2020qvm,Mertens:2022irh, Turiaci:2023wrh}, as suggested implicitly by earlier work~\cite{Preskill:1991tb}. These developments motivate us to ask what the
unattainability statement itself becomes when the black hole is treated
quantum mechanically.

\paragraph{Main results.} We begin by reviewing the thin shell collapse process and showing that it
reproduces most of the features of the smooth solutions discussed
above. This suggests that the thin shell models capture more of the relevant
physics than their distributional nature might suggest. In asymptotically flat space, the mechanism for reaching extremality with charged shells was developed by Boulware, Pr\'oszy\'nski, and Farrugia and H\'aj\'{\i}\v{c}ek \cite{Boulware1973,Proszynski:1978,FarrugiaHajicek1979,
Proszynski:1983}. An overcharged pressureless shell is sent inward with enough kinetic energy to overcome electrostatic repulsion and form an exactly extremal RN exterior in finite advanced time. We refer to the inequality ensuring horizon crossing in finite advanced time as the \emph{Pr\'oszy\'nski condition}, since it is most clearly stated in~\cite{Proszynski:1983}. We then extend this
construction to asymptotically AdS$_4$ spacetimes. This will be useful
for our later quantum gravity analysis and puts the third law violation in the more precise context of AdS/CFT~\cite{Maldacena:1997re,Gubser:1998bc,Witten:1998qj}. 

In classical gravity, we propose a simple two-stage process that violates the third law. We start with a near-extremal black hole whose charge is close to that of the target extremal solution. This geometry can be chosen as the initial state, or formed by collapse in a possibly long but finite time. We then send in a shell that drives the black hole to extremality. This construction has two advantages. First, for a sufficiently weakly
charged shell, the nontrivial final stage can be studied entirely within
a single near-horizon throat, which takes the form AdS$_2 \times X$ or,
more generally, a nontrivial fibration over AdS$_2$. We reduce the dynamics to Jackiw-Teitelboim (JT) gravity~\cite{Teitelboim:1983ux, Jackiw:1984je} coupled to a charged matter field describing the shell. This simplifies the classical analysis and allows us to solve the collapse dynamics in quantum gravity. Second, the description depends only on the near-horizon geometry, suggesting that the mechanism applies to a broad class of black holes with AdS$_2$ throats \cite{Kunduri:2007vf,Figueras:2008qh,PandoZayas:2026vbg}. In particular, the same classical approach applies to Kerr if the shell is made of rapidly rotating gravitons.

For Kerr, the effective charged field in AdS$_2$ comes from a suitable rapidly rotating partial wave of the graviton. We believe this can be a promising direction to rigorously prove the existence of third-law violations for Kerr and more complicated spacetimes. Previous constructions of rotating shells and rotating collapse either treated 4d rotation perturbatively or exploited special symmetries available in higher dimensions~\cite{DeLaCruzIsrael:1968, LindblomBrill:1974, Kegeles:1978, Delsate:2014iia, Rocha:2017uwx, Pereira:2023wnz}. Our proposal instead uses the near-horizon AdS$_2$ dynamics of a rapidly rotating
near-extremal Kerr black hole, a regime not covered by these constructions.

Next, we extend our analysis to quantum gravity. An important consequence of the classical collapse condition becomes apparent in the AdS$_2$ description. For a semiclassical initial black
hole, the \pro condition places the effective charged matter channel below the AdS$_2$ BF bound~\cite{Breitenlohner:1982bm, Breitenlohner:1982jf}, in a principal-series representation
of SL$(2,\mathbb R)$. In flat space it implies $q>m$, which might point to an interesting connection with the weak gravity conjecture \cite{Arkani-Hamed:2006emk}. Instead of the standard Dirichlet boundary conditions, we show that the collapse problem naturally fixes the incoming component of this field at the boundary of the throat. From there we construct the corresponding fixed-incoming holographic dictionary and derive the transition amplitude between near-extremal black hole energy eigenstates. 

The quantum description changes how we should think about the third law near extremality. Classically, an extremal black hole appears to retain
a macroscopic entropy even as the temperature vanishes, in apparent tension
with the Nernst statement concerning the zero-temperature entropy. Quantum
gravity substantially modifies this extrapolation
\cite{Iliesiu:2020qvm}; see~\cite{Turiaci:2023wrh} for a brief review. The classical analysis becomes unreliable below the breakdown scale $T_{ q}$, which is of order $1/(r_0S_0)$, with $r_0$ the extremal radius and $S_0$ the extremal entropy. This scale was identified in  \cite{Preskill:1991tb} as the point
where the classical thermodynamic description breaks down, and
the quantum theory shows that the low-temperature entropy and spectrum
differ qualitatively from their semiclassical continuation. The density of black hole states vanishes as we approach extremality. For this reason, the analysis in \cite{Iliesiu:2020qvm} implies that extremal black holes do not exist in quantum gravity, at least without supersymmetry. This does not rule out extremal states with very low degeneracy, but any such states would not be described by the original extremal black hole geometry at all. The classical picture of a black hole carrying a
macroscopic entropy all the way to zero temperature is therefore misleading.

These developments might suggest that the unattainability question
simply disappears in quantum gravity. If there is no extremal black hole state,
there is no final state to collapse to. However, this does not settle the operational question. It remains important to determine whether we can prepare, in finite time, a near-extremal black hole in the quantum gravity regime, below the breakdown scale. If
such a preparation were possible, the classical third-law-violating
process could still provide a novel operational probe of quantum gravity. To make it precise, we consider a finite-time process
tuned so that its classical limit ends at extremality. The difference in quantum theory is that the endpoint is no longer described by a single deterministic geometry, and this statement
must be formulated \emph{probabilistically}. The natural question is therefore how the final-state probability is distributed near the extremal edge. Our techniques allow us to answer this question quantitatively.

The classical collapse process is modified
in quantum gravity in the following way. Localizing the shell insertion in time necessarily introduces a finite energy spread, so we describe the shell by a wavepacket of duration $\delta t$. We take the initial state to be a near-extremal black hole energy eigenstate in the semiclassical regime, at a temperature well above the quantum scale $T_q$. For a source tuned to reach extremality in the classical limit, the final energy distribution vanishes at extremality and peaks at strictly positive energy due to quantum gravity effects. At fixed
$\delta t$ in the semiclassical limit, the peak remains parametrically
above the quantum crossover scale, even though its energy becomes
parametrically small compared with that of the initial black hole. For example in flat space, for a Gaussian wavepacket with an exactly
extremal classical target, we find that the temperature corresponding
to the peak of the final energy probability density is bounded
parametrically by
\beq
T \gtrsim \left( \frac{1}{8\pi r_0 S_0 (\delta t)^2}\right)^{1/3}.
\label{eq:intro_temperature}
\eeq
The minimum temperature achievable
corresponding to the peak remains much larger than $T_{q}$, and therefore we cannot prepare, in finite time, a near-extremal black hole in the regime where quantum gravity effects are important. We can show this conclusion is independent of the details of the wavepacket, although the precise form of \eqref{eq:intro_temperature} does depend on it. Preparing a final distribution concentrated well within the quantum regime instead requires a long source duration, $\delta t \gg T_{q}^{-1} \sim r_0 S_0$. At fixed
classical geometry, this timescale grows linearly with the black hole
entropy, and therefore diverges in the classical limit. Operationally, one might naively expect the quantum state corresponding to a classically extremal endpoint to remain indistinguishable from an extremal black hole for a parametrically long time, until the observer becomes sensitive to the quantum gravity time scale $T^{-1}_q$. Instead, \eqref{eq:intro_temperature} implies that a positive final temperature can be resolved well before the quantum gravity regime is reached.

The suppression of the extremal edge is much more general. Under reasonable assumptions, we extend the
analysis to general normalizable superpositions of black hole energy
eigenstates and to mixed initial states. For a smooth wavepacket representing the shell, the final energy probability density always vanishes at extremality. Generically, it exhibits the same behavior as the density of states, while destructive interference in specially tuned coherent preparations can only suppress the edge further. The quantum gravitational spectral factor therefore implies that the probability of finding the final black hole within a
small energy window $0<E_f<\delta$ scales generically as $\delta^{3/2}$, where $E_f= M_f - M_{\rm ext}(Q_f)$ is the final energy above extremality. Without incorporating quantum gravity, the  same source only gives a near-edge probability linear in $\delta$. No normalizable choice of initial state
can compensate for the vanishing density of states and produce a nonzero
probability density at the extremal endpoint.

\paragraph{Comments on supersymmetric black holes.} Finally, although the focus will be on generic theories of gravity, we collect some comments on black holes whose extremal limit
preserves supersymmetry. The situation is qualitatively different from
the generic non-supersymmetric case considered above.  We begin with the entropic statement. For these black holes, the macroscopic BPS degeneracy survives the quantum gravity corrections and reproduces the leading
Bekenstein-Hawking entropy \cite{Heydeman:2020hhw, Iliesiu:2022onk}. Let $\mathcal H_{\text{BPS}}$ denote the relevant Hilbert space of BPS states. We define $S_{\text{BPS}}$ as the logarithm of the total number of these states, and $S_{\text{index}}$ as the logarithm of the absolute value of their supersymmetric index. The total count adds bosonic and fermionic contributions, whereas the index subtracts them. Thus, unitarity gives the simple bound
\beq
\left|\text{Tr}_{\mathcal{H}_{\text{BPS}}}(-1)^{\sf F}\right|
\leq \text{Tr}_{\mathcal H_{\text{BPS}}}1
 ,\qquad \Rightarrow \qquad \exp{\left( S_{\text{index}}\right)} \leq \exp{\left(S_{\text{BPS}}\right)},
\eeq
where $(-1)^{\sf F}$ is fermion parity.\footnote{This can be generalized when the protected quantity is a
helicity supertrace or a refined index.} Both the Bekenstein-Hawking entropy $S_{\text{BPS}}$ and the index $S_{\text{index}}$ can be computed independently using the gravitational path integral \cite{Cabo-Bizet:2018ehj,Iliesiu:2021are}. Short multiplets whose contributions cancel in the index can
recombine into long multiplets and be lifted above the BPS bound
without changing the index. This suggests a supersymmetric version of the entropic third law: the index accounts for the entire leading extremal entropy. More precisely, we conjecture that
\beq
\lim_{G_N \to 0} \frac{S_{\text{index}}}{S_{\text{BPS}}} =1.
\eeq
This holds in concrete black hole examples known to us, although we do not have a general proof. In holographic settings, it need not hold in the weak coupling regime of the dual quantum system, where the index can exhibit
large cancellations. At strong coupling, when the geometric description applies, this implies that all such states are lifted by interactions and only the minimal set of states compatible with the index remains. In this sense, the special property of black hole quantum systems is the fact that they have a large index.

We now turn to unattainability in the supersymmetric case. The surviving BPS degeneracy gives
nonzero spectral weight at the BPS threshold, and the suppression
at threshold discussed above does not apply. In this case, the obstruction to extremal collapse already arises at the classical level. Indeed, it was shown that a supersymmetric RN black hole
cannot form in finite-time collapse when the matter satisfies an
appropriate local mass-charge inequality \cite{Reall:2024njy}. This is clear in the thin-shell model since it requires $q>m$ and supersymmetric matter satisfies $m \geq q$.  This
result has subsequently been extended to supersymmetric Kerr-Newman-AdS black holes and to the evolution of an initially nonextremal black hole
\cite{McSharry:2025iuz}. It is tempting to conjecture that appropriate
supersymmetric positivity conditions provide such a protection more
generally in supergravity.

\smallskip

The organization of this paper is as follows. In Section \ref{sec:classical_thin_shell}, we review classical thin shell collapse to extremality in asymptotically flat and AdS spacetimes, derive its near-horizon description in JT gravity, and explain the extension to Kerr black holes. In Section \ref{sec:dictionary_BF_violation}, we quantize the BF-violating matter field describing the shell, couple it to Schwarzian-Maxwell dynamics, and derive the quantum transition amplitude. In Section \ref{sec:disk-level-protection}, we analyze the final energy distribution, establish the suppression of the extremal edge, and study how closely the final state can approach extremality by tuning the source. Section \ref{sec:discussion} examines the implications of our results, possible extensions, and additional
dynamical effects that can compete with the source-driven collapse,
including Hawking radiation and Schwinger discharge. Appendix \ref{app:Ori_discussion} relates the massless shell limit to bouncing charged null matter, while Appendix \ref{app:fixed_incoming_kernel} derives the effective action for the charged scalar field on a fixed AdS$_2$ background.

\section{Classical thin-shell violation of the third law and the AdS$_2$ throat} 
\label{sec:classical_thin_shell}

In this section, we study a simple classical process that violates the unattainability formulation of the third law. Recent constructions with smooth matter~\cite{Kehle:2022,Marin:2024rxt,Gadioux:2025unn} motivate revisiting earlier thin shell models, in which the collapse dynamics can be worked out explicitly. We review the Farrugia-H\'aj\'{\i}\v{c}ek-\pro process~\cite{Proszynski:1978,FarrugiaHajicek1979,Proszynski:1983} and extend the asymptotically flat construction to AdS black holes.

We then specialize to an initially near-extremal black hole and a shell whose charge is small compared with that of the black hole. In this regime, the final approach to extremality can be described entirely within the AdS$_2$ throat using JT gravity coupled to charged matter. This provides the starting point for our quantum gravity analysis. It also suggests a more general two-stage construction: first form a near-extremal black hole, then drive it to extremality through a process that can be analyzed simply within its throat. We illustrate the broader applicability of this construction by extending it to Kerr black holes.

\subsection{Extremal black hole formation in asymptotically flat and AdS spaces}
\label{subsec:classical-thin-shell}

We first review some general facts on the dynamics of thin shells with electric charge, originally developed in \cite{Israel1958,Israel:1966rt,deLaCruzIsrael1967,
Kuchar:1968,Chase:1970}. We then apply these results to extremal collapse.

\subsection*{Charged thin-shell collapse}

We work in 4d Einstein-Maxwell theory with negative cosmological constant
$\Lambda=-3/\ell^2$ and action
\beq
S
=
\frac{1}{16 \pi}
\int \d^4x\sqrt{-g}
\left(
R
+
\frac{6}{\ell^2}
-
F^2
\right).
\eeq
The shell $\Sigma$ divides spacetime into an interior
region ${\cal M}_-$ and an exterior region ${\cal M}_+$. Assuming spherical symmetry, the geometry on each side is 
\be
\mathrm{d}s_\pm^2
=
-f_\pm(r)\, \mathrm{d}t_\pm^2
+
{\mathrm{d}r^2\over f_\pm(r)}
+
r^2\mathrm{d}\Omega_2^2,
\qquad
f_\pm(r)
=
1+{r^2\over\ell^2}
-
{2M_\pm\over r}
+
{Q_\pm^2\over r^2}.
\label{eq:rnads-metric}
\ee
The Maxwell field is the
corresponding Coulomb field on each side. The geometry inside the shell is characterized by a mass and charge $(M_-,Q_-)$, while
the exterior after the shell has passed is 
$(M_+,Q_+)$. For configurations corresponding to black holes, we denote the radial position of the outer horizon on each side by $r_{h,\pm}$, and the surface gravity on each side by $\kappa_\pm$ such that 
\be
f_\pm(r_{h,\pm})
=
0,
\qquad \text{and}\qquad 
\kappa_\pm
=
{1\over2}|f_\pm'(r_{h,\pm})| \geq 0.
\label{eq:initial-subextremal}
\ee
This is the location of the horizon in the maximally extended geometry. Of course, depending on parameters, the horizon might not be part of the physical patch on each side. Notice that, for example, $r_{h,-}$ is the outer horizon of $\mathcal{M}_-$ and not an inner horizon.

The shell is a spherically symmetric pressureless charged dust shell
with proper rest mass $m$ and charge $q$.  In Gaussian normal
coordinates adapted to $\Sigma$, its distributional stress tensor is
\be
T^{\mu\nu}_{\Sigma}
=
{m\over4\pi R^2}\, u^\mu u^\nu \, \delta(\eta),
\label{eq:shell-stress-distribution}
\ee
where $\eta$ is signed proper distance from the shell and $u^\mu$ is
its unit timelike velocity. We choose the sign of $\eta$ so that
$\eta>0$ on ${\cal M}_+$. The unit normal
$n^\mu=(\partial/\partial\eta)^\mu$ therefore points from
${\cal M}_-$ to ${\cal M}_+$, and we denote its limiting values on the
two sides of the shell by $n_\pm^\mu$. It is easy to check that such a shell satisfies the WEC. The Maxwell junction condition fixes the charge jump,
\be
Q_+-Q_-
=
q.
\label{eq:maxwell-junction}
\ee
We take $Q_+>0$ from now on for notational simplicity. The physics corresponding to the opposite sign follows by
charge conjugation. 

The shell trajectory is parametrized by the radial location $r=R(\tau)$ as a function of its proper time $\tau$. 
We introduce the signed kinetic factors
\be
\beta_\pm(R)
\equiv
n_\pm^\mu\nabla_\mu r,
\qquad
\beta_\pm(R)^2=\dot R^2+f_\pm(R).
\label{eq:beta-def}
\ee
The sign is fixed by the
embedding and the normal orientation. In a static patch where the coordinates $t_\pm$ are regular and
$\partial_{t_\pm}$ is future directed, this convention gives $\dot t_\pm=\beta_\pm/f_\pm$ away from a horizon. With the normal orientation above, the angular
Israel junction condition \cite{Israel:1966rt} reduces to
\be
\beta_+
-
\beta_-
=
-{m\over R}.
\label{eq:israel-angular}
\ee
We can solve for $\beta_\pm$ as a function of the position $R$. First eliminate $\dot{R}$ by computing 
$\beta_-^2-\beta_+^2$ and combine this with the Israel junction condition, leading to
\be
\beta_+(R)
=
{\Delta M\over m}
-
{\Delta(Q^2)+m^2\over2mR},
\label{eq:beta-plus-basic}
\ee
where $\Delta M
\equiv
M_+-M_-$ and $\Delta(Q^2)
\equiv
Q_+^2-Q_-^2$. This determines the signed kinetic
factor.  On the physical exterior collapsing branch
relevant to the horizon crossing one has $\beta_+>0$, while the algebraic radial continuation of
\eqref{eq:beta-plus-basic} toward smaller $R$ may change sign, as
discussed below.

\begin{figure}[t!]
    \centering

    \begin{subfigure}[t]{0.48\linewidth}
        \centering
        \includegraphics[width=\linewidth]{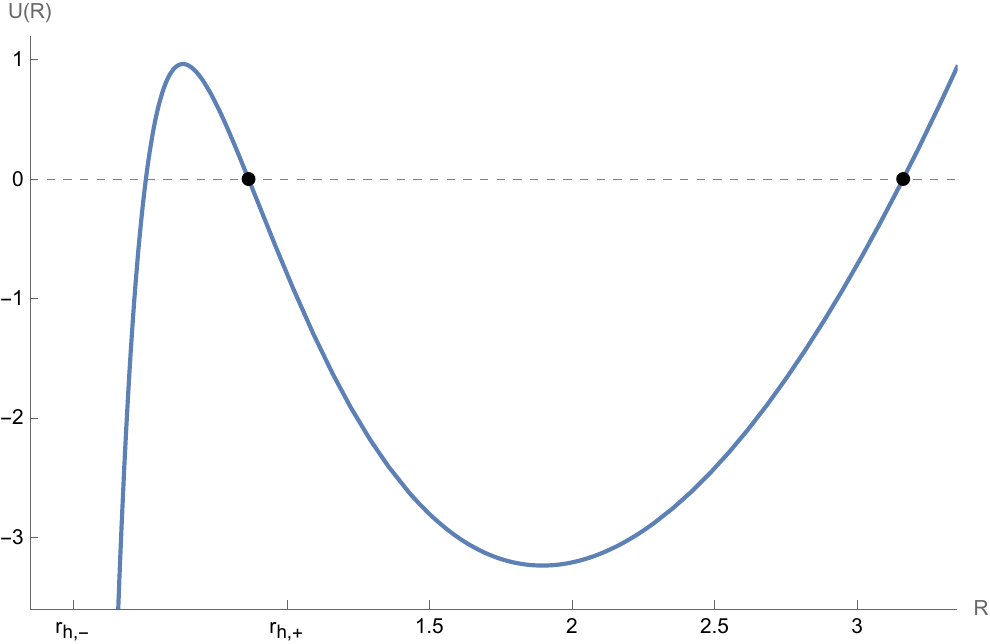}
        \caption{(a) Shell potential in AdS$_4$}
        \label{fig:shell-potential-ads}
    \end{subfigure}
    \hfill
    \begin{subfigure}[t]{0.48\linewidth}
        \centering
        \includegraphics[width=\linewidth]{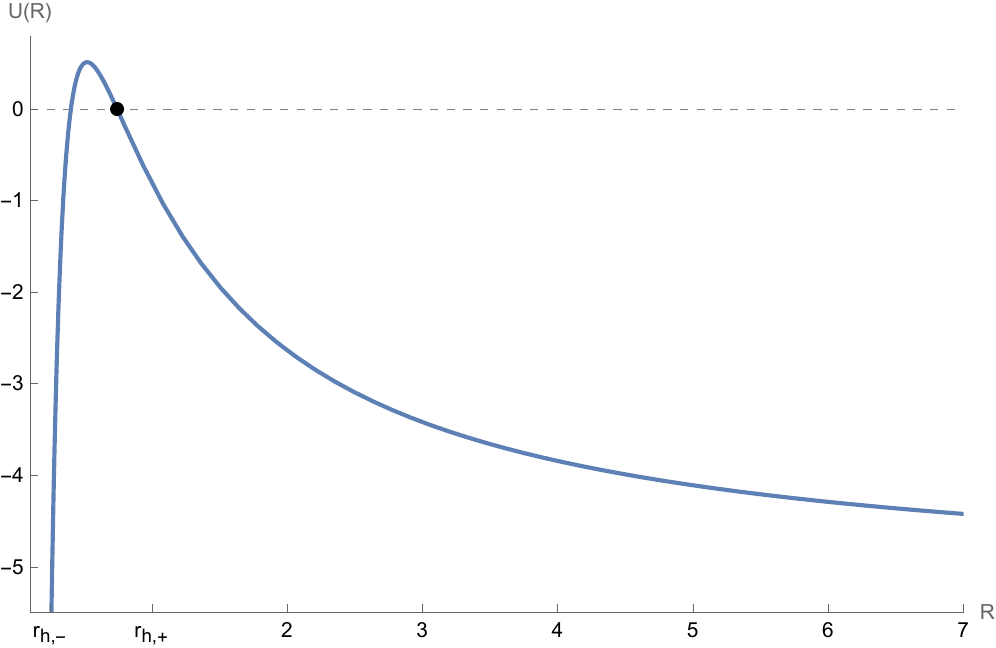}
        \caption{(b) Shell potential in flat space}
        \label{fig:shell-potential-flat}
    \end{subfigure}

    \caption{Shell potential for a specific choice of parameters in (a) asymptotically AdS space with $\ell=1$  and in (b) flat space with $\ell \to \infty$. When the \pro condition is satisfied, the inner turning point is to the left of the final outer horizon. The AdS case has an outer turning point while the flat space case does not.}
    \label{fig:shell-potential}
\end{figure}

We can describe the shell trajectory in the following way. Combining \eqref{eq:beta-def} and
\eqref{eq:beta-plus-basic}, we can find a relationship between $\dot R$ and
$R$
\be
\dot R^2+ U(R)=0,\qquad U(R)
\equiv
f_+(R)-\beta_+(R)^2
\label{eq:radial-potential}.
\ee
This can be interpreted as an energy conservation equation with
potential $U(R)$, with $\beta_+(R)$ understood as the signed radial
function given by \eqref{eq:beta-plus-basic}.  The shell trajectory is
allowed in regions where $U(R)\leq0$, and turning points satisfy
$U(R)=0$. We can determine important qualitative features of the trajectory from the shape of the potential.

Using~\eqref{eq:beta-plus-basic}, the large-radius limit of the potential is
\be
U(R)
=
\frac{R^2}{\ell^2}
+
1
-
\left(
\frac{\Delta M}{m}
\right)^2
+
O(R^{-1}).
\ee
For finite $\ell$, $U(R)\to\infty$, so the AdS boundary lies
outside the allowed region $U\leq0$. Any collapsing branch that is
allowed at smaller radius therefore has a finite outer endpoint.
When $\Delta M/m\gg1$ and this endpoint lies in the asymptotic AdS
region, we have
\be
R_{\text{out}}
\simeq
\ell
\sqrt{
\left(
\frac{\Delta M}{m}
\right)^2
-
1
}
\simeq
\ell\frac{\Delta M}{m}
\gg
\ell.
\ee
Thus a finite-energy massive shell in AdS must be prepared at finite
radius rather than launched from the boundary. If
$R_{\text{out}}$ is a simple root of $U(R)$, the shell may in
particular be released from rest there. In the asymptotically flat
limit, $U(R)=1-(\Delta M/m)^2+O(R^{-1})$, so spatial infinity
belongs to the allowed region only for $\Delta M\geq m$. For
$\Delta M>m$, the incoming shell has asymptotic Lorentz factor
$\gamma_\infty=\Delta M/m>1$ and must be fired inward with finite
kinetic energy instead of being released from rest at infinity. An
example of the potential for AdS and flat space is given in
Figure~\ref{fig:shell-potential}.

\subsection*{Collapse to extremality and the Pr\'oszy\'nski condition}

The analysis so far is generic. We now require the initial geometry to be nonextremal with $\kappa_->0$ and the final geometry after the shell has passed to be
exactly extremal $\kappa_+=0$. Moreover, we require the shell to end up behind the final extremal horizon such that it is part of the physical spacetime.  If $r_e$ denotes the final degenerate horizon, then
\be
r_e = r_{h,+},\qquad \text{such that} \qquad f_+(r_e)
=
0,
\qquad
f_+'(r_e)
=
0.
\label{eq:target-extremality}
\ee
The RN-AdS extremal curve for $M$ versus $Q$ is implicitly parametrized by the extremal radius
\be
Q_{\rm ext}(r)^2
=
r^2
\left(
1+{3r^2\over\ell^2}
\right),
\qquad
M_{\rm ext}(r)
=
r
\left(
1+{2r^2\over\ell^2}
\right).
\label{eq:rnads-extremal-curve}
\ee
The condition that the final geometry is extremal is
\be
Q_+
=
Q_{\rm ext}(r_e),
\qquad
M_+
=
M_{\rm ext}(r_e).
\ee
In the asymptotically flat limit,
$Q_+=M_+=r_e$ for $Q_+>0$.

The shell forms the extremal black hole in the operational sense
relevant to the third law if it crosses the newly formed horizon
$R=r_e$ with nonzero inward proper velocity. That way the extremal horizon is part of the physical $\mathcal{M}_+$ region. Since
$f_+(r_e)=0$, the radial equation gives
$|\dot R(r_e)|=|\beta_+(r_e)|$.  On the physical infalling branch,
$f_+>0$ and $\dot t_+>0$, and therefore $\beta_+>0$, while
$\dot R<0$.  Taking the horizon limit on this connected branch gives
$\dot R(r_e)=-\beta_+(r_e)$.  Finite and nonzero crossing therefore
requires
$\beta_+(r_e)>0$.
The limiting case $\beta_+(r_e)=0$ makes the advanced time at crossing
singular. Using \eqref{eq:beta-plus-basic} together with the extremality
relations gives
\be
\beta_+(r_e)
=
{\Delta M\over m}
-
{\Delta(Q^2)+m^2\over2m r_e}
>
0,\qquad
\Delta M > \frac{\Delta(Q^2)}{2 r_e} + \frac{m^2}{2 r_e}.
\ee
This can be interpreted as demanding that, in order for the shell to fall behind the extremal horizon, the change in total energy is larger than the change in electromagnetic energy plus the self-energy of the shell. This relation can be put into another form using $f_+(r_e) =0$, namely
\be
\beta_+(r_e)
=
{r_e^2f_-(r_e)-m^2\over2mr_e}
>
0,
\qquad
m^2
<
r_e^2f_-(r_e).
\label{eq:generalized-proszynski}
\ee
This is the RN-AdS generalization of the Pr\'oszy\'nski condition~\cite{Proszynski:1983}, and is the most useful form of the classical third-law-violating condition. If the shell is too heavy, or if $r_e$ lies too close
to the old outer horizon so that $f_-(r_e)$ is too small, the shell
turns around before crossing the would-be extremal horizon. When the initial black hole is near extremal and the process can be described within an AdS$_2$ throat, this inequality has a simple interpretation in terms of $\mathrm{SL}(2,\mathbb{R})$ that we will explain in Section~\ref{subsec:near-horizon-jt}. Before doing that, we summarize some important properties of the process: 

\paragraph{Constraint on Matter Field.} Consider the asymptotically flat limit. The final state is extremal and $M_+ = Q_+$. In terms of the initial parameters this is equivalent to $M_- + \Delta M = Q_- + q$. Rearranging this gives the inequality $q = \Delta M + (M_--Q_-)>\Delta M$ since the initial state is nonextremal. By energy conservation $\Delta M$ is the total energy of the shell and therefore $\Delta M > m$. This implies that $q > m$. Thus the shell must be overcharged and supplied with enough kinetic
energy to overcome its electrostatic repulsion. Related charge bounds have recently been derived for smooth
charged scalar collapse to extremality \cite{Schneider:2026zta}. It would be interesting to explore if there is any connection between third-law violations and the weak gravity conjecture \cite{Arkani-Hamed:2006emk}. 

\paragraph{Second Law.} Since the shell satisfies the weak energy condition, Hawking's area law holds \cite{Hawking:1971vc, Bardeen:1973gs}. This does not always imply that $r_{h,-} < r_{h,+}$, which holds only if the outer horizon of the maximally extended spacetime $\mathcal{M}_+$ is part of the physical geometry. In general, this condition requires an inequality similar to \eqref{eq:generalized-proszynski}, with $r_e$ replaced by $r_{h,+}$. When the final horizon is visible, \eqref{eq:generalized-proszynski} together with energy positivity $\Delta M>0$ implies that $r_{h,-}<r_{h,+}$. The thin shell collapse always increases the Bekenstein-Hawking entropy of the black hole~\cite{Bekenstein:1973ur, Hawking:1974sw}.

\begin{figure}[t!]
\centering

\begin{subfigure}[t]{0.49\linewidth}
\centering
\resizebox{\linewidth}{!}{%
\begin{tikzpicture}[
  x=1cm,y=1cm,
  line cap=round,
  line join=round,
  scale=0.8
]
\useasboundingbox (-4.8,-3.8) rectangle (4.7,4.8);

\coordinate (hmL)      at (-4.05,-0.15);
\coordinate (hmSwitch) at (-1.48, 2.13);
\coordinate (hmR)      at ( 1.38, 4.67);

\coordinate (hpPastL)  at (-1.95,-2.65);
\coordinate (Cplus)    at ( 0.46,-0.50);
\coordinate (hpR)      at ( 3.80, 2.48);

\coordinate (Rb)       at (0.34,0.90);
\coordinate (shellIn)  at (2.85,-3.35);
\coordinate (shellOut) at (3.05, 4.48);

\draw[line width=0.70pt] (hmL)--(hmR);
\draw[line width=2.10pt] (hmL)--(hmSwitch);

\draw[line width=0.70pt] (hpPastL)--(Cplus);
\draw[line width=2.10pt] (Cplus)--(hpR);

\coordinate (CplusOutCtrl) at (0.3124,-0.0335);
\coordinate (RbOutCtrl)    at (0.4856, 2.2922);

\draw[
  line width=1.05pt,
  postaction={decorate},
  decoration={
    markings,
    mark=at position 0.17 with {
      \arrow{Latex[length=2.2mm]}
    },
    mark=at position 0.84 with {
      \arrow{Latex[length=2.2mm]}
    }
  }
]
  (shellIn)
  .. controls
       (2.15,-2.75)
       and ($(Cplus)!-2.647!(CplusOutCtrl)$)
     .. (Cplus)
  .. controls
       (CplusOutCtrl)
       and ($(Rb)!-0.313!(RbOutCtrl)$)
     .. (Rb)
  .. controls
       (RbOutCtrl)
       and (1.28,3.05)
     .. (shellOut);

\fill (Rb) circle (1.45pt);

\node[anchor=east] at (-2.78,1.15) {$r_{h,-}$};
\node[anchor=west] at ( 2.15,0.45) {$r_e$};
\node[anchor=east] at ( 0.10,1.05) {$R_b$};
\node[anchor=west] at ( 2.25,-2.10) {$\Sigma$};

\node at (0.00,-3.05) {$\mathcal{M}_-$};
\node at (3.35,-1.10) {$\mathcal{M}_+$};

\end{tikzpicture}%
}
\caption{Finite mass, $m>0$.}
\end{subfigure}
\hfill
\begin{subfigure}[t]{0.49\linewidth}
\centering
\resizebox{\linewidth}{!}{%
\begin{tikzpicture}[
  x=1cm,y=1cm,
  line cap=round,
  line join=round,
  scale=0.8
]
\useasboundingbox (-4.8,-3.8) rectangle (4.7,4.8);

\coordinate (hmL)      at (-4.05,-0.15);
\coordinate (hmSwitch) at (-1.48, 2.13);
\coordinate (hmR)      at ( 1.38, 4.67);

\coordinate (hpPastL)  at (-1.95,-2.65);
\coordinate (Cplus)    at ( 0.64871,-0.33150);
\coordinate (hpR)      at ( 3.80, 2.48);

\coordinate (Rb)       at (-0.45,0.90);
\coordinate (shellIn)  at ( 3.34174,-3.35);
\coordinate (shellOut) at ( 3.56267, 4.48);

\draw[line width=0.70pt] (hmL)--(hmR);
\draw[line width=2.10pt] (hmL)--(hmSwitch);

\draw[line width=0.70pt] (hpPastL)--(Cplus);
\draw[line width=2.10pt] (Cplus)--(hpR);

\draw[
  line width=1.05pt,
  line join=miter,
  postaction={decorate},
  decoration={
    markings,
    mark=at position 0.17 with {
      \arrow{Latex[length=2.2mm]}
    },
    mark=at position 0.84 with {
      \arrow{Latex[length=2.2mm]}
    }
  }
]
  (shellIn)--(Rb)--(shellOut);

\fill (Rb) circle (1.45pt);

\node[anchor=east] at (-2.78,1.15) {$r_{h,-}$};
\node[anchor=west] at ( 2.15,0.45) {$r_e$};
\node[anchor=east] at (-0.69,1.05) {$R_b$};
\node[anchor=west] at ( 2.45,-2.10) {$\Sigma$};

\node at (0.00,-3.05) {$\mathcal{M}_-$};
\node at (3.35,-1.10) {$\mathcal{M}_+$};

\end{tikzpicture}%
}
\caption{Massless limit, $m\to0$.}
\end{subfigure}

\caption{Shell trajectories for $m>0$ and in the massless limit.
In both panels, the shell crosses the final horizon
$r_{h,+}=r_e$ and reaches $R_b$, with
$r_{h,-}<R_b<r_e$. For $m>0$, the shell turns smoothly.
In the $m\to0$ limit, the future-directed continuation joins
the ingoing and outgoing null branches at $R_b$, giving the
sharp turn shown on the right. Thick lines indicate the
corresponding horizon jump.}
\label{fig:bounce}
\end{figure}
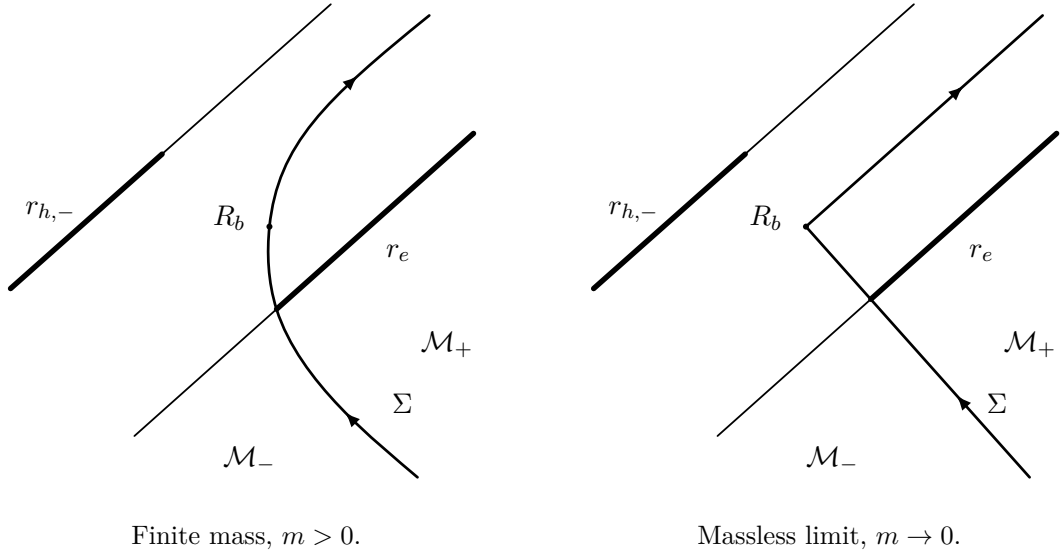

\paragraph{Location of the Bounce.} For the process we are interested in, we have $r_{h,+}=r_e$ and hence $f_+(R)>0$ for $R \neq r_e$. With $r_{h,-}<r_e$, we can use~\eqref{eq:beta-plus-basic} to show that $\beta_+(r_{h,-})<0$. However, the \pro condition gives $\beta_+(r_e)>0$. Since $\beta_+(R)$ is linear in $1/R$, there is a unique zero between the two horizons $(r_{h,-},r_e)$ given by $R_\beta=[\Delta(Q^2)+m^2]/(2\Delta M)$. At this point $U(R_\beta)=f_+(R_\beta)>0$, while $U(r_e)=-\beta_+(r_e)^2<0$. The physical turning point $R_{b}$ connected to the exterior is the largest zero of $U$ in $(R_\beta,r_e)$. This is because it should be the first endpoint of the allowed region encountered by the infalling shell after crossing $r_e$, and $\beta_+>0$ throughout this interval.\footnote{Since
$U(r_{h,-})=-\beta_-(r_{h,-})^2\leq0<U(R_\beta)$, there is also at
least one smaller zero in $[r_{h,-},R_\beta)$. It is
separated from $R_b$ by a forbidden region and is therefore not part
of the collapsing branch connected to the exterior. Generically this root lies
strictly outside $r_{h,-}$, although it can be fine-tuned to merge with $r_{h,-}$.} We therefore find
\be
r_{h,-}<R_b<r_e=r_{h,+}.
\label{eq:physical-bounce-ordering}
\ee
The shell trajectories are illustrated in Figure~\ref{fig:bounce}. For $m>0$, the shell is
timelike and turns smoothly at $R_b$. In the massless limit, the
ingoing and outgoing portions become null and meet in the sharp
turn shown in the right panel.

\paragraph{Elapsed proper and advanced times.} The shell reaches the newly formed
extremal horizon in finite proper and advanced time. Introduce the
ingoing coordinate of the exterior geometry,
$v=t_++R_*$, where
$\mathrm{d}R_*/\mathrm{d}R=f_+(R)^{-1}$. For a shell starting at an allowed radius $R_0\geq r_e$, the elapsed
proper and advanced times are therefore
\be
\Delta\tau=
\int_{r_e}^{R_0}
\frac{\mathrm{d}R}{-\dot{R}(R)},
\qquad
\Delta v=
\int_{r_e}^{R_0}
\mathrm{d}R \frac{\dot{v}(R)}{
-\dot{R}(R)}.
\label{eq:elapsed-shell-times}
\ee
At the horizon, the \pro condition precisely guarantees that $\dot{R}$ is nonzero and negative at $R = r_e$. Using that $\dot{v} = 1/(\beta_+-\dot R)$ one concludes that $\dot{v}$ is also finite at $r_e$. Consequently, the two integrands have finite limits, and there is no divergence arising from integration around $r_e$. If $R_0$ lies strictly inside the allowed region, the two integrals are finite. In AdS, the only additional
case to consider is a shell released from the outer
turning point $R_0 = R_{\text{out}}$, such that $U(R_{\text{out}})=0$. It is easy to see from \eqref{eq:radial-potential} that close to $R_{\text{out}}$ we have $|\dot R|\sim\sqrt{R_{\text{out}}-R}$ while $\dot{v}$ has a finite limit, which leads to an integrable singularity near $R_{\text{out}}$. Therefore $\Delta \tau, \Delta v < \infty$ for any initial condition that violates the third law. For flat space, there is no outer turning point but the integral can diverge if we take $R_0\to\infty$. This divergence reflects that the massive shell is
being traced back to past timelike infinity.

\paragraph{Connection to the Ori model.} The massive shell construction has a useful connection to Ori's charged null dust picture \cite{Ori:1991}. We develop this connection in Appendix \ref{app:Ori_discussion}, beginning with a controlled $m\to0$ limit and then resolving the finite jump into infinitesimal layers. The same discussion relates this limit to the smooth Vlasov construction of Kehle and Unger \cite{Kehle:2024vyt}. The upshot is that the most physically relevant features found for smooth shells can also be reproduced by thin shells. One particular feature invisible in the thin shell is the fact that violating the third law requires the black hole to become superextremal for a brief period of time, which we also explain in that appendix. The region where this happens goes away in the thin shell limit.

\subsection{Initial near-extremal regime: collapse in JT gravity }
\label{subsec:near-horizon-jt}

Although the target final state is extremal, the initial state could in principle be completely generic. With the ultimate goal of studying the fate of the classical process in quantum gravity, we restrict to an initial near-extremal black hole with a nearly
AdS$_2\times S^2$ throat. This allows us to use an effective JT description of the classical process which serves as the starting point for the quantum gravity analysis. The relation between 4d near-extremal black holes and JT quantum gravity can be found in
\cite{Iliesiu:2020qvm,Mertens:2022irh}. From this point of view, the distinction between the asymptotically flat and AdS cases is how this throat is embedded into the asymptotic 4d geometry. We will work mostly in asymptotically AdS space and mention special features that appear in the flat space case.

We take the initial charge to be $Q_-=Q_0$. By near-extremality, we mean that the initial surface gravity, or equivalently the Hawking temperature $T$, is small. In this regime, the geometry sufficiently close to the horizon is well approximated by $\mathrm{AdS}_2\times S^2$. If we denote by $r_0$ the extremal horizon radius at charge $Q_0$, it is convenient to introduce dimensionless coordinates
$
r = r_0 + {\sf L}_2 {\sf r},
$
and $t = {\sf L_2} {\sf t}$, where ${\sf L_2}$ is the AdS$_2$ curvature radius computed below. The geometry close to the horizon $|r-r_0|\ll r_0$ becomes
\beq
\d s^2 \approx {\sf L}_2^2 \left[
-\left({\sf r}^2-(2\pi {\sf T})^2\right)\d {\sf t}^2
+
\frac{\d{\sf r}^2}{{\sf r}^2-(2\pi {\sf T})^2}
\right] + r_0^2 \d \Omega^2,\qquad \mathsf L_2^2
\equiv
{r_0^2
\over
1+6r_0^2/\ell^2},
\eeq
where $T= {\sf T}/{\sf L}_2$. All sans-serif coordinates are naturally associated with the throat. Both radii of curvature $r_0$ and ${\sf L}_2$ are determined by $Q_0$ through~\eqref{eq:rnads-extremal-curve}. In asymptotically flat space, $r_0=\mathsf L_2=Q_0$, whereas in asymptotically AdS space the two radii are generally different. At low temperatures, the ADM energy~\cite{Arnowitt:1962hi, Abbott:1981ff, Henneaux:1984xu} admits the expansion
\beq
M_-
\approx
M_{\text{ext}}(Q_0)
+
2\pi^2r_0\mathsf L_2^2T^2,\qquad S_- \approx S_{\text{ext}}(Q_0) + 4 \pi^2 r_0 {\sf L}_2^2 T,
\label{eq:4d-near-ext-MS}
\eeq
where $M_{\text{ext}}(Q_0)$ and $S_{\text{ext}}(Q_0)$ are the extremal energy and entropy at charge $Q_0$. This expansion is reliable provided that
\beq
T
\ll
\frac{r_0}{\mathsf L_2^2},
\eeq
or equivalently ${\sf T} \ll r_0/{\sf L}_2$. We take this condition as our definition of the near-extremal regime. In flat space this condition is that the temperature is much smaller than $1/Q_0$.

The third-law-violating process we are interested in consists of the collapse of a thin shell that lowers the black hole temperature from a small initial value $T$ to zero. In general, this process may also involve a substantial change in charge, $Q_+\neq Q_-$, and consequently a large increase in the ADM energy. Since the geometries before and after the collapse both possess an approximately $\mathrm{AdS}_2\times S^2$ throat, one might expect the entire process to admit a description in JT gravity. 

\begin{figure}[t]
\centering
\resizebox{0.98\linewidth}{!}{%
\begin{tikzpicture}[
  x=1cm,y=1cm,
  line cap=round,
  line join=round,
  throat fill/.style={fill=black!8,draw=none}
]
\useasboundingbox (-4.95,-4.20) rectangle (15.25,5.35);

\def\panelshiftB{10.10}
\def\HAngle{41.6}

\coordinate (AhmL)      at (-4.05,-0.15);
\coordinate (AhmSwitch) at (-1.48, 2.13);
\coordinate (AhmR)      at ( 1.38, 4.67);

\coordinate (AhpPastL)  at (-2.2865557,-2.1362211);
\coordinate (ACplus)    at ( 0.46,-0.50);
\coordinate (AhpCross)  at ( 0.3294637, 0.1864601);
\coordinate (AhpR)      at ( 3.1429443, 2.6843789);

\coordinate (ARb)       at (0.34,0.90);
\coordinate (AshellIn)  at (2.85,-3.35);
\coordinate (AshellOut) at (3.05,4.48);

\begin{scope}[shift={(AhmL)},rotate=\HAngle]
  \fill[
    throat fill,
    rounded corners=11pt
  ]
    (-0.14,-3.00) rectangle (7.40,0.34);
\end{scope}

\draw[line width=0.70pt]
  (AhmL)--(AhmR);

\draw[line width=2.10pt]
  (AhmL)--(AhmSwitch);

\draw[line width=0.70pt]
  (AhpPastL)--(AhpCross);

\draw[line width=2.10pt]
  (AhpCross)--(AhpR);

\draw[
  line width=1.05pt,
  postaction={decorate},
  decoration={
    markings,
    mark=at position 0.17 with {
      \arrow{Latex[length=2.2mm]}
    },
    mark=at position 0.84 with {
      \arrow{Latex[length=2.2mm]}
    }
  }
]
  (AshellIn)
  .. controls
       (2.15,-2.75)
       and (0.8507,-1.7348)
     .. (ACplus)
  .. controls
       (0.3124,-0.0335)
       and (0.2944,0.4643)
     .. (ARb)
  .. controls
       (0.4856,2.2922)
       and (1.28,3.05)
     .. (AshellOut);

\fill (ARb) circle (1.45pt);

\node[anchor=east]
  at (-2.90,1.20)
  {$r_{h,-}$};

\node[anchor=west]
  at (1.62,0.80)
  {$r_e$};

\node[anchor=east]
  at (0.10,1.05)
  {$R_b$};

\node
  at (0.10,-3.12)
  {$\mathcal{M}_-$};

\node
  at (3.58,-0.95)
  {$\mathcal{M}_+$};

\begin{scope}[shift={(\panelshiftB,0)}]

  \coordinate (hmL)      at (-4.4152,0.2613);
  \coordinate (hmSwitch) at (-1.8452,2.5413);
  \coordinate (hmR)      at ( 1.0148,5.0813);

  \coordinate (hpPastL)  at (-1.95,-2.65);
  \coordinate (Cplus)    at ( 0.46,-0.50);
  \coordinate (hpR)      at ( 3.80,2.48);

  \coordinate (Rb)       at (0.34,0.90);
  \coordinate (shellIn)  at (2.85,-3.35);
  \coordinate (shellOut) at (3.05,4.48);

  \coordinate (shellEntry) at (0.7378974,-1.1224897);

  \begin{scope}[shift={(hmL)},rotate=\HAngle]
    \fill[
      throat fill,
      rounded corners=11pt
    ]
      (-0.18,-0.62) rectangle (7.44,0.62);
  \end{scope}

  \fill[throat fill]
    (shellEntry)
    .. controls
         (0.6231234,-0.9223007)
         and (0.5278575,-0.7144625)
       .. (Cplus)
    .. controls
         (0.3124,-0.0335)
         and (0.2944,0.4643)
       .. (Rb)
    .. controls
         (0.4856,2.2922)
         and (1.28,3.05)
       .. (shellOut)
    .. controls
         (3.2369,4.6460)
         and (4.9993,2.6610)
       .. (4.8124,2.4950)
    -- (shellEntry)
    -- cycle;

  \draw[line width=0.70pt]
    (hmL)--(hmR);

  \draw[line width=2.10pt]
    (hmL)--(hmSwitch);

  \draw[line width=0.70pt]
    (hpPastL)--(Cplus);

  \draw[line width=2.10pt]
    (Cplus)--(hpR);

  \draw[
    line width=1.05pt,
    postaction={decorate},
    decoration={
      markings,
      mark=at position 0.17 with {
        \arrow{Latex[length=2.2mm]}
      },
      mark=at position 0.84 with {
        \arrow{Latex[length=2.2mm]}
      }
    }
  ]
    (shellIn)
    .. controls
         (2.15,-2.75)
         and (0.8507,-1.7348)
       .. (Cplus)
    .. controls
         (0.3124,-0.0335)
         and (0.2944,0.4643)
       .. (Rb)
    .. controls
         (0.4856,2.2922)
         and (1.28,3.05)
       .. (shellOut);

  \fill (Rb) circle (1.45pt);

  \node[anchor=east]
    at (-3.00,1.65)
    {$r_{h,-}$};

  \node[anchor=west]
    at (2.05,0.45)
    {$r_e$};

  \node[anchor=east]
    at (0.10,1.05)
    {$R_b$};

  \node[anchor=west]
    at (2.25,-2.10)
    {$\Sigma$};

  \node
    at (0.15,-3.12)
    {$\mathcal{M}_-$};

  \node
    at (3.55,-1.05)
    {$\mathcal{M}_+$};

\end{scope}

\node[font=\large]
  at (-0.10,-3.90)
  {(a)};

\node[font=\large]
  at (10.00,-3.90)
  {(b)};

\end{tikzpicture}%
}
\caption{
Two near-horizon configurations. The shaded regions indicate the corresponding throat
regions. In panel~(a), $r_{h,-}$, $R_b$, and
$r_e$ lie within a common throat region. In panel~(b), corresponding to a shell with a large charge, the initial and
final throats are separated, and the shell $\Sigma$ probes only the final
extremal throat. The case that can be fully described by JT gravity corresponds
to panel~(a).
}
\label{fig:two_panel_throats}
\end{figure}
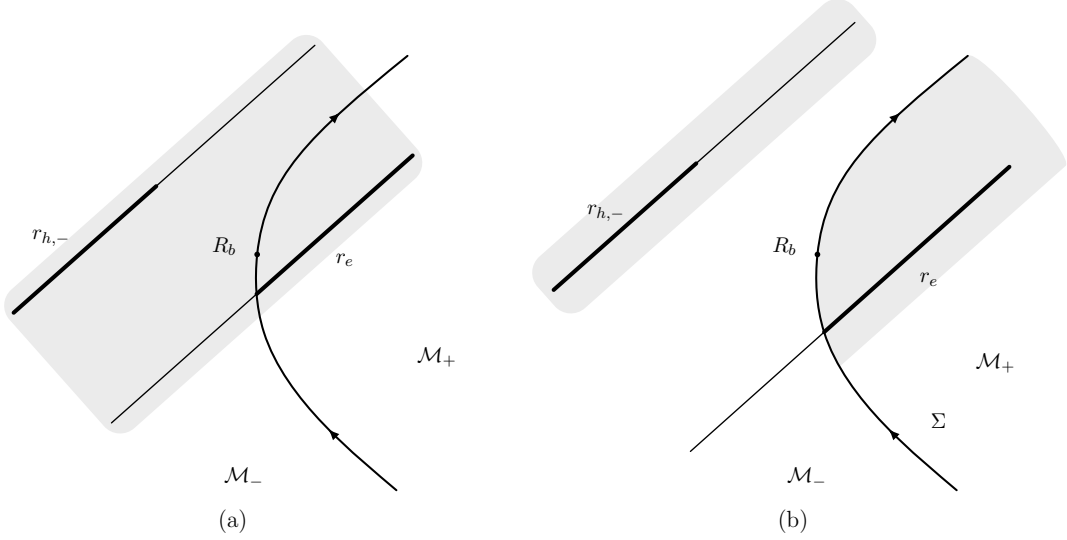

This is not generally the case. Notice first that the bouncing radius $R_b$ is located inside the final horizon but outside the initial horizon $r_{h,-}<R_b$. When the jump in the charge is not small, $R_b$ can lie very far from the initial near-extremal throat. Therefore from the shell's viewpoint only the final geometry is described by an AdS$_2 \times S^2$ throat. Put another way, let $r_e(Q)$ denote the extremal horizon radius at charge $Q$ and define
$
\Delta r_e\equiv r_e(Q_+)-r_e(Q_-),
$ with $r_e(Q_-)=r_0$.
If the jump in the horizon radius is comparable with its initial value $\Delta r_e\gtrsim r_0$, the initial and final $\mathrm{AdS}_2$ throats are separated by an intermediate region not approximated by $\mathrm{AdS}_2\times S^2$. Indeed, the initial throat approximation is valid only for $|r-r_0|\ll r_0$. The full collapse cannot then be formulated within a single $\mathrm{AdS}_2$ throat, and JT gravity cannot be used reliably to describe the complete classical process nor to incorporate its quantum corrections. This distinction is shown in Figure~\ref{fig:two_panel_throats}.

The way out of this issue is clear. We not only have to consider the regime of low initial temperatures but we also need to work with weakly charged shells. If $q$ is small then the change in the black hole charge is also small $Q_+ = Q_0 + \delta Q$, with $q=\delta Q \ll Q_0$. One can easily check that $q\ll Q_0$ guarantees that $\Delta r_e \ll r_0$ and therefore the jump in the horizon location is small. The jump in $r_0$ has a very natural expression when written in terms of the jump in the extremal entropy $S_{\text{ext}} = \pi r_e^2$, namely
\beq
S_{\text{ext}}(Q+\delta Q)\approx S_{\text{ext}}(Q) + 2\pi {\sf e}_0 \delta Q ,~~~~{\sf e}_0 \equiv \frac{Q_0 {\sf L}_2^2}{r_0^2},
\eeq
where ${\sf e}_0$ is the electric field in the throat. For black holes in flat space, this is equivalent to the electric charge ${\sf e}_0 = Q_0$. More precisely, it is the coefficient appearing in the gauge potential close to the throat 
\beq
A \approx {\sf e}_0 ({\sf r}-2\pi {\sf T}) \d {\sf t}.
\eeq
The connection between the charge-dependent extremal entropy and electric field is a universal property of AdS$_2$ throats as explained using the quantum entropy function by Sen \cite{Sen:2005wa, Sen:2008vm}. The process therefore reduces the temperature by a small amount, from $T$ to $0$, and increases the charge by a small amount as well.

But energy conservation also prevents  $q$ from being too small. In the limiting case of a neutral shell with $q=0$ for example, reducing temperature from $T$ to $0$ would reduce the ADM mass and require a negative shell energy $\Delta M <0$. Relatedly, this process would also decrease the entropy of the black hole, violating the area law. Therefore some small amount of charge is necessary in order to violate the third law in a physical way. To analyze this point, it is instructive to expand the extremal energy around $Q_0$ as follows 
\be
M_{\rm ext}(Q_0+\delta Q)
\approx
M_0
+
\mu_0\,\delta Q
+
{(\delta Q)^2\over2K_0},\qquad \mu_0
=
\sqrt{
1+{3r_0^2\over\ell^2}
},
\qquad
{1\over K_0}
=
{3\mathsf L_2^2\over\ell^2r_0},
\label{eq:Mext-charge-expansion}
\ee
where $M_0$ is the extremal mass at charge $Q_0$. This expansion is accurate as long as $\delta Q \ll Q_0$. For black holes in flat space $\mu_0=1$ and $1/K_0 = 0$, since the extremal mass is linear in the charge. The change in the ADM energy $\Delta M$ between the initial near-extremal and final extremal states, which is equal to the energy of the shell, is given by
\beq
\label{eq:Mext-charge-expansion2}
\Delta M
\approx
\mu_0 q
+
{q^2\over 2K_0}
-
2\pi^2 r_0 \mathsf L_2^2 T^2.
\eeq
For $\Delta M$ to be positive it is clear that $q$ cannot be too small. We can also look at the change in horizon radius. The initial and final outer horizon radii are given by 
\beq
r_{h,-} \approx r_0 +2\pi\mathsf L_2^2 T,\qquad r_e \approx r_0 + \frac{{\sf e}_0}{r_0} q.
\eeq
The latter radius is extremal and the simple form of the correction of order $q$ is a consequence of Sen's relation mentioned earlier. Hawking's area law requires that $\frac{{\sf e}_0}{r_0} q \geq 2\pi\mathsf L_2^2 T$ placing a lower bound on $q$. A convenient regime is when $q$ scales with $T$ and these two terms are comparable. In this case it is also true that the second and third terms on the RHS of~\eqref{eq:Mext-charge-expansion2} are of the same order of magnitude. If the lower bound on $q$ is satisfied, then the ADM energy also increases in the process.

For a near-extremal initial black hole and a weakly charged shell, expanding the \pro condition to leading order in $T$ and $q$ gives a stronger constraint than the area-law bound above. Introducing the throat-adapted charge and mass,
\beq
{\sf q} = q {\sf e}_0,~~~~{\sf m}=m {\sf L}_2,
\eeq
the bound becomes
\beq
\mathsf q^2 - \mathsf m^2
>
\left( \frac{\delta S}{2\pi} \right)^2, \qquad \delta S \equiv S_--S_0=4 \pi^2r_0 \mathsf L^2_2 T,
\label{eq:pro-bound}
\eeq
where $S_0 \equiv S_{\rm ext}(Q_0)$. Thus the initial nonextremality barrier on the RHS is quadratic in $T$. This inequality in particular implies that ${\sf e}_0 q > 2\pi r_0 {\sf L}_2^2 T$ which guarantees that the area law is satisfied $r_{e}>r_{h,-}$, but it is strictly stronger when $m>0$. For black holes in flat space ${\sf q}= q Q_0$ and ${\sf m} = m Q_0$, and therefore the quantity on the LHS of the inequality is proportional to $q^2-m^2$, leading to $q>m$. The optimal bound in AdS depends on the black hole through the ratio ${\sf e}_0/{\sf L}_2$, but the condition $q>m$ is still necessary.

The trajectory of the shell has a simple description in this regime, since the potential simplifies into a quadratic function of the radial coordinate
\beq
U(R)
\approx
\left({\sf r}-\frac{{\sf q}}{r_0 {\sf L}_2}\right)^2
-
\left[
\frac{{\sf q}}{{\sf m}}\,\left({\sf r}-\frac{{\sf q}}{r_0 {\sf L}_2}\right)
+
\frac{
{\sf q}^2-{\sf m}^2-\left(\delta S/2\pi\right)^2
}{
2{\sf m}r_0\mathsf L_2
}
\right]^2.
\eeq
This approximates the positive bump close to the horizon in Figure~\ref{fig:shell-potential}. Close to the conformal boundary of the AdS$_2$ throat the potential looks like
\beq
U(R)
\sim
-\Big(
{{\sf q}^2 \over {\sf m}^2}
-1
\Big)
\mathsf {\sf r}^2,
\eeq
so indeed the shape is that of an inverted parabola. The \pro condition implies that the potential becomes arbitrarily negative near the conformal boundary of AdS$_2$, since it requires ${\sf q} > {\sf m}$. This indicates the shell has to be able to propagate freely into the throat. Although there are no turning points close to the boundary of AdS$_2$, there are internal turning points where $U(R)=0$. The result, written in terms of the original radial coordinate, is
\beq
R_b
=
r_0
+
\frac{1}{2r_0}
\Big[
{\sf q}+{\sf m}
+
\frac{
\left(\delta S/2\pi\right)^2
}{
{\sf q}+{\sf m}
}
\Big].\label{eq:ADS2RTURN}
\eeq
We can use the \pro condition to show that this is between the initial and final horizons $r_{h,-}< R_b<r_e$. This is the largest solution for $U(R)=0$. A second solution, which is not a physical branch connected to the exterior, can be shown to be smaller than \eqref{eq:ADS2RTURN}, while generically larger than $r_{h,-}$ unless fine-tuned such that it merges with $r_{h,-}$. This is consistent with what we found near the end of Section~\ref{subsec:classical-thin-shell}.

\subsection*{JT-Maxwell throat description}

In the rest of the section we reinterpret the process in JT gravity. This description of near-extremal black holes has received a lot of attention in recent years and therefore we will be brief and follow the approach in \cite{Iliesiu:2020qvm}. We reproduce the \pro condition as a consequence of conservation of $\mathrm{SL}(2,\mathbb{R})$ charges developed in \cite{Maldacena:2016upp}. This will set up the problem for the quantum gravity analysis.

Consider first the massless modes which take the form $\d s^2
=
\Phi^{-1/2}g_{ab}(x)\,\d x^a \d x^b
+
\Phi(x)\,\d\Omega^2,$ and $A= a_a(x) \d x^a$ where $a,b=t,r$. There is also a massless $p$-wave mode associated with rotations, but for simplicity we work in an ensemble with $J=0$ so it can be ignored. Other partial waves in the metric and photon appear as massive matter in the 2d description. The action of this sector is
\beq
S
=
{1\over4}
\int \d^2x\sqrt{-g}
\left[
\Phi R
+
2\Phi^{-1/2}
+
{6\over\ell^2}\Phi^{1/2}
-
\Phi^{3/2}F_{ab}F^{ab}
\right]+ S_{\text{bdy}}.
\eeq
Now consider the Maxwell field to be fixed to its background value $F^2=-2Q_0^2/\Phi^3$ with charge $Q_0$. Write $\Phi
=
r_0^2+2\phi,$ and $g_{ab}
\to g_{ab}/
(r_0\mathsf L_2^2)$, where $r_0$ and ${\sf L}_2$ are the values associated to charge $Q_0$, and expand at small $\phi$. At $\phi=0$, the power of $r_0$ simply cancels the $\Phi^{-1/2}$ prefactor in the $x^a$ component of the metric. Therefore the new rescaled metric is unit-radius AdS$_2$. The action becomes
\beq
S = {S_{\text{ext}}\over4\pi}
\int\d^2x\sqrt{- g}\,
R + {1\over2}
\int \d^2x\sqrt{-g}\,
\phi\left(R+2\right)+ S_{\text{bdy}}.
\eeq
We continue to denote the rescaled 2d metric by $g_{ab}$. A classical solution with a constant dilaton has $\phi=0$, and the equations of motion set $R=-2$. Both of these are consistent with the 4d extremal throat.

To complete the analysis, we need to incorporate charge fluctuations since we are interested in the collapse of charged matter. We expand around the background field strength as $F \to F + f$ with $f=\d a$. It is also convenient to rescale $a={\sf e_0} {\sf a}$ and $f = {\sf e}_0 {\sf f}$. The resulting action is 
\beq
S = {S_{\text{ext}}\over4\pi}
\int\d^2x\sqrt{- g}\,
R + {1\over2}
\int \d^2x\sqrt{-g}\,
\phi\left(R+2\right)-{Q_0{\sf e}_0\over4}
\int \d^2x\sqrt{-g}\,
{\sf f}_{ab} {\sf f}^{ab}+ S_{\text{bdy}}.\label{eq:JT+maxwell}
\eeq
The simple form of~\eqref{eq:JT+maxwell} is obtained after a field
redefinition. If the Maxwell sector is expanded using fields defined
with respect to the fixed reference charge $Q_0$, it generates terms
linear in ${\sf f}$ as well as mixed terms of the form
$\phi{\sf f}$. To the order relevant here, these terms can be
absorbed, up to boundary terms, by a Maxwell-dependent shift of
$\phi$ together with a Weyl rescaling of the 2d metric. On a solution
with charge $Q$, the dilaton shift moves its origin from
$r_0=r_e(Q_0)$ to the extremal radius $r_e(Q)$, while the Weyl
rescaling changes the unit of the 2d metric to the corresponding
AdS$_2$ radius ${\sf L}_2(Q)$. The resulting variables are therefore charge adapted. In each fixed
charge sector, $\phi$ measures the deviation from the extremal throat
at that charge. In particular, $\phi_h=0$ at extremality and
$S-S_{\rm ext}(Q)=2\pi\phi_h$. We can consequently work directly
with~\eqref{eq:JT+maxwell}, with the understanding that the map
between these variables and the original 4d fields is charge
dependent.

We consider the nearly-AdS boundary conditions introduced in \cite{Almheiri:2014cka, Jensen:2016pah, Maldacena:2016upp, Engelsoy:2016xyb}. These conditions imply that, while the geometry is asymptotically AdS$_2$, a source for the dilaton is turned on, breaking the conformal isometries. Concretely, sufficiently far from the horizon,
${\sf r}\gg2\pi{\sf T}$, we impose $\phi \sim \phi_r {\sf r}$. The precise value of the source $\phi_r$ has to be found from matching the throat geometry to the ambient AdS or flat space. This can be done easily from the RN solution since the size of the transverse sphere is proportional to $\Phi = r^2 \approx  r_0^2 + 2 r_0 {\sf L}_2 {\sf r}$, and therefore classically $\phi = r_0 {\sf L}_2 {\sf r}$, leading to the source $\phi_r = r_0 {\sf L}_2$. The on-shell action of JT gravity with this choice of boundary condition reproduces exactly the near-extremal energy and entropy found in 4d in~\eqref{eq:4d-near-ext-MS}.

Finally we need to include the action of the shell itself. In 2d language the shell is simply a charged particle and therefore has an action
\beq
S_{\text{shell}} = - {\sf m} \int \d s + {\sf q} \int \Big[({\sf r}-2\pi{\sf T})\d {\sf t} + {\sf a}\Big],
\eeq
where the rescaling of $a$ by the electric field produces ${\sf q}$ and the rescaling by the AdS$_2$ radius of the metric produces ${\sf m}$. Notice that the charge couples to the full gauge field. The combinations ${\sf m}$ and ${\sf q}$ are precisely  the mass and charge in a throat with unit AdS$_2$ radius and unit electric field. 

One can derive the equations of motion of this theory, and from there the junction conditions which match with the 4d analysis. Here we instead use the $\mathrm{SL}(2,\mathbb{R})$ symmetry of AdS$_2$ to give a more illuminating derivation of the \pro condition. To make the isometries manifest, we use the embedding formalism following~\cite{Maldacena:2016upp}. We can parametrize AdS$_2$ by $Y^A$ with $A=0,1,2$, with signature $(--+)$ subject to the constraint $Y_A Y^A = -1$. The classical solution for the dilaton at each side of the shell has a very simple form 
\beq
\phi(Y) = Z \cdot Y,~~~\phi^2_h = - Z^2,
\eeq
in terms of an arbitrary vector $Z^A$. We also indicated that its norm gives the value of the dilaton at the horizon $\phi_h$. This can be derived by first computing $(\nabla \phi)^2 = Z^2 + \phi^2$ and then recognizing that $(\nabla \phi)^2=0$ at the horizon. The dilaton, through $Z^A$, selects one unbroken generator of $\text{SL}(2,\mathbb{R})$ which acts as $\delta_\xi Y = Z \times Y$. This vector is timelike outside the horizon since $\xi^2 = \phi_h^2-\phi^2$ and can therefore be identified with time translations in the 4d geometry.

We denote the solution by $Z_-$ before the shell and $Z_+$ after the shell. If we start from a near-extremal black hole at finite temperature we go from $Z_-^2 = - \phi_{h,-}^2$ to extremality with $Z_+^2= 0$. Notice that in our conventions extremality is always at $\phi_{h,+}=0$. In a sector of fixed charge, the entropy above extremality at that charge is given by 
\beq
\delta S = 2 \pi \phi_h,\quad \Rightarrow \quad \delta S_- = 2 \pi \phi_{h,-} \quad \text{and} \quad \delta S_+ = 0.
\eeq
It is shown in \cite{Maldacena:2016upp} that we can identify $Z_\pm$ with the $\text{SL}(2,\mathbb{R})$ charge of the spacetime. Implementing the Noether procedure, the $\text{SL}(2,\mathbb{R})$ charge of the shell is given by
$
Z_{\text{shell}} = {\sf m} Y \times \dot{Y} - {\sf q} Y$, where $\dot Y$ denotes a derivative with respect to the shell proper
time. The norm
$Z_{\rm shell}^2={\sf m}^2-{\sf q}^2$ is the quadratic Casimir of
the classical shell orbit. It is also the invariant that labels the corresponding quantum matter channel. In particular, the scaling dimension of a particle in AdS$_2$ with mass ${\sf m}$ and charge ${\sf q}$ is given by 
\beq
\Delta = \frac{1}{2} \pm \sqrt{ {\sf m}^2 - {\sf q}^2 + \frac{1}{4}}.
\eeq
Conservation of the $\mathrm{SL}(2,\mathbb{R})$ charge across the shell gives\footnote{If we generalize this to smooth shells, we can use this formalism to show that in order for the geometry to interpolate from $Z^2<0$ to $Z^2=0$ there should be an intermediate region with $Z^2>0$. A branch with $Z^2>0$ is the JT gravity analog of superextremality.
The portion of the solution with $Z^2>0$ goes away in the thin shell limit.}
\beq
Z_- = Z_+ + Z_{\text{shell}},\qquad \Rightarrow \qquad
 - \phi_{h,-}^2 = {\sf m}^2- {\sf q}^2  +2 Z_+ \cdot Z_{\text{shell}},
\eeq
where on the right we just took the inner product of the charge conservation equation with itself. We can write $Z_+ \cdot Z_{\text{shell}} = {\sf m} \xi \cdot \dot{Y} - {\sf q} \phi$. If we evaluate this on the extremal horizon we get $Z_+ \cdot Z_{\text{shell}} = {\sf m} \xi \cdot \dot{Y}|_h$ and the RHS is the kinetic energy of the shell when it crosses the horizon which has to be positive $Z_+ \cdot Z_{\text{shell}}>0$. Combining $\text{SL}(2,\mathbb{R})$ conservation with the positivity of the energy of the shell leads to 
\beq
{\sf q}^2 - {\sf m}^2 > \phi_{h,-}^2 = \left(\frac{\delta S}{2\pi}\right)^2,
\eeq
reproducing the near-extremal limit of the \pro bound \eqref{eq:pro-bound}. 
For an initial near-extremal black hole, $\delta S \ll S_0$, while a classical description also requires  $\delta S \gg 1$. Thus, $1 \ll \delta S \ll S_0$. The \pro condition in the form \eqref{eq:pro-bound} therefore leads to ${\sf q}^2 - {\sf m}^2 \gg 1$, implying that third-law violation is necessarily associated with BF-violating fields in the AdS$_2$ throat.

\subsection{Third law violation with Kerr black holes}

Once we understand the process from the throat perspective, we can also apply it to Kerr black holes. For simplicity here we focus on Kerr black holes in asymptotically flat 4d space.
The analogous near-horizon reduction in AdS$_4$ is discussed
in~\cite{Mariani:2025hee}. The first stage is to form a Kerr black hole close enough to extremality, but not extremal, by collapse.

The Kerr black hole is parametrized by $M$ and $J$ and it becomes extremal when $M=a$ with $a=J/M$. For simplicity we consider $J>0$. The near-horizon geometry of a near-extremal black hole can be written as \cite{Bardeen:1999px, Castro:2019crn, Castro:2021csm}
\begin{equation}
\d s_4^2 =
r_0^2(1+\cos^2\theta)
\left[
\d s_2^2
+\d\theta^2
\right]+\frac{4r_0^2\sin^2\theta}{1+\cos^2\theta}
\left[\d\varphi+{\sf A}\right]^2.
\end{equation}
Here $r_0$ is the extremal Kerr horizon radius. The 2d metric and gauge potential are
\beq
\d s_2^2 = -\bigl({\sf r}^2-(2\pi{\sf T})^2\bigr)\,\d{\sf t}^{\,2}
+\frac{\d{\sf r}^{\,2}}{{\sf r}^2-(2\pi{\sf T})^2},~~~{\sf A}=({\sf r}-2\pi{\sf T})\,\d{\sf t}.
\eeq
In the derivation of this near-horizon geometry one can extract the precise relation between the throat-adapted coordinates $({\sf t}, {\sf r})$ and the original asymptotically flat coordinates $(t,r)$ such as the Boyer-Lindquist choice. 

Consider a massless scalar field $\Phi$ in 4d. The conclusions here can be extended to the graviton but the presentation for a scalar field is much simpler. When rewriting the 4d physics in the throat language, it is useful to expand the scalar field in partial waves, specifically in spheroidal harmonics, 
\begin{equation}
\Phi =
\sum_{n\in\mathbb{Z}}\sum_{\ell=|n|}^{\infty}
\phi_{\ell n}({\sf t},{\sf r})\,
S_{\ell n}(\theta)\,
\frac{e^{\i n\varphi}}{\sqrt{2\pi}}.
\end{equation}
More details of this decomposition for scalars and gravitons can be found in \cite{Amsel2009NoDynamics, Dias2009KerrCFT}. The fibration makes derivatives in AdS$_2$ act as $\partial_a \to \partial_a - {\sf A}_a \partial_\varphi$. Therefore this immediately implies ${\sf q} = n$, with $n$ the azimuthal angular momentum number. The harmonic equation is \begin{equation}
-\frac{1}{\sin\theta}\frac{\d}{\d\theta}
\left(\sin\theta\frac{\d S_{\ell n}}{\d\theta}\right)
+
\left[
\frac{n^2}{\sin^2\theta} -\frac{n^2}{4}\cos^2\theta
\right]S_{\ell n}=A_{\ell n} S_{\ell n}.
\end{equation}
Upon reduction the action of the partial wave takes the form of a 2d charged scalar field with parameters
\begin{equation}
{\sf q}^2=n^2,
\qquad
{\sf m}^{\,2}
=A_{\ell n}-\frac{3}{4}n^2.
\end{equation}
An analysis of the eigenvalue problem for the spheroidal harmonics in the large $n$ limit can be found in \cite{Amsel2009NoDynamics,Hod2015Spheroidal}. To leading order the eigenvalues have a simple behavior
$
A_{|n|+p,n} \simeq n^2 + O(n),$ for $p$ fixed. Therefore, at least in this regime, there exist modes that can satisfy the \pro bound
\beq
{\sf q}^2 - {\sf m}^2 \simeq \frac{3n^2}{4} \gg 1.
\eeq
This is intended to be just a proof of principle that at least these waves can be used to produce a collapse from a near-extremal Kerr to an extremal one in finite time. If we have a mode with ${\sf q}^2 - {\sf m}^2>1$ we could take a shell made of a collection of them large enough to overcome the initial $\delta S$ needed in the \pro bound. This can be combined with the analysis of gravitons in \cite{Amsel2009NoDynamics, Dias2009KerrCFT} to show that there are also graviton partial waves that can satisfy the \pro bound, see Section 4.1 of \cite{Amsel2009NoDynamics}, allowing a violation of the third law within vacuum solutions.

\section{BF-violating holographic dictionary and transition amplitude}
\label{sec:dictionary_BF_violation}

We have constructed a third-law-violating process where an initial near-extremal black hole evolves to extremality when a weakly charged shell falls in. In the effective JT description, the main subtlety is that the 2d field describing the shell has to violate the BF bound. In this section we analyze how to quantize such a matter field on AdS$_2$.

\subsection{Supercritical charged shell and violation of the AdS$_2$ BF bound}
\label{subsec:supercritical_shell_BF_bound}

We use the same coordinates introduced in Section~\ref{subsec:near-horizon-jt}.  Near the UV end
of the throat, the unit-radius AdS$_2$ metric and the rescaled Maxwell field are
\be
\mathrm{d}\mathsf s_2^2
=
-\mathsf r^2\mathrm{d}\mathsf t^2
+
{\mathrm{d}\mathsf r^2\over\mathsf r^2},
\qquad {\sf A} = {\sf r} \, \d {\sf t},
\label{eq:unit-ads2-rho-metric}
\ee
as $\mathsf r\to\infty$. The physical 4d metric is rescaled by the AdS$_2$ radius,  and the gauge field has been rescaled by the throat electric field ${\sf e}_0$.  The throat is
glued to the 4d exterior at a large but finite cutoff $\mathsf r
=
\mathsf r_c$ such that $2\pi {\sf T}
\ll
\mathsf r_c
\ll
{r_0\over\mathsf L_2}.$
The upper bound keeps the cutoff inside the near-horizon region.   The boundary condition at this
cutoff is supplied by the 4d gluing problem, as explained in Section~\ref{subsec:near-horizon-jt}.

We represent the spherically symmetric shell channel by an effective
charged scalar in the throat.  This is a local completion of the same
U$(1)$ charge and SL$(2,\mathbb R)$ representation data.  In the localized WKB
limit, its propagator reduces to the charged worldline description.  A
convenient quadratic action is
\be
S_\psi
=
-\int\mathrm{d}^2x\sqrt{-g}\,
\left[|D\psi|^2
+
\mathsf m^2
|\psi|^2
\right],
\label{eq:charged-scalar-action}
\ee
where $D_a
=
\mathsf\nabla_a
-
\mathrm{i}\mathsf q\mathsf {\sf A}_a$ is the covariant derivative.   Varying the action gives the equation of motion
$
(D^2
-
\mathsf m^2
)\psi
=
0$. All derivatives and contractions are taken with the metric
\eqref{eq:unit-ads2-rho-metric}.

We will first solve the equation on rigid AdS$_2$. To do this we expand into fixed-frequency modes
$
\psi(\mathsf t,\mathsf r)
=
e^{-\mathrm{i}\omega\mathsf t}
\, \psi_\omega(\mathsf r),
$
 and the radial equation is
\be
\partial_{\mathsf r}
\left(
\mathsf r^2
\partial_{\mathsf r}\psi_\omega
\right)
+
\left[
{(\omega+\mathsf q\mathsf r)^2\over\mathsf r^2}
-
\mathsf m^2
\right]
\psi_\omega
=
0.
\label{eq:radial-equation-rho}
\ee
At large $\mathsf r$, the leading equation gives the indicial relation
\be
\Delta(\Delta-1)
=
\mathsf m^2
-
\mathsf q^2,
\label{eq:indicial-delta}
\ee
and hence
\be
\Delta_\pm
=
{1\over2}
\pm
\nu,
\qquad
\nu^2
=
{1\over4}
+
\mathsf m^2
-
\mathsf q^2.
\label{eq:Delta-pm-general}
\ee
For vanishing electric coupling,
\eqref{eq:Delta-pm-general}
reduces to the standard AdS dictionary for a scalar field
\cite{Breitenlohner:1982bm,Breitenlohner:1982jf}.
The electric background shifts the effective mass squared by
$-\mathsf q^2$ and can drive the indicial exponent into a
complex, oscillatory regime.  Complex infrared weights of this type
are familiar in charged near-horizon AdS$_2$ systems
\cite{Faulkner:2009wj,Anninos:2019oka}. The analogous formula for a charged fermion leads to $\nu_{\text{fermion}}^2 = {\sf m}^2 - {\sf q}^2$~\cite{Brown:2024ajk}. 

For a BF-satisfying matter field, $\nu$ is real and the two radial branches
have different real falloffs.  The slower and faster branches may then
be identified with the usual Dirichlet source and response. To study the matter field that describes the shell we instead are interested in the regime
\be
\mathsf q^2
>
\mathsf m^2
+
{1\over4}.
\label{eq:principal-series-condition}
\ee
The same supercritical combination controls charged pair production in
electric AdS$_2$
\cite{Pioline:2005pf,Kim:2008xv}, which will be discussed separately
in Section~\ref{sec:discussion}. We therefore write
\be
\nu
=
\mathrm{i}\lambda,
\qquad
\lambda
=
\sqrt{\mathsf q^2
-
\mathsf m^2
-
{1\over4}},
\label{eq:lambda-principal-series}
\ee
so that
\be
\Delta_\pm
=
{1\over2}
\pm
\mathrm{i}\lambda.
\label{eq:Delta-principal-series}
\ee
Notice that in our conventions $\lambda$ is always positive. The two branches now have the same real falloff and differ by an
oscillation in the logarithmic radial coordinate.  The parameter
$\lambda$ can be interpreted as their logarithmic radial momentum, or as labeling a principal-series representation of $\text{SL}(2,\mathbb{R})$. The \pro condition can be rewritten as $\lambda^2 > (\delta S / 2\pi)^2 - 1/4$. The requirement $\delta S  \gg 1$ for initial semiclassical geometries implies that $\lambda \gg 1$, and in particular the $1/4$ term is negligible. Hence the shell is well inside this principal-series
regime. Nevertheless, it is worth pointing out that the fact that $\text{Re}(\Delta)=1/2$ will be important later.

\subsection*{Boundary conditions at the throat}

To study the quantum dynamics of the BF-violating matter field, we need to specify boundary conditions. For BF-satisfying matter, imposing Dirichlet boundary conditions
is equivalent to selecting one of the two solutions of the indicial relation. This is particularly subtle for BF-violating fields. For each fixed frequency, the principal-series solution has the
large-$\mathsf r$ form
\be
\psi_\omega(\mathsf r)
=
J_{\rm in}(\omega)
\mathsf r^{-1/2-\mathrm{i}\lambda}
+
R_{\rm out}(\omega)
\mathsf r^{-1/2+\mathrm{i}\lambda}.
\label{eq:rho-principal-branches}
\ee
Because both branches decay as $\mathsf r^{-1/2}$, imposing Dirichlet boundary conditions is not the same as selecting one over the other.  

To decide on the most natural choice of boundary condition, we can go back to the higher-dimensional picture where we glue the throat to the asymptotic region. In that case, we want to quantize the field such that turning on a source corresponds, in the semiclassical limit, to inserting a particle falling into the black hole. Components of the solution which are purely outgoing or ingoing are distinguished by radial flux, defined as
\be
{\cal F}
=
-\mathrm{i}\mathsf r^2
\left(
\psi^*\partial_{\mathsf r}\psi
-
\psi\partial_{\mathsf r}\psi^*
\right).
\label{eq:radial-kg-flux}
\ee
For the two basis functions,
\be
\begin{aligned}
u_{\rm in}(\mathsf r)
&=
\mathsf r^{-1/2-\mathrm{i}\lambda},
&
{\cal F}[u_{\rm in}]
&=
-2\lambda,
\\
u_{\rm out}(\mathsf r)
&=
\mathsf r^{-1/2+\mathrm{i}\lambda},
&
{\cal F}[u_{\rm out}]
&=
+2\lambda.
\end{aligned}
\label{eq:radial-flux-branches}
\ee
We take positive flux to point toward increasing $\mathsf r$, namely
toward the 4d exterior.  The first branch therefore carries flux from
the exterior into the throat, while the second carries flux back toward
the exterior. The problem we are interested in, a shell falling into the black hole, fixes $J_{\rm in}$ as the
boundary data and determines $R_{\rm out}$ dynamically. 

The boundary condition described in the previous paragraph can be realized as a mixed Dirichlet/Neumann boundary condition where the combination
\be
J_{\rm in}
=
-{\mathsf r^{1-\Delta_{\text{in}}}
\over
2\mathrm{i}\lambda}
\left(
{\sf r} \partial_{{\sf r}} \psi
+
\Delta_{\text{in}}\psi
\right)\bigg|_{{\mathsf r}\to\infty},\qquad \Delta_{\text{in}} = \frac{1}{2} - \i \lambda,
\label{eq:J-in-projector}
\ee
is held fixed at the boundary. For a BF-satisfying field, both $\psi$ and $\psi^*$ would share the same boundary condition, for example Dirichlet. Since $\Delta_{\text{in}}$ is complex, complex conjugation $\psi \to \psi^*$ maps the incoming radial polarization to the outgoing one, flipping the boundary condition. Using
\eqref{eq:rho-principal-branches}, it exchanges $J_{\rm in}$ with
$R_{\rm out}^*$ and is therefore incompatible with fixing both
$J_{\rm in}$ and $J_{\rm in}^*$. 

We instead complexify the field $\psi$ into two independent fields $\psi$ and $\widetilde \psi$ and impose ingoing boundary conditions on both fields. This gives the
incoming quantization used below. This is nonunitary in the sense that the usual reality condition is absent. The Euclidean functional can still be defined. A similar situation appears in the representation of the Schwarzian theory as a particle in an imaginary magnetic field, where the Euclidean path integral is well defined although the continued Lorentzian Hamiltonian is not Hermitian~\cite{Iliesiu:2019xuh}.

Given our choice of mixed boundary conditions, it is important to be clear on the choice of boundary terms. For the boundary conditions $\delta( {\sf r} \partial_{\sf r} \psi + \Delta_{\text{in}} \psi)=0$ and $\delta( {\sf r} \partial_{\sf r} \widetilde\psi + \Delta_{\text{in}} \widetilde\psi)=0$ to be compatible with the variational problem we need to add a boundary term
\be
S_{\psi,\rm bdy}
=
\int\mathrm{d}\mathsf t\,
\left[
{\sf r} \widetilde\psi {\sf r}\partial_{\sf r} \psi
+
{\sf r} \psi {\sf r}\partial_{\sf r} \widetilde\psi
+
{\sf r}\Delta_{\text{in}}\psi\widetilde\psi
\right]_{\mathsf r=\mathsf r_c},
\label{eq:incoming-boundary-term}
\ee
at the cut-off radius ${\sf r}={\sf r}_c$, which at the end of the day we will take to infinity. This boundary condition would not be compatible with the usual reality condition for the fields. 

In a path integral formulation of the problem, the integration is not done over two complex variables $(\psi, \widetilde\psi)$ but over some surface of complex dimension one. One way to choose this contour is to expand fluctuations of $\psi$ and $\widetilde\psi$ around a classical solution in modes $\delta \psi({\sf x}) = \sum_n c_n u_n({\sf x})$ and $\delta \widetilde \psi = \sum_n \widetilde c_n \widetilde u_n({\sf x})$, such that asymptotically $u_n \sim {\sf r}^{-1/2+\i \lambda}R_{\text{out}}$ and $\widetilde u_n \sim {\sf r}^{-1/2+\i \lambda} \widetilde{R}_{\text{out}}$ so that the fluctuations preserve the boundary conditions. We can integrate over the surface $|c_n| = | \widetilde c_n|$ and choose the relative phase such that the integral follows the steepest descent contour of the action \eqref{eq:charged-scalar-action}. We will not attempt to make this construction explicit here since we will not need it.

For BF-satisfying fields, the incoming condition is equivalent to Dirichlet. For BF-violating fields, the two are different and we do not impose Dirichlet boundary conditions. Although Dirichlet conditions might be a reasonable choice of
AdS$_2$ boundary conditions, it is not the natural one arising when the throat is glued into the external geometry. Similar considerations can be found in \cite{Anninos:2019oka}.

\subsection*{Effective action on rigid AdS$_2$}

We can now derive the effective action obtained upon integrating out the matter field in rigid
AdS$_2$.  The details are straightforward and can be found in Appendix \ref{app:fixed_incoming_kernel}. The result is a function of $J({\sf t})$ and $\widetilde J({\sf t})$
\beq
\begin{aligned}
S_{\text{matter}}
={}&
\log Z_{\text{one-loop}} \, + \,
\int_{-\infty}^{\infty} \d {\sf t}_2
\int_{-\infty}^{\infty} \d {\sf t}_1\,
\widetilde J({\sf t}_2)\,J({\sf t}_1) K_{\rm in}({\sf t}_1,{\sf t}_2).
\end{aligned}
\eeq
We denote by $\log Z_{\text{one-loop}}$ the logarithm of the partition function for the matter field with the sources turned off.  The kernel $K_{\rm in}$ is the nonlocal part of the
on-shell matter action after local cutoff-dependent contact terms have
been subtracted. The result is the
principal-series conformal kernel
\beq
K_{\rm in}({\sf t}_1,{\sf t}_2) =  e^{-2\pi \i {\sf T}{\sf q}({\sf t}_2-{\sf t}_1)}
\left(
\frac{\pi{\sf T}}{
\sinh\!\left[
\pi {\sf T}
\left(
{\sf t}_2-{\sf t}_1-\i\epsilon
\right)
\right]
}
\right)^{2\Delta_{\text{in}}}.
\eeq
As explained in Appendix \ref{app:fixed_incoming_kernel}, there is a normalization factor we can absorb by a rescaling of the sources. Notice that the source-dependence of the effective action has the same conformal structure as the one corresponding to matter fields with real $\Delta$. In order to get this result, it is important that the boundary conditions are not Dirichlet. 

We can see metastability of the matter field in two ways. First, the term $\log Z_{\text{one-loop}}$ captures effects such as the Schwinger pair production when $\Delta$ is complex. When the Schwinger channel is open, the amplitude $ Z_{\text{one-loop}}$ need not have unit modulus. The corresponding
one-loop effective action acquires an imaginary part which measures
the decay of the charged background through pair production
\cite{Pioline:2005pf,Kim:2008xv, Brown:2024ajk} and also \cite{WIP_SR}. This gives the near-horizon
contribution to the decay of the charged black hole, though the discharge rate also depends on whether the produced charges propagate into the exterior, as discussed in Section~\ref{sec:discussion}.

The second concerns the source-dependent term. The kernel is complex, and the two endpoints are assigned $\Delta_{\rm in}$. They are not related by Hermitian conjugation. If $J$ couples to $O_{{\rm in},\sf q}$, then $\widetilde J$ couples to a charge-lowering operator $\widetilde{O}_{{\rm in},-\mathsf q}$ which lowers the charge by $\sf q$ and has the same scaling dimension $\Delta_{\rm in}$. This operator is not $O_{{\rm in},\mathsf q}^{\dagger}$, whose scaling dimension is $\Delta^\ast_{\rm in}=1-\Delta_{\rm in}$. At the level of the
principal-series representation, a conjugate with the same scaling dimension can
be related to the ordinary adjoint by a shadow transform
\cite{Anninos:2019oka}. We will not need the explicit form of this
relation.

\subsection{Schwarzian-Maxwell dressing and the energy-basis vertex}
\label{subsec:schwarzian-maxwell-dressing}

Next, we couple the BF-violating matter field of the previous section to 2d gravity. Because JT gravity and a Maxwell field reduce to a boundary Schwarzian mode and a boundary gauge mode, this section determines the precise coupling between the matter field and these boundary degrees of freedom.

\subsection*{Boundary Schwarzian-Maxwell dynamics}

The Schwarzian mode $f(u)$ relates the physical boundary time $u$ to the
fixed-frame AdS$_2$ time $\mathsf t$~\cite{Maldacena:2016upp,Engelsoy:2016xyb, Stanford:2017thb,Mertens:2022irh},
$
\mathsf t
=
f(u),
$
and is defined modulo the SL$(2,\mathbb R)$ isometries of the rigid
AdS$_2$ frame. We denote the Schwarzian coupling by
$\mathsf C$.  The Lorentzian action is
\be
S_{\rm Sch}
=
-\mathsf C
\int \mathrm{d}u\,\{f,u\},
\qquad {\sf C} = r_0 {\sf L}_2,
\label{eq:schwarzian-action}
\ee
where $\{f,u\}
=
{f'''\over f'}
-
{3\over2}
(
{f''\over f'}
)^2 $ is the Schwarzian derivative. As shown in \cite{Maldacena:2016upp} the Schwarzian coupling is equal to the source for the dilaton $\phi_r$ whose value we gave in Section \ref{subsec:near-horizon-jt}. Since we work in units with $G_N=1$, the Schwarzian coupling for a macroscopic black hole is large ${\sf C} \gg 1$. On a static solution, the corresponding geometric energy is
$\mathsf E=-\mathsf C\{f,u\}$ and reproduces both the on-shell action of JT gravity with appropriate boundary terms, as well as the energy of the 4d black hole in the near-extremal regime. The Schwarzian mode is weakly coupled when ${\sf T} \gg 1/{\sf C}$, or equivalently when $T \gg 1/(r_0 {\sf L}_2^2)$.

The 2d Maxwell field also has no bulk local propagating degree of
freedom.  After solving Gauss's law, its reduced phase space is
described by the electric flux $\mathsf Q$, which is constant between
charged insertions. This is related to the 4d definition of the charge by 
\beq
{\sf Q}=(Q-Q_0) {\sf e}_0,\qquad \Rightarrow \qquad {\sf Q}_-=0~~~\text{and}~~~{\sf Q}_+ = {\sf q},
\eeq
where on the right we indicated the charges for the classical process we are interested in. It is  conjugate to the boundary phase $\varphi(u)$~\cite{Mertens:2019tcm, Iliesiu:2019lfc, Kapec:2019ecr, Iliesiu:2020qvm}, such that 
$
[\varphi,\mathsf Q]
=
\mathrm{i}. 
$
The first-order Maxwell boundary action is
\be
S_{\rm Max}
=
\int \mathrm{d}u
\left[
\mathsf Q\varphi'
-
\frac{{\sf Q}^2}{2 {\sf K}}
\right],\qquad {\sf K}
=
\frac{
\ell^2 r_0 {\sf e}_0^2
}{
3{\sf L}_2^3
}.
\label{eq:maxwell-boundary-action}
\ee
The quantization of this mode is equivalent to a free compact scalar in 1d which is straightforward. For flat space ${\sf K} \to \infty$ and the second term disappears completely, making the dynamics of the phase mode completely trivial. 

Let us parametrize the total Hamiltonian of the system by ${\sf H}' = (M-M_{\text{ext}}(Q_0)){\sf L}_2$. This is the excitation energy above the initial extremal value, in units adapted to the AdS throat. It has three contributions in the JT description. The first is the charge-dependent energy due to the zero-temperature chemical potential. The second contribution is from the Schwarzian mode. The third is from the boundary gauge mode. 
\beq
{\sf H}' = \upmu_0 {\sf Q} + {\sf H},\qquad {\sf H}=  {\sf H}_{\text{Sch}} + {\sf H}_{\text{gauge}},\quad\text{with} \quad {\sf H}_{\text{Sch}}= - {\sf C} \{ f,u\},\quad {\sf H}_{\text{gauge}}= \frac{{\sf Q}^2}{2 {\sf K}}.
\label{eq:boundary-total-H}
\eeq
In the first term we have rescaled the zero-temperature chemical potential $\upmu_0 = \mu_0 {\sf L}_2 / {\sf e}_0$ due to both energy and charge rescalings. While ${\sf H}'$ measures the total ADM energy above the reference
extremal energy $M_{\rm ext}(Q_0)$, it is ${\sf H}$ that plays the role of the AdS$_2$ Hamiltonian. Notice that in the flat space limit this transformation is trivial since ${\sf L}_2 = {\sf e}_0 = Q_0$ and $\upmu_0 = \mu_0 = 1$. In flat space, the gauge-mode contribution ${\sf H}_{\text{gauge}}$ is also absent.

\subsection*{Gauge and Schwarzian dressing of the incoming kernel}

We can now take the matter field effective action on rigid AdS$_2$ and couple it to the Schwarzian and gauge mode. This can be obtained in a way completely analogous to the one in \cite{Maldacena:2016upp}. The effective action amounts to the modification of the kernel, written in terms of physical time $u$, given by
\be
K_{\text{in}}(u_1,u_2) = e^{-\mathrm{i}\mathsf q\varphi(u_2)}
e^{+\mathrm{i}\mathsf q\varphi(u_1)}
\left[
{f'(u_2)f'(u_1)
\over
\left(
f(u_2)-f(u_1)-\mathrm{i}0
\right)^2}
\right]^{\Delta_{\rm in}} .
\label{eq:dressed-incoming-bilocal}
\ee
The two-point function obtained by evaluating this bilocal in the
Schwarzian theory has the same form as the exact correlators computed
in \cite{Mertens:2017mtv}, with complex scaling dimension. For a BF-satisfying field the two-point function computes the thermal expectation value of $\langle O^\dagger O \rangle$. For the fixed-incoming principal-series theory under consideration, it instead computes $\left\langle
\widetilde{O}_{{\rm in},-\mathsf q}(u_2)
O_{{\rm in},\mathsf q}(u_1)
\right\rangle $. Next, we discuss how to extract matrix elements of these operators from the exact answers.

\subsection*{Schwarzian energy basis and the principal-series vertex}

We can combine our quantization of the matter field in the BF-violating regime with the quantization of the Schwarzian and gauge mode. The Schwarzian spectrum is labeled by $\mathsf E\geq0$, or
equivalently by the momentum $\mathsf k\geq0$, with
\be
\mathsf E
=
{\mathsf k^2\over2\mathsf C},
\qquad
\mathsf k
=
\sqrt{2\mathsf C\mathsf E},
\qquad
\rho(\mathsf E)
=e^{S_0}
{\mathsf C\over2\pi^2}
\sinh(2\pi\mathsf k).
\label{eq:density}
\ee
The factor of $e^{S_0}$ cancels
from the normalized conditional distributions considered in
Section~\ref{sec:disk-level-protection}, but should not be omitted
when discussing absolute state counts or microscopic level
statistics.

The density \eqref{eq:density} is the Schwarzian Plancherel density
written with respect to $\mathrm{d}\mathsf E$~\cite{Stanford:2017thb,Mertens:2017mtv,Mertens:2022irh}.  The continuous
energy representation is therefore
\be
{\bf 1}
=
\sum_{\mathsf Q}
\int_0^\infty
\mathrm{d}\mathsf E\,
\rho(\mathsf E)
|\mathsf E,\mathsf Q\rangle
\langle\mathsf E,\mathsf Q|.
\label{eq:spectral-resolution}
\ee
The boundary Hamiltonian acts as 
\be
\mathsf H
=
\mathsf H_{\rm Sch}
+
\mathsf H_{\rm gauge}(\mathsf Q),
\qquad
\mathsf H_{\rm Sch}
|\mathsf E,\mathsf Q\rangle
=
\mathsf E
|\mathsf E,\mathsf Q\rangle.
\label{eq:boundary-Hamiltonian}
\ee
The Schwarzian dressing of an operator of weight $\Delta$ gives the
energy-basis kernel~\cite{Mertens:2017mtv}
\be
{\cal W}_\Delta(\mathsf k_f,\mathsf k_i)
=
\frac{
\Gamma(\Delta+\mathrm{i}\mathsf k_f+\mathrm{i}\mathsf k_i)
\Gamma(\Delta+\mathrm{i}\mathsf k_f-\mathrm{i}\mathsf k_i)
\Gamma(\Delta-\mathrm{i}\mathsf k_f+\mathrm{i}\mathsf k_i)
\Gamma(\Delta-\mathrm{i}\mathsf k_f-\mathrm{i}\mathsf k_i)
}{
(2\mathsf C)^{2\Delta}
\Gamma(2\Delta)
}.
\label{eq:four-gamma-energy-kernel}
\ee
This four-Gamma structure can be obtained directly in the 1d Liouville quantum mechanics representation of the Schwarzian theory
\cite{Bagrets:2016cdf,Mertens:2017mtv}. In terms of the positive Liouville coordinate $x>0$, a Schwarzian energy
eigenstate is represented by
\be
\langle x|\mathsf k\rangle
\propto
K_{2\mathrm{i}\mathsf k}(x),
\qquad
K_{2\mathrm{i}\mathsf k}(x)
=
K_{-2\mathrm{i}\mathsf k}(x),
\label{eq:Liouville-energy-state}
\ee
where $K_{2\mathrm{i}\mathsf k}(x)$ is the $K$-Bessel function. The labels $\mathsf k$ and $-\mathsf k$ therefore describe the same
physical energy state. The Liouville overlap which gives the bilocal spectral kernel is 
\be
{\cal I}_{\Delta}(\mathsf k_f,\mathsf k_i)
=
\int_0^\infty
\mathrm{d}x\,
x^{2\Delta-1}
K_{2\mathrm{i}\mathsf k_f}(x)
K_{2\mathrm{i}\mathsf k_i}(x).
\label{eq:Liouville-KK-overlap1}
\ee
The standard Macdonald product identity is
\be
\begin{aligned}
\int_0^\infty
\mathrm{d}x\,
x^{s-1}
K_\mu(x)K_\nu(x)
={}&
{2^{s-3}\over\Gamma(s)}
\Gamma\left(
{s+\mu+\nu\over2}
\right)
\Gamma\left(
{s+\mu-\nu\over2}
\right)
\\
&\times
\Gamma\left(
{s-\mu+\nu\over2}
\right)
\Gamma\left(
{s-\mu-\nu\over2}
\right).
\end{aligned}
\label{eq:Macdonald-product-identity-main}
\ee
The integral on the LHS is absolutely convergent in the domain
\be
\operatorname{Re}s
>
\left|
\operatorname{Re}\mu
\right|
+
\left|
\operatorname{Re}\nu
\right|.
\label{eq:Macdonald-direct-domain}
\ee
For the Schwarzian overlap, $s=2\Delta$,
$\mu=2\mathrm{i}\mathsf k_f$, and
$\nu=2\mathrm{i}\mathsf k_i$.  Since the momenta are real, the
condition reduces to $\operatorname{Re}\Delta>0$.  The physical value
$\Delta_{\rm in}=1/2-\mathrm{i}\lambda$ lies strictly inside this
domain. Moreover, at the physical value,
$x^{2\Delta_{\rm in}-1}=x^{-2\mathrm{i}\lambda}$ has unit modulus. Possible divergences can arise only at $x=0$ and $x=\infty$. However, each $K_{ 2 \mathrm{i} \sf k}(x)$ decays exponentially at large $x$. Near $x=0$, it is bounded for nonzero momentum and grows logarithmically when the momentum vanishes. The overlap is therefore absolutely
convergent for every physical pair
$\mathsf k_i,\mathsf k_f\geq0$. Hence no analytic continuation from
real $\Delta$ is needed, and we see the product identity precisely gives the four-Gamma structure in
\eqref{eq:four-gamma-energy-kernel}. In particular, the arguments of the four numerator Gamma functions have real part $1/2$, while
${\rm Re}(2\Delta_{\rm in})=1$ for the denominator, so none of the Gamma-function arguments encounters a
nonpositive integer pole at the physical point.

For real BF-satisfying $\Delta$, the kernel is understood as a positive spectral
weight.  At complex $\Delta_{\rm in}$ it is generically complex and
cannot itself be interpreted in that way.  For real momenta and
positive $\mathsf C$, complex conjugation gives $\overline{
{\cal W}_{\Delta}(\mathsf k_f,\mathsf k_i)
}
=
{\cal W}_{\Delta^*}(\mathsf k_f,\mathsf k_i)$. Since
$\Delta_{\rm in}^*=1-\Delta_{\rm in}$, this is generally different
from ${\cal W}_{\Delta_{\rm in}}$. In our case, the spectral decomposition of the fixed-incoming bilocal determines
the product of its two endpoint coefficients ${\cal W}_{\Delta_{\rm in}}(\mathsf k_f,\mathsf k_i)
=
\widetilde{\cal V}_{\rm in}(\mathsf k_i,\mathsf k_f)
{\cal V}_{\rm in}(\mathsf k_f,\mathsf k_i)$. Here ${\cal V}_{\rm in}$ is the charge-raising coefficient and
$\widetilde{\cal V}_{\rm in}$ is the charge-lowering coefficient at the other
endpoint. The two coefficients are not related by complex conjugation, since the operator $\widetilde{O}_{{\rm in},-\mathsf q}$ is shadow related to
$O_{{\rm in},\mathsf q}^{\dagger}$ and has dimension
$\Delta_{\rm in}$, while
$O_{{\rm in},\mathsf q}^{\dagger}$ has $\Delta_{\rm in}^*$. 

Hence, we see that the bilocal fixes only the product of the two coefficients. But we can determine the remaining freedom by matching the
vertex at positive real $\Delta$ and requiring a continuous and holomorphic dependence
on $\Delta$ in the same energy basis. We therefore consider a symmetric factorization and define
\be
\widetilde{\cal V}_{\rm in}(\mathsf k_i,\mathsf k_f)
=
{\cal V}_{\rm in}(\mathsf k_f,\mathsf k_i)
=
\left[
{\cal W}_{\Delta_{\rm in}}(\mathsf k_f,\mathsf k_i)
\right]^{1/2}.
\label{eq:principal-series-vertex}
\ee
To justify this choice, first note that for positive real $\Delta$, the square root of the four-Gamma kernel
is the standard Schwarzian energy-basis vertex assigned to each
endpoint of the bilocal
\cite{Mertens:2017mtv,Blommaert:2018oro}. For fixed real momenta,
${\cal W}_{\Delta}$ is analytic and nonzero throughout
$\text{Re}(\Delta)>0$. Its positive root at real $\Delta$ can
therefore be followed continuously to $\Delta_{\rm in}$. As we have shown, the
Liouville overlap is absolutely convergent at $\Delta_{\rm in}$ and
defines the kernel there directly. Continuity from real $\Delta$ is
used only to fix the endpoint factorization and the phase of the
square root.

Second, the energy-basis representation also explains why the complete
four-Gamma kernel appears under the square root. A black hole energy
eigenstate is represented by the complete wavefunction
$K_{2\mathrm{i}\mathsf k}(x)$. It is invariant under
$\mathsf k\to-\mathsf k$ and contains both asymptotic components
$x^{2\mathrm{i}\mathsf k}$ and
$x^{-2\mathrm{i}\mathsf k}$ near $x=0$. The overlap of two complete
energy eigenstates gives the four-Gamma kernel in
\eqref{eq:Liouville-KK-overlap1}. Replacing one of the complete
wavefunctions by a single asymptotic component produces a two-Gamma
expression and introduces a choice of Liouville orientation. Such an
expression is a mixed-basis overlap and cannot be identified as the vertex between the
initial and final black hole energy states used here.

Therefore, for a
transition between fixed initial and final energy eigenstates, we use the following
positive partial transition weight 
\be
\left|
{\cal V}_{\rm in}(\mathsf k_f,\mathsf k_i)
\right|^2
=
\left|
{\cal W}_{\Delta_{\rm in}}(\mathsf k_f,\mathsf k_i)
\right|.
\label{eq:vertex-modulus-kernel}
\ee
This prescription does not impose a unitary completeness relation. We will discuss the case for a coherent superposition of initial energies in Section~\ref{subsec:disk-transition-amplitude}. In this normalization, the matrix element is
\be
\langle
\mathsf E_f,\mathsf Q_f
|
O_{{\rm in},\mathsf q}(0)
|
\mathsf E_i,\mathsf Q_i
\rangle
=
\delta_{\mathsf Q_f,\mathsf Q_i+\mathsf q}
{\cal V}_{\rm in}(\mathsf k_f,\mathsf k_i).
\label{eq:energy-basis-matrix-element}
\ee
This gives the operator dictionary needed for the charge-raising transition.\footnote{We comment on the
gravitational SL$(2,\mathbb R)$ Wilson line interpretation here. In the conventional JT
construction, operator insertions are described by Wilson lines in
discrete-series representations.  For a principal-continuous
representation, the Bessel function has no boundary limit which
isolates a single factor $x^{2\Delta_{\rm in}}$
\cite{Blommaert:2018oro}.  The fixed-incoming boundary condition used
here specifies this Liouville insertion directly, and its overlap is
convergent as shown above.  A conventional principal-series Wilson line realization of this insertion has not been constructed yet.}

\subsection{Transition amplitude and probability}
\label{subsec:disk-transition-amplitude}

We now use the energy-basis matrix element
\eqref{eq:energy-basis-matrix-element}
to define the transition amplitude for general initial preparations
and physical sources.  No condition of final extremality is imposed
at this stage. A general normalizable initial state is 
\be
|\Psi_i\rangle
=
\sum_{\mathsf Q_i}
\int_0^\infty
\mathrm{d}\mathsf E_i\,
\rho(\mathsf E_i)
\psi_i(\mathsf E_i,\mathsf Q_i)
|\mathsf E_i,\mathsf Q_i\rangle.
\label{eq:initial-state}
\ee
With the spectral convention in
\eqref{eq:spectral-resolution}, this state is normalized when
\be
\sum_{\mathsf Q_i}
\int_0^\infty
\mathrm{d}\mathsf E_i\,
\rho(\mathsf E_i)
\left|
\psi_i(\mathsf E_i,\mathsf Q_i)
\right|^2
=
1.
\label{eq:initial-normalization}
\ee
We write the charge-sector decomposition as a discrete sum.  When a charge packet is broad compared with the
elementary charge spacing, the sum may instead be approximated by a
continuum integral. 

The corresponding physical probability density with respect to the
ordinary energy measure $\mathrm{d}\mathsf E_i$ is
\be
p_i(\mathsf E_i,\mathsf Q_i)
\equiv
\rho(\mathsf E_i)
\left|
\psi_i(\mathsf E_i,\mathsf Q_i)
\right|^2.
\label{eq:pi-definition}
\ee
The transition formula below only requires normalizability in
this measure. Note that it only normalizes the initial preparation before the source is applied and places no normalization condition on the partial transition weight below. Furthermore, finiteness of its integral also depends on the frequency decay of the source. We will see in Section~\ref{subsec:finite-time-targeting} that a smooth finite-width source supplies the required decay.

The Hamiltonian without the source \eqref{eq:boundary-Hamiltonian} defines the energy basis used above.
The source
deformation is
\be
\mathsf H_J(u)
=
J(u)O_{{\rm in},\mathsf q}(u)
+
\widetilde{J}(u)\widetilde{O}_{{\rm in},-\mathsf q}(u).
\label{eq:source-Hamiltonian}
\ee
We will not need to specify $\widetilde{J}$ further for our problem, but in general one should. The operator $\widetilde O_{{\rm in},-\mathsf q}$ is the
charge-lowering partner in the fixed-incoming problem. The source deformation is therefore not
Hermitian in the usual sense. We work to leading order in $J$ and focus on the charge-raising sector
$\mathsf Q_f=\mathsf Q_i+\mathsf q$. Only
$J O_{{\rm in},\mathsf q}$ contributes at first order to this sector.

We evaluate the final spectral weight after the source profile has
passed. For a localized source, its Fourier transform is
\be
J(\omega)
\equiv
\int_{-\infty}^{\infty}
\mathrm{d}u\,
J(u)e^{\mathrm{i}\omega u}.
\label{eq:source-fourier-general}
\ee
The integration variable $u$ here is the boundary time at which the source acts. If the final energy were measured while the source was still active, the upper limit should be replaced by the measurement time. But once the source has passed, subsequent free evolution would only change the phases in the energy basis and would not be relevant to the final spectral weight. We work in a regime where such a description is appropriate.

For a transition between energy-charge basis states, time evolution selects the frequency
\be
\omega_{fi}
=
\left[
\mathsf E_{f}
+
\mathsf E_{\rm gauge}(\mathsf Q_f)
\right]
-
\left[
\mathsf E_{i}
+
\mathsf E_{\rm gauge}(\mathsf Q_i)
\right].
\label{eq:nu-general}
\ee
This is the boundary-energy difference for the pair of
states appearing in the amplitude. The corresponding total boundary-energy
difference is $\omega_{fi}+\upmu_0\mathsf q$. This transition frequency should be distinguished from the central frequency $\omega_\ast$ chosen for a particular source profile, where a classical trajectory approximation selects one value of $\omega_\ast$. Finally, the quantum amplitude integrates over all possible
$\mathsf E_i$ and produces a transition amplitude for every
$\mathsf E_f$.

Using
$O_{{\rm in},\mathsf q}(u)
=
e^{\mathrm{i}\mathsf H u}
O_{{\rm in},\mathsf q}(0)
e^{-\mathrm{i}\mathsf H u}$,
the energy-basis matrix element becomes
\be
\begin{split}
\langle
\mathsf E_f,\mathsf Q_f
|
O_{{\rm in},\mathsf q}(u)
|
\mathsf E_i,\mathsf Q_i
\rangle
={}&
e^{\mathrm{i}\omega_{fi}u}
\delta_{\mathsf Q_f,\mathsf Q_i+\mathsf q}
{\cal V}_{\rm in}(\mathsf k_f,\mathsf k_i).
\end{split}
\label{eq:time-dependent-matrix-element}
\ee
The first-order charge-raising amplitude is therefore
\be
\begin{split}
{\cal A}_{J}(\mathsf E_f,\mathsf Q_f)
={}&
-\mathrm{i}
\int_{-\infty}^{\infty}
\mathrm{d}u\,
J(u)
\langle
\mathsf E_f,\mathsf Q_f
|
O_{{\rm in},\mathsf q}(u)
|
\Psi_i
\rangle
\\
={}&
-\mathrm{i}
\sum_{\mathsf Q_i}
\int_0^\infty
\mathrm{d}\mathsf E_i\,
\rho(\mathsf E_i)
\psi_i(\mathsf E_i,\mathsf Q_i)
\delta_{\mathsf Q_f,\mathsf Q_i+\mathsf q}
{\cal V}_{\rm in}(\mathsf k_f,\mathsf k_i)
\int_{-\infty}^{\infty}
\mathrm{d}u\,
J(u)e^{\mathrm{i}\omega_{fi}u}
\\
={}&
-\mathrm{i}
\sum_{\mathsf Q_i}
\int_0^\infty
\mathrm{d}\mathsf E_i\,
\rho(\mathsf E_i)
\psi_i(\mathsf E_i,\mathsf Q_i)
J(\omega_{fi})
\delta_{\mathsf Q_f,\mathsf Q_i+\mathsf q}
{\cal V}_{\rm in}(\mathsf k_f,\mathsf k_i).
\end{split}
\label{eq:master-amplitude}
\ee
We define the differential transition probability density in the specified final charge sector by
\be
P(\mathsf E_f,\mathsf Q_f)
=
\rho(\mathsf E_f)
\left|
{\cal A}_{J}(\mathsf E_f,\mathsf Q_f)
\right|^2.
\label{eq:master-probability-opening}
\ee
Thus
$P(\mathsf E_f,\mathsf Q_f)\mathrm{d}\mathsf E_f$
is the leading-order transition probability into the interval
$[\mathsf E_f,\mathsf E_f+\mathrm{d}\mathsf E_f]$. When finite, its integral over $\mathsf E_f$ is the partial transition weight into that charge sector and retains the overall source
and operator normalization. Since it includes no contributions from other charge sectors or from the case
in which no transition occurs, it is not normalized to unity as a distribution in $\mathsf E_f$.  Normalized
conditional distributions, in which the overall fixed-channel
factors cancel, will be introduced when we specialize to the
classically targeted source in
Section~\ref{subsec:finite-time-targeting}.

Finally, let us briefly comment on the role of ${\cal W}_{\Delta_{\rm in}}(\mathsf k_f,\mathsf k_i)$ here. For a specified final charge, the charge constraint fixes $\sf Q_i=\sf Q_f- \sf q$ already, so expanding the modulus square in
\eqref{eq:master-probability-opening} gives
\be
\begin{split}
P(\mathsf E_f,\mathsf Q_f)
={}&
\rho(\mathsf E_f)
\int_0^\infty
\mathrm{d}\mathsf E_i\,
\mathrm{d}\mathsf E_i'\,
\rho(\mathsf E_i)
\rho(\mathsf E_i')\,
\overline{
\psi_i(\mathsf E_i,\mathsf Q_i)
}
\psi_i(\mathsf E_i',\mathsf Q_i)
\\
&\times
\overline{
J(\omega_{fi})
}
J(\omega_{fi'})
\overline{
{\cal V}_{\rm in}(\mathsf k_f,\mathsf k_i)
}
{\cal V}_{\rm in}(\mathsf k_f,\mathsf k_i').
\end{split}
\label{eq:master-probability-expanded}
\ee
Here $\omega_{fi'}$ and $\mathsf k_i'$ are obtained by replacing
$\mathsf E_i$ with $\mathsf E_i'$ in their definitions. These overlines denote ordinary complex conjugation, not the
independent source $\widetilde J$ or the coefficient
$\widetilde{\mathcal V}_{\rm in}$ in the fixed-incoming bilocal. They just come from complex conjugating the charge-raising amplitude. For a coherent initial state, $\mathsf E_i$ and $\mathsf E_i'$ are
independent integration variables. The off-diagonal terms retain the
relative phase of the square-root vertex and cannot in general be
written using $|{\cal W}_{\Delta_{\rm in}}|$ alone. Individual
off-diagonal terms need not be positive, while the full expression is
real and nonnegative because it comes from a modulus square.

When the initial energy is fixed, the two energy arguments coincide
and the vertex factor reduces to
\be
\overline{
{\cal V}_{\rm in}(\mathsf k_f,\mathsf k_i)
}
{\cal V}_{\rm in}(\mathsf k_f,\mathsf k_i)
=
\left|
{\cal V}_{\rm in}(\mathsf k_f,\mathsf k_i)
\right|^2
=
\left|
{\cal W}_{\Delta_{\rm in}}(\mathsf k_f,\mathsf k_i)
\right|.
\label{eq:fixed-energy-vertex-product}
\ee
The same situation occurs for an initial density matrix that is
diagonal in the energy basis. In Section~\ref{sec:disk-level-protection}, we adopt this fixed initial energy
specialization.

\section{Quantum protection of the extremal edge}
\label{sec:disk-level-protection}

The preceding construction allows us to study the fate in quantum gravity of the
classical third law violation with thin shells. In this section we use that formalism to compute the probability distribution over energies of the final state, and study how quantum effects suppress the extremal edge.

\subsection{The classical targeted thin shell as a quantum transition}
\label{subsec:finite-time-targeting}

We now apply the general transition formula of
Section~\ref{subsec:disk-transition-amplitude}
to the process which motivated the construction. Let us briefly recall some features of the connection to the classical 4d setup. We denote the initial energy above extremality, measured in throat units, by
${\sf E}_- = 2\pi^2 {\sf C}{\sf T}^2$, where ${\sf C}=r_0{\sf L}_2$ is the Schwarzian coupling and ${\sf T}$ is the initial black hole temperature.

For simplicity, for most of the discussion we consider an initial energy eigenstate with energy ${\sf E}_i = {\sf E}_-$ and charge ${\sf Q}_-=0$, 
\beq
| \Psi_i \rangle = | {\sf E}_-, {\sf Q}_-=0\rangle.
\eeq
With this choice, both the integral over ${\sf E}_i$ and the sum over ${\sf Q}_i$ collapse to a single term in~\eqref{eq:master-amplitude}. To ensure that the initial state is semiclassical, we take ${\sf C}{\sf E}_- \gg 1$. When relevant, we will also comment on how the results change for initial states that are not energy or charge eigenstates.

Given the initial state $|\Psi_i\rangle$ we can turn on a source $J(u)$ and evaluate the probability that at late times the black hole will be found in a state of energy ${\sf E}_f$. The general expression can be found in \eqref{eq:master-probability-expanded} and in our case it simplifies to
\be
P(\mathsf E_f)
=
\rho(\mathsf E_f) \, |J(\omega_{fi})|^2\,  | \mathcal{V}_{\rm in}({\sf k}_f, {\sf k}_-)|^2,\qquad \omega_{fi} = {\sf E}_f + \frac{{\sf q}^2}{2 {\sf K}} - {\sf E}_-.
\label{eq:master-probability-opening-sec4}
\ee
The spectral factor on the RHS, written in terms of ${\sf k}$ such that the Schwarzian energy is ${\sf k}^2/2 {\sf C}$, can be simplified into
\be
\begin{aligned}
\rho({\sf E}_f)\left|
{\cal V}_{\rm in}(\mathsf k_f,\mathsf k_i)
\right|^2
={}&
{\sinh(2\pi {\sf k}_f) \over 4}
\sqrt{
{\sinh(2\pi\lambda)\over 2\pi\lambda}
}
\\
&\times
\Big[
\cosh\!\left(\pi(\lambda+\mathsf k_f+\mathsf k_i)\right)
\cosh\!\left(\pi(\lambda+\mathsf k_f-\mathsf k_i)\right)
\\
&\hspace{2.5em}\times
\cosh\!\left(\pi(\lambda-\mathsf k_f+\mathsf k_i)\right)
\cosh\!\left(\pi(\lambda-\mathsf k_f-\mathsf k_i)\right)
\Big]^{-1/2},
\end{aligned}
\label{eq:principal-series-vertex-absolute-square-sec4}
\ee
with $e^{S_0}$ omitted. With this choice, we evaluate the transition at fixed initial energy and charge. Any overall normalization of the continuum energy eigenstate cancels from the conditional distribution in ${\sf E}_f$. Notice that although we fix the initial energy to be ${\sf E}_-$, the final energy ${\sf E}_f$ remains undetermined. A shell insertion localized in boundary time necessarily has a finite frequency width, and therefore produces a distribution over final energies. We are thus led to consider a wavepacket source for the shell.  

\subsection*{Smeared shell insertion}

We next give a description of the wavepacket source that in the classical limit reproduces the process in Section~\ref{sec:classical_thin_shell}. Assume that we want to insert the shell at time $u_*$ with a central frequency $\omega_*$.  We set $\hbar=1$ for now, so that frequency and
energy have the same units. In particular, 
$\omega$ is numerically equal to the corresponding boundary-energy
transfer. Consider a profile 
\be
J(u)
=
J_0\,
g_\sigma(u-u_\ast)
e^{-\mathrm{i}\omega_\ast u},
\label{eq:J-source-profile}
\ee
where $g_\sigma(x)$ is a function with a peak at $x=0$ and a width parametrized by $\sigma$ that is normalized $\int \d x | g_\sigma(x)|^2=1$. A convenient choice for explicit calculations is a Gaussian profile given by
\be
g_\sigma(u-u_\ast)
=
{1\over(\pi\sigma^2)^{1/4}}
\exp\left[
-{(u-u_\ast)^2\over2\sigma^2}
\right],
\label{eq:g-sigma-def}
\ee
where $\sigma$ gives a measure of the temporal spread of the wavepacket. The Fourier transform of the source for this profile is
\be
\begin{aligned}
J(\omega)
=
J_0 \sqrt{2\sigma}\pi^{1/4}
\exp\left[
-{\sigma^2\over2}
(\omega-\omega_\ast)^2
\right]
e^{\mathrm{i}(\omega-\omega_\ast)u_\ast},
\end{aligned}
\label{eq:source-fourier}
\ee
where we follow the conventions of the Fourier transform given in~\eqref{eq:source-fourier-general}. The source consequently has a frequency width given by
$\Delta\omega\sim1/\sigma$. The overall normalization of the source can be fixed from the incoming radial flux discussed in Section~\ref{subsec:supercritical_shell_BF_bound}. For the incoming component we have ${\cal F}[J(u)u_{\rm in}]
=
|J(u)|^2{\cal F}[u_{\rm in}]
=
-2\lambda |J(u)|^2$. Since
$\int \mathrm{d}u\,|g_\sigma(u)|^2=1$, normalizing the incoming packet
to unit number flux gives
\be
-\int \mathrm{d}u\,{\cal F}[J(u)u_{\rm in}]
=
2\lambda |J_0|^2=1 \quad \implies \quad
|J_0|^2
=
{1\over 2\lambda}.
\ee
This fixes the normalization of the incoming shell packet.\footnote{In Section~\ref{subsec:supercritical_shell_BF_bound} we absorbed the coefficient of
the rigid AdS$_2$ kernel into the source. Here we fix this convention
by matching the source to the incoming flux. One can
keep the normalization factor displayed in Appendix~\ref{app:fixed_incoming_kernel} explicit and
compensate it by the corresponding rescaling of the source and
energy-basis vertex. This does not affect the discussion about the normalized final energy
distributions below.
} 

The frequency $\omega$ is conjugate to the boundary time $u$. We use the Hamiltonian \eqref{eq:boundary-Hamiltonian} in throat units with
$\upmu_0{\sf Q}$ subtracted.
In the classical approximation, its central value $\omega_\ast$ is
chosen to equal the corresponding boundary-energy transfer between the
initial near-extremal and the classically targeted final black hole, namely 
\be
\omega_\ast
\equiv 
 \Big[{\sf E}_{+} + {\sf E}_{\text{gauge}}({\sf Q}_+) \Big] - \Big[ {\sf E}_{-} + {\sf E}_{\text{gauge}}({\sf Q}_-)\Big].
\label{eq:total_bdy_energy_transfer}
\ee
Notice that on the RHS we have the classical target ${\sf E}_+$, and not ${\sf E}_f$, which is the argument in the quantum probability distribution. For the process we are interested in, with ${\sf Q}_-=0$ and ${\sf Q}_+={\sf q}$ and with ${\sf E}_+=0$, we have 
\beq
\omega_\ast = \frac{{\sf q}^2}{2{\sf K}} - {\sf E}_-.
\eeq
$\omega_\ast$ does not need to be positive. For example, in flat space it is given by $\omega_\ast = - {\sf E}_-$ since ${\sf K} \to \infty$. Instead, the total ADM energy transfer measured asymptotically includes the contribution from the background chemical potential and remains positive for the classical process, as explained in Section~\ref{subsec:near-horizon-jt}.

It is instructive to take the limit $\sigma\to0$ with a fixed
nonzero value of $\int \mathrm{d}u\,J(u)$, so that the wavepacket
becomes fully localized in time and see what goes wrong. For the Gaussian profile this would instead
require $J_0\propto\sigma^{-1/2}$.
In this limit $|J(\omega)|^2$ is frequency independent and therefore
introduces no dependence on ${\sf E}_f$ when evaluated at
$\omega_{fi}$. The high-energy behavior of the instantaneous
transition is then controlled by the density of states and matrix
elements
\be
F(\mathsf E_f)
\equiv
\rho(\mathsf E_f)
\left|
{\cal V}_{\rm in}(\mathsf k_f,\mathsf k_-)
\right|^2,
\label{eq:high-energy-control}
\ee
Using the Schwarzian density of states and the explicit expression for $| \mathcal{V}_{\text{in}}|^2$ given in \eqref{eq:principal-series-vertex-absolute-square-sec4}, we can determine the behavior of the probability at large final energy, where
${\sf k}_f\gg{\sf k}_-,\lambda$, namely
\be
\left|
{\cal V}_{\rm in}(\mathsf k_f,\mathsf k_-)
\right|^2
\sim
{2\pi^2\over \mathsf C}
\sqrt{
{\sinh(2\pi\lambda)\over 2\pi\lambda}
}
e^{-2\pi\mathsf k_f},
\qquad
\rho(\mathsf E_f)
\sim
{\mathsf C\over4\pi^2}
e^{2\pi\mathsf k_f},\qquad {\sf k}_f \to \infty.
\label{eq:rho-UV}
\ee
The exponential factors cancel and the full probability distribution $P({\sf E}_f)$ 
becomes independent of the final energy ${\sf E}_f$ at the UV end of the spectrum, which we refer to as a UV plateau. The probability distribution is therefore non-normalizable. This is generically expected for a local insertion.

The same behavior occurs for generic initial superpositions of energy eigenstates. For example, consider a normalizable wavefunction $\psi({\sf E}_i)$ with compact support in energy, so the large-${\sf k}_f$ expansion is uniform over the initial states. The combination of amplitudes that appears in~\eqref{eq:master-probability-expanded} is 
\beq
\overline{{\cal V}_{\rm in}(\mathsf k_f,\mathsf k_i)}\,  {\cal V}_{\rm in}(\mathsf k_f,\mathsf k_i')
\eeq
Using the Stirling approximation for the Gamma functions in $\mathcal{V}_{\rm in}({\sf k}_f, {\sf k}_i)$, we find that its leading large-${\sf k}_f$ behavior is independent of ${\sf k}_i$
\beq
{\cal V}_{\rm in}(\mathsf k_f,\mathsf k_i)
\, \sim \,
\frac{2\pi}{\sqrt{2\mathsf C}}\,
\frac{e^{-\pi\mathsf k_f}}
{\sqrt{\Gamma(1-2\mathrm{i}\lambda)}}\,
\left(
\frac{\mathsf k_f^2}{2\mathsf C}
\right)^{-\mathrm{i}\lambda}.
\label{eq:limit-of-Vin}
\eeq
Therefore the leading dependence on ${\sf k}_f$ cancels against the
density of states as in \eqref{eq:rho-UV}, even when we take the product with different ${\sf k}_i$. Notice that for a localized source $\overline{J(\omega_{fi})}J(\omega_{fi'})
\propto
e^{\mathrm{i}({\sf E}_i-{\sf E}_i')u_\ast}$, so this term also does not introduce extra dependence on final energy. Using the source profile in \eqref{eq:source-fourier}, its phase gives
$J(\omega_{fi})\propto e^{-\mathrm{i}{\sf E}_i u_\ast}$ up to a
factor independent of the initial energy. The coefficient of the resulting UV behavior is proportional to
\be
\left|
\int_0^\infty
\mathrm{d}{\sf E}_i\,
\rho({\sf E}_i)\,
\psi_i({\sf E}_i)\,
e^{-\mathrm{i}{\sf E}_i u_\ast}
\right|^2.
\ee
This can still vanish for special coherent initial states through destructive interference, but it is generically nonzero, making the final distribution non-normalizable.

As expected, taking $\sigma>0$ resolves this issue. This happens because now we have a nontrivial profile for $|J(\omega)|^2$ in \eqref{eq:master-probability-opening-sec4}. Since $\omega_{fi}-\omega_\ast={\sf E}_f-{\sf E}_+$, \eqref{eq:source-fourier} gives a multiplicative factor
$e^{-\sigma^2({\sf E}_f-{\sf E}_+)^2}$, which suppresses the large ${\sf E}_f$ tail, making the probability distribution normalizable.

\subsection*{Relation to the localized shell}

The finite-$\sigma$ transition constructed above is therefore well defined for
any $0<\sigma<\infty$. But we can ask what range of $\sigma$ is physically reasonable for the collapse process. Additional conditions enter if the source is
also required to approximate the localized classical shell of
Section~\ref{sec:classical_thin_shell}.

First, if we require the boundary wavepacket be localized near $u_\ast$ while
remaining sharply peaked in frequency, then
\be
\Delta \omega \sim \frac{1}{\sigma} \ll |\omega_\ast|.
\label{eq:wavepacket-frequency-window}
\ee
Equivalently, the envelope is wide compared with the oscillation
period $1/|\omega_\ast|$, so that the source selects a well-defined
central boundary-energy transfer.

However, there is a second notion of resolution which is more directly tied to
the classical cooling process. The classical trajectory specifies a
definite change in the Schwarzian energy, from $\mathsf E_-$ to
$\mathsf E_+$. If the frequency width of the source is comparable to this change, it cannot sharply distinguish the target
transition from processes which remove noticeably more or less
Schwarzian energy, and can no longer be viewed
as sharply resolving the classical trajectory. For an exactly extremal target, $\mathsf E_{+}=0$, so we require
\be
{1\over\sigma}
\ll
\mathsf E_-.
\label{eq:cooling-resolution}
\ee
These are distinct conditions because the central frequency contains
both the geometric cooling and the change in the charge-sector energy. That is,
\be
\omega_\ast
=
{\mathsf q^2\over2\mathsf K}
-
\mathsf E_-.
\label{eq:extremal-carrier-repeat}
\ee
In particular, the two terms can in principle nearly cancel, so that
$|\omega_\ast|\ll\mathsf E_-$ even though both $\mathsf E_-$
and $\mathsf q^2/2\mathsf K$ are individually large. Requiring both gives
\be
\sigma
\gg
\max\left(
{1\over|\omega_\ast|},
{1\over\mathsf E_-}
\right).
\label{eq:source-resolution-lower-bound}
\ee
In the asymptotically flat limit $\mathsf K\to\infty$, the charge-sector
term vanishes and $\omega_\ast=-\mathsf E_-$, so the two resolution
conditions coincide.

Temporal localization gives a separate condition. The initial
near-extremal throat has an intrinsic thermal time scale set by the
inverse surface gravity $(2\pi\mathsf T_-)^{-1}=\mathsf C/\mathsf k_-$. If we want the source to approximate a shell inserted at a definite boundary time, it is natural to require its duration to
be short compared with this scale. We therefore can also require
\be
\sigma
\ll
{\mathsf C\over\mathsf k_-}.
\label{eq:localized-shell-time}
\ee
A source which is both resolved and
localized in this sense then satisfies
\be
\max\left(
{1\over|\omega_\ast|},
{1\over\mathsf E_-}
\right)
\ll
\sigma
\ll
{\mathsf C\over\mathsf k_-}.
\label{eq:localized-shell-window}
\ee
These are preparation conditions for retaining the localized-shell
interpretation. They are not fundamental bounds on the quantum transition. With different choices of $\sigma$, we are probing
different parts of the final energy distribution. We return to this in Section~\ref{subsec:disk_prob}. Throughout Section~\ref{sec:disk-level-protection}, we
do not incorporate Hawking radiation and spontaneous Schwinger discharge.
Requiring these effects to remain negligible during the source
introduces additional upper bounds on $\sigma$, which depend on the
exterior and microscopic completion and will be discussed in
Section~\ref{sec:discussion}.

\subsection{Universal suppression of the extremal edge}
\label{subsec:disk_prob}

In this section, we study general properties of $P({\sf E}_f)$ near the extremal edge. We show that it is impossible to prepare the shell so that the final-state probability distribution is peaked at extremality. The analysis will also allow us to identify where the localized classical
shell of Section~\ref{subsec:finite-time-targeting} lies within the full spectral problem.

\subsubsection*{Probability distribution near the extremal edge}

As the final energy ${\sf E}_f$ approaches extremality from above, the density of states vanishes with the square-root edge
\be
\rho(\mathsf E_f)
\approx
{\sqrt{2}\mathsf C^{3/2}\over\pi}
\sqrt{\mathsf E_f}.
\label{eq:density-edge-expansion}
\ee
The expansion is controlled for
$\mathsf C\mathsf E_f\ll1$ which lies precisely in the regime where quantum gravity effects near the black hole horizon are important. Since the final probability density has an overall prefactor of $\rho({\sf E}_f)$ this might suggest that the probability of landing at extremality vanishes. To make this claim we still need to verify that the transition amplitude contains no inverse
power capable of canceling this zero.

At extremality we have $\mathsf k_f=0$ and the exact squared amplitude $| \mathcal{V}_{\text{in}}|^2$ that appears in the probability density simplifies to
\be
\left|
{\cal V}_{{\rm in}}(0,\mathsf k_-)
\right|^2
=
{\pi^2\over2\mathsf C}
\sqrt{
{\sinh(2\pi\lambda)
\over
2\pi\lambda}
}
{1
\over
\cosh\left[
\pi(\mathsf k_--\lambda)
\right]
\cosh\left[
\pi(\mathsf k_-+\lambda)
\right]}.
\label{eq:W-edge-abs}
\ee
The expression is nonzero and finite at ${\sf E}_f=0$. The four-Gamma kernel is even in ${\sf k}_f$ and nonzero at the edge, so the chosen square root is locally analytic in ${\sf E}_f$, namely
\be
\left|
{\cal V}_{{\rm in}}({\sf k}_f,\mathsf k_-)
\right|^2
=
\left|
{\cal V}_{{\rm in}}(0,\mathsf k_-)
\right|^2 + O({\sf k}^2_f).
\ee
The edge value at ${\sf E}_f=0$ is finite and nonzero for every fixed ${\sf k}_-$ and $\lambda$. The source factor $|J(\omega_{fi})|^2$ is also regular at the final edge. Generically, a smooth source at a corresponding transition frequency has a regular expansion in ${\sf E}_f$ and therefore cannot compensate the zero in $\rho({\sf E}_f)$. For the extremal target here, we have $\omega_{fi} = \omega_\ast + {\sf E}_f$. If $|J(\omega)|^2$ has a smooth maximum at $\omega_\ast$, as for the
Gaussian source above, then
$
|J(\omega_{fi})|^2
=
|J(\omega_\ast)|^2
+
O({\sf E}_f^2)$.
The linear term vanishes because $\omega_\ast$ is the location of the maximum.

For the Gaussian source, from the square-root edge in the density of states combined with the finite limit of the amplitude we can conclude that
\be
P(\mathsf E_f)
\approx
\frac{\sqrt{2}\mathsf C^{3/2}}{\pi}
|J(\omega_\ast)|^2
\left|
{\cal V}_{\text{in}}(0,\mathsf k_-)
\right|^2
\mathsf E_f^{1/2}
e^{-\sigma^2\mathsf E_f^2},
\qquad
{\sf C}{\sf E}_f\ll1.
\label{eq:edge-probability-explicit}
\ee
The leading correction to this expression is of order $\mathsf E_f^{3/2}$ and comes from the next term in the Schwarzian density and
the smooth energy dependence of the vertex. It follows in particular that
\be
P({\sf E}_f =0)
=
0.
\label{eq:Pzero}
\ee
This is stronger than the statement that a single point has zero
measure in a continuous distribution.  The density itself vanishes at
the endpoint.  For a generic channel with non-vanishing amplitude 
${\cal A}_{J}(0,\mathsf Q_f)\neq0$,
\be
\begin{aligned}
{\rm Prob}
\left(
0<\mathsf E_f<\delta
\right)
&=
\int_0^\delta
\mathrm{d}\mathsf E_f\,
P(\mathsf E_f)
\\
&=
{2\sqrt{2}\mathsf C^{3/2}\over3\pi}
\left|
{\cal A}_{J}(0,\mathsf q)
\right|^2
\delta^{3/2}
+
O(\delta^{5/2}).
\end{aligned}
\label{eq:small-window-probability}
\ee
An endpoint with finite nonzero density would instead give a
probability proportional to $\delta$.  The
$\delta^{3/2}$ law is the additional Schwarzian suppression. At this level of approximation, there is no separate term proportional to
$\delta(\mathsf E_f)$ and no singularity capable of compensating the
vanishing density of states.

For a general normalizable initial superposition the conclusion is unchanged. 
The product of energy-basis vertices appearing in the probability is regular at the final edge,
\beq
\overline{
{\cal V}_{\rm in}(\mathsf k_f,\mathsf k_i)
}
\,{\cal V}_{\rm in}(\mathsf k_f,\mathsf k_i')
=
\overline{
{\cal V}_{\rm in}(0,\mathsf k_i)
}
\,{\cal V}_{\rm in}(0,\mathsf k_i')
+
O(\mathsf E_f).
\eeq
For a smooth finite-width source, the source factors are likewise regular in 
$\mathsf E_f$. The coefficient multiplying the square-root edge from the density of states is
\be
\left|
\int_0^\infty
\mathrm{d}\mathsf E_i\,
\rho(\mathsf E_i)\,
\psi_i(\mathsf E_i)\,
J(\omega_{0i})\,
{\cal V}_{\rm in}(0,\mathsf k_i)
\right|^2.
\ee
By Cauchy-Schwarz and the normalization of the initial state, this is bounded by
\be
\int_0^\infty
\mathrm{d}\mathsf E_i\,
\rho(\mathsf E_i)\,
|J(\omega_{0i})|^2
\left|
{\cal V}_{\rm in}(0,\mathsf k_i)
\right|^2.
\ee
By symmetry of the vertex, $\rho(\mathsf E_i)|{\cal V}_{\rm in}(0,\mathsf k_i)|^2$ approaches the same finite UV plateau as in~\eqref{eq:rho-UV}. Since $\omega_{0i}$ differs from $-\mathsf E_i$ in a fixed charge sector only by an $\mathsf E_i$-independent constant and
$\int \mathrm{d}\omega\,|J(\omega)|^2<\infty$, the integral is finite. This convergence
remains uniform for ${\sf E}_f$ sufficiently close to zero. Thus the initial energy integral remains regular at the edge and cannot compensate the zero in the Schwarzian density of states. For a generic initial state, the resulting edge coefficient is nonzero and
\beq
P(\mathsf E_f)
\propto
\mathsf E_f^{1/2}.
\eeq
Special coherent initial states may make the leading coefficient vanish 
through destructive interference, in which case the endpoint is even more 
strongly suppressed.\footnote{One can verify from the
double-energy expression in Section~\ref{subsec:disk-transition-amplitude} that $u_\ast$
appears in the probability through relative phases between different
initial energies. It can therefore change the finite coefficient of
the edge distribution. This does not affect the regularity argument
and cannot compensate the zero in the Schwarzian density of states. The same conclusion holds for normalizable
mixed states.
} Thus, no normalizable choice of initial wavefunction can generate a singular 
contribution that restores a nonzero probability density at the extremal edge. 

The same conclusion holds if the initial state has a finite charge
width. At fixed final charge, the selection rule fixes
$\mathsf Q_i=\mathsf Q_f-\mathsf q$, and each charge sector retains
the same final Schwarzian density factor. If the final charge is not
measured, the final-energy distribution is obtained by summing the
probabilities over the orthogonal charge sectors. Provided this sum
converges near the edge, this marginal distribution still vanishes as
$\mathsf E_f^{1/2}$. Charge fluctuations only change the shape and width
of the final-energy distribution.\footnote{
For example, consider a Gaussian initial charge distribution of width
$\sigma_Q$ centered at $\mathsf Q_i=0$, while keeping the initial
Schwarzian energy fixed at $\mathsf E_-$. Since
$\mathsf E_{\rm gauge}(\mathsf Q)=\mathsf Q^2/(2\mathsf K)$, we have
$\omega_{fi}-\omega_\ast
=
\mathsf E_f+(\mathsf q/\mathsf K)\mathsf Q_i$.
In our single-throat discussion, integrating over the
Gaussian charge distribution therefore gives
\be
p(\mathsf E_f)
\propto
F(\mathsf E_f)
\exp\left[
-{\mathsf E_f^2\over2V_{\rm eff}}
\right],
\qquad
V_{\rm eff}
=
{1\over2\sigma^2}
+
\left(
{\mathsf q\over\mathsf K}
\right)^2
\sigma_Q^2.
\ee
Thus a finite charge width broadens the final-energy distribution,
while the overall Schwarzian edge factor remains unchanged.
}

Furthermore, there is at least one global maximum at strictly positive $\mathsf E_f>0$ for the simple reason that the transition density is continuous and tends to zero both at $\mathsf E_f=0$ and as $\mathsf E_f \to \infty$, provided that the transition is not identically zero. For the fixed initial energy and Gaussian source considered below, one can further show that the exact distribution has a unique maximum.

\subsection*{Peak of the final energy distribution}

Let us now specialize to the fixed initial energy and Gaussian source
introduced in Section~\ref{subsec:finite-time-targeting} and compare the most likely final energy with
the classical target $\mathsf E_+$. For an initial state that is an energy eigenstate, the final probability distribution as a function of ${\sf E}_f$ involves the source $|J(\omega)|^2$ evaluated at $\omega_{fi}$. For a Gaussian wavepacket this is a function of 
$
\omega_{fi}-\omega_\ast
=
\mathsf E_f-\mathsf E_+.
$
The final energy probability density therefore takes
the form
\be
P(\mathsf E_f)
=
{\cal N}\,
F(\mathsf E_f)\,
\exp\left[
-\sigma^2
\left(
\mathsf E_f-\mathsf E_+
\right)^2
\right],
\label{eq:probability-around-classical-target}
\ee
where $F(\mathsf E_f)$ is given in~\eqref{eq:high-energy-control} and ${\cal N}$ is a prefactor independent of $\mathsf E_f$.  
The energy 
${\sf E}_f^{\text{peak}}$ at which $P({\sf E}_f)$ is maximized can be found via the saddle-point equation
\be
\mathsf E_f^{\rm peak}
=
\mathsf E_+ + \frac{1}{2\sigma^2}\left.
{\partial\over\partial\mathsf E_f}
\log F(\mathsf E_f)
\right|_{\mathsf E_f=\mathsf E_f^{\rm peak}}.
\label{eq:peak-condition-general}
\ee
For the exact spectral factor in \eqref{eq:principal-series-vertex-absolute-square-sec4},
one finds
$\partial_{\mathsf E_f}\log F>0$ and
$\partial_{\mathsf E_f}^2\log F<0$ for $\mathsf E_f>0$.
The distribution for the Gaussian source is therefore strictly
log-concave and has a unique maximum. In particular,
\eqref{eq:peak-condition-general} implies
$\mathsf E_f^{\rm peak}>\mathsf E_+$ for $\mathsf E_+\geq0$.

The initial black hole is always taken to be at least semiclassical, such that $\mathsf k_- \gg 1$. Since $\delta S/(2\pi)={\sf k}_-$, the
\pro condition implies
$\lambda^2+1/4>{\sf k}_-^2$ and hence
$\lambda\gtrsim{\sf k}_-\gg1$. It is useful to distinguish several regimes as the target final state is moved toward extremality. The purpose is to separate the broadening of the classical transition from the effects of the semiclassical spectrum and finally from the Schwarzian edge.

We first consider a classical ``cooling'' process in which both the initial
and final black holes remain classical. We keep the initial
black hole fixed and require $0<{\sf T}_+<{\sf T}_-$. A convenient way
to state this regime is by the limit
\be
\mathsf C\to\infty,
\qquad
{\mathsf E_-\over\mathsf C},
\quad
{\mathsf E_f\over\mathsf C},
\quad
{\mathsf E_+\over\mathsf C}
\quad\text{fixed},
\label{eq:semiclassical-energy-scaling}
\ee
with the corresponding shell parameters scaling so that
$\lambda/{\sf C}$ remains fixed. In this limit $\mathsf k_f, \mathsf k_{\pm}$ are large and of order ${\sf C}$. It is evident from \eqref{eq:principal-series-vertex-absolute-square-sec4} that the spectral factor takes the semiclassical form
\be
F(\mathsf E_f)
=
\exp\left[
\mathsf C\,
{\cal F}\left(
{\mathsf E_f\over\mathsf C}
\right)
+
O(\log\mathsf C)
\right],
\label{eq:F-semiclassical-form}
\ee
since it involves evaluating the hyperbolic functions when their arguments are large. In this regime,  $\log F \sim O({\sf C})$ and $\partial_{{\sf E}_f} \log F \sim O(1)$. The finite width of the source already broadens the classical target, while the spectral factor can shift the most likely final energy away from $\mathsf E_+$.

We next lower the target temperature. Before the dominant final
distribution reaches the Schwarzian edge, the relevant final energies
can still be semiclassical with $1\ll {\sf k}_f \ll {\sf C}$. For the classical initial scaling ${\sf k}_-=O({\sf C})$, an upper crossover occurs at
${\sf k}_f\sim\lambda+{\sf k}_-$, where the exponential growth
of the spectral factor is canceled and $F({\sf E}_f)$ approaches its UV
plateau. Since the \pro condition implies ${\sf k}_f =O({\sf C})$, this upper crossover lies parametrically outside the semiclassical cooling regime. The only relevant crossover is therefore controlled by
$\lambda-{\sf k}_--{\sf k}_f$. Indeed, throughout this regime, the other three combinations appearing in
the hyperbolic functions,
$\lambda+{\sf k}_-+{\sf k}_f$,
$\lambda+{\sf k}_--{\sf k}_f$, and
$\lambda-{\sf k}_-+{\sf k}_f$,
remain parametrically large and positive, while
$\lambda-{\sf k}_--{\sf k}_f$ can pass through zero. We therefore keep
this combination exact while using the large-argument form for the other three hyperbolic functions, which gives
\be
{\partial\over\partial{\sf E}_f}\log F({\sf E}_f)
\simeq
{\pi{\sf C}\over2{\sf k}_f}
\left[
3+\tanh\!\left(\pi(\lambda-{\sf k}_--{\sf k}_f)\right)
\right].
\label{eq:semiclassical-spectral-slope}
\ee
If $\lambda-{\sf k}_--{\sf k}_f\gg1$, this reduces to
$1/{\sf T}_f$. In this regime, using $\log \cosh x \sim |x|$ as $|x| \to \infty$, we get that  $\log |\mathcal{V}_{\text{in}}({\sf k}_f,{\sf k}_-)|^2 \sim - \pi \lambda$ and is independent of ${\sf k}_f$ at leading order, and the growth comes entirely from
$\rho({\sf E}_f)\sim e^{2\pi{\sf k}_f}$. On the other side of this
crossover, with
${\sf k}_f-(\lambda-{\sf k}_-)\gg1$ while
${\sf k}_f\ll\lambda+{\sf k}_-$, the vertex contributes an additional
factor $e^{-\pi{\sf k}_f}$, and we get $1/(2 {\sf T}_f)$. We see that it is important to keep the exact vertex factor as it reduces, but does not cancel, the spectral growth throughout the semiclassical cooling regime.

Substituting these two limiting behaviors into the saddle-point
condition~\eqref{eq:peak-condition-general}, we find that for an exactly extremal target
${\sf E}_+=0$,
\be
\left(
{\pi^2{\sf C}\over8\sigma^4}
\right)^{1/3}
\lesssim
{\sf E}_f^{\rm peak}
\lesssim
\left(
{\pi^2{\sf C}\over2\sigma^4}
\right)^{1/3}.
\ee
Still, this result is in the regime
$1\ll{\sf k}_f^{\rm peak}\ll{\sf C}$, and the scaling is the same on both sides of the crossover. At first sight, this may seem surprising. For fixed $\sigma$,
the peak energy grows as ${\sf E}_f^{\rm peak}
\sim
{\sf C}^{1/3}\sigma^{-4/3}$. This does not mean that the final black hole remains at a large
macroscopic energy. For the classical initial scaling
${\sf E}_-=O({\sf C})$, we have ${{\sf E}_f^{\rm peak}/{\sf E}_-}
\sim
{\sf C}^{-2/3}\sigma^{-4/3}
\longrightarrow 0$. Thus, at fixed $\sigma$, the most likely final state becomes
parametrically colder and closer to extremality as ${\sf C}$ increases,
even though ${\sf E}_f^{\rm peak}$ itself increases with ${\sf C}$.
It remains semiclassical as long as ${\sf k}_f^{\rm peak}\gg1$, and has
not yet entered the Schwarzian edge where ${\sf k}_f=O(1)$. Indeed, at the same time we obtain the scaling relations 
\be
\pi^{1/3}
\left(
{{\sf C}\over\sigma}
\right)^{2/3}
\lesssim
{\sf k}_f^{\rm peak}
\lesssim
(2\pi)^{1/3}
\left(
{{\sf C}\over\sigma}
\right)^{2/3},
\ee
\be
{1\over2\pi^{2/3}}
{\sf C}^{-1/3}\sigma^{-2/3}
\lesssim
{\sf T}_f^{\rm peak}
\lesssim
{1\over(2\pi)^{2/3}}
{\sf C}^{-1/3}\sigma^{-2/3}.
\label{eq:semiclassical-temperature-range}
\ee
In asymptotically flat space, using ${\sf L}_2=r_0$,
$S_0=\pi r_0^2$, and $\delta t=r_0\sigma$, the lower bound gives the 4d estimate
quoted in~\eqref{eq:intro_temperature}. We see that even though the classical target is extremal ${\sf E}_+=0$, the peak of the quantum distribution is positive ${\sf E}_f^{\rm peak}>0$. This nonzero peak energy is large compared with the Schwarzian edge
scale. In the classical initial scaling considered above, it nevertheless
remains parametrically small compared with the initial black hole energy.

The precise expressions derived above depend on the choice of a Gaussian wavepacket. Nevertheless, the fact that the peak ${\sf k}_f^{\text{peak}}$ lies above the quantum regime $1 \ll {\sf k}_f^{\text{peak}}$ is universal for any regular, normalizable wavepacket centered at a positive energy and on the classical extremal target with $\sigma$ fixed in the large ${\sf C}$ limit. To see this, go back to \eqref{eq:probability-around-classical-target}. If ${\sf E}_+=0$ and ${\sf E}_f \sim 1/{\sf C}$, then the wavepacket profile is evaluated at its saddle point. To find a global saddle point for $P({\sf E}_f)$ would require an independent saddle point within $F({\sf E}_f)$ itself. But $F({\sf E}_f)$ is a monotonic function, as shown in Figure~\ref{fig:spectral_regimes} below. For $n>1$, we can check that if $|J(\omega_{fi})|^2
\sim
\exp\!\left[-\sigma^n|{\sf E}_f-{\sf E}_+|^n\right]$, the peak is at ${\sf k}_f^{\text{peak}} \sim {\sf C}^{n/(2n-1)}$ which is always much larger than one and much smaller than the classical order ${\sf C}$ regime.

\begin{figure}[t]
    \centering
    \makebox[\textwidth][c]{%
        \includegraphics[width=0.95\textwidth]{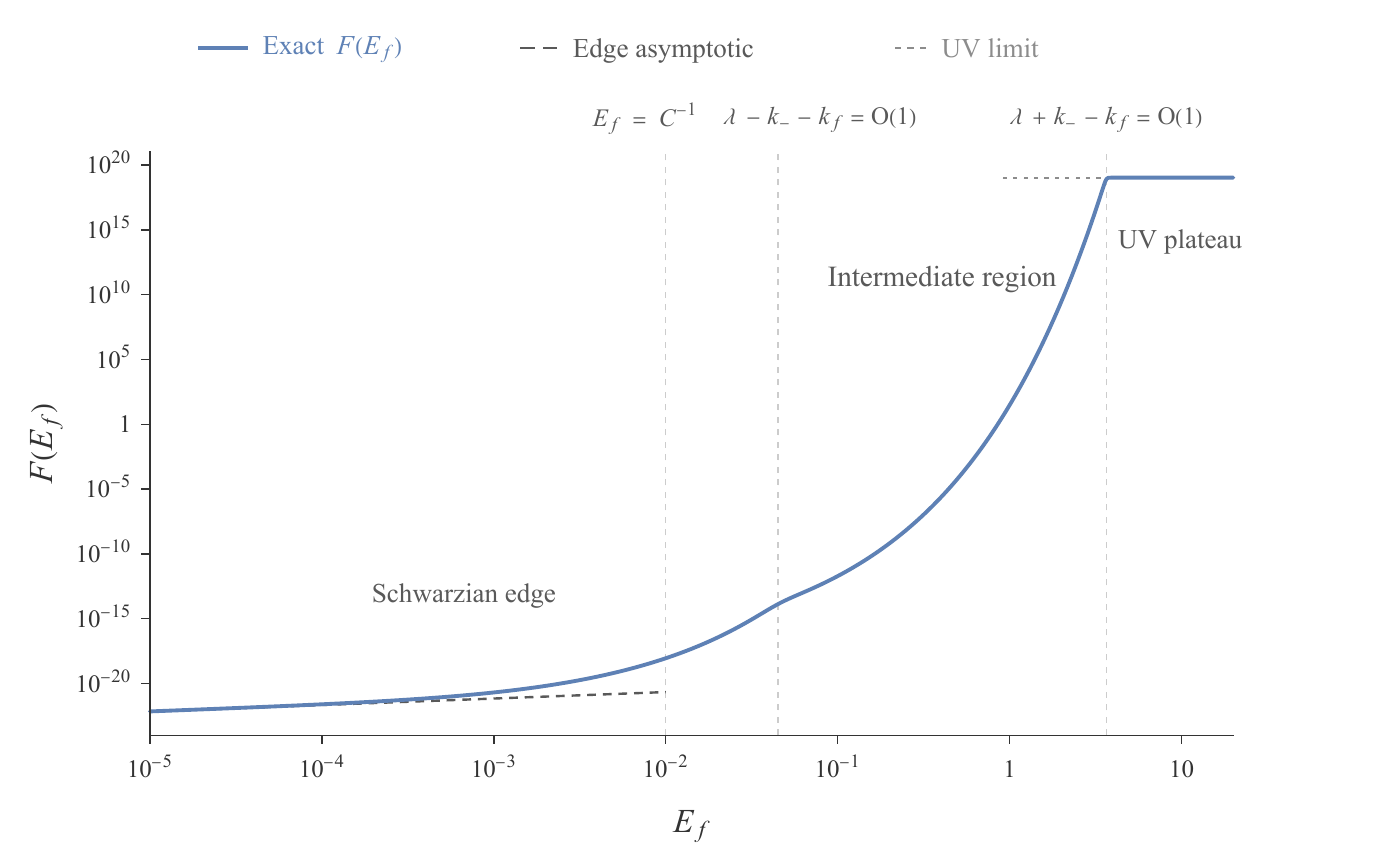}%
    }
    \caption{The blue curve represents the spectral factor
$F(\mathsf E_f)=\rho(\mathsf E_f)
|{\cal V}_{\rm in}(\mathsf k_f,\mathsf k_-)|^2$ obtained from~\eqref{eq:principal-series-vertex-absolute-square-sec4}. The dashed
grey curve at small $\mathsf E_f$ shows the Schwarzian edge behavior
$F(\mathsf E_f)\propto\mathsf E_f^{1/2}$, while the horizontal dotted
line shows the analytic UV limit.  The vertical dashed lines indicate different
characteristic scales. The first marks the scale
$\mathsf E_f=\mathsf C^{-1}$, where the edge approximation
is no longer reliable and the full spectral factor
must be retained. The lower matter-dependent
crossover occurs when
$\lambda-\mathsf k_- -\mathsf k_f=O(1)$. Across this region one of the
hyperbolic factors in the vertex changes its large-$\mathsf k_f$
behavior, and the growth of $F$ changes from approximately
$e^{2\pi\mathsf k_f}$ to $e^{\pi\mathsf k_f}$ when the corresponding
asymptotic regimes are well separated. The upper crossover occurs when
$\lambda+\mathsf k_- -\mathsf k_f=O(1)$. Beyond it the remaining
exponential growth in the numerator is canceled by the vertex
denominator and $F$ rapidly approaches the UV plateau.}
    \label{fig:spectral_regimes}
\end{figure}

We finally consider the local behavior of the distribution near the Schwarzian edge ${\sf C}{\sf E}_f\ll1$. In this region the
semiclassical approximation to $F({\sf E}_f)$ breaks down and we must
use the square-root edge derived earlier in this section,
$
F(\mathsf E_f)\propto\mathsf E_f^{1/2},
$
with a finite nonzero coefficient. The near-edge probability is therefore
\be
P(\mathsf E_f)
\propto
\mathsf E_f^{1/2}
\exp\left[
-\sigma^2
\left(
\mathsf E_f-\mathsf E_+
\right)^2
\right].
\label{eq:near-edge-target-probability}
\ee
Within this edge approximation, the formal stationary point is located at
\be
\mathsf E_{f,\rm edge}^{\rm peak}
=
{1\over2}
\left[
\mathsf E_+
+
\sqrt{
\mathsf E_+^2+{1\over\sigma^2}
}
\right].
\ee
For this point to lie within the regime of validity of the
edge approximation, self-consistency requires $\mathsf C\mathsf E_+\ll1$ and $\sigma\gg\mathsf C$. For an exactly extremal target this reduces to
$\mathsf E_{f,\rm edge}^{\rm peak}=1/(2\sigma)$, so that the peak of the edge approximation remains at an energy of order $1/\sigma$, and we only need $\sigma \gg \mathsf C$. If this consistency
condition is not satisfied, the stationary point lies outside
the region where the edge approximation can be used, and the peak must
instead be determined from the full spectral factor. However, if we relax the requirement for an exactly extremal target, then this conclusion can change. We return to this possibility in
Section~\ref{subsec:operational-third-law}. The local edge behavior itself requires no such restriction on
$\sigma$. For any fixed $0<\sigma<\infty$, the
probability density still vanishes as
$\mathsf E_f^{1/2}$ as $\mathsf E_f\to0$.

\subsubsection*{The localized-shell regime}

We can now place the localized-shell preparation of Section~\ref{subsec:finite-time-targeting} within the regimes described above. If the dominant final energies do not lie in the Schwarzian edge, the exact
spectral factor must be retained. As we can see from Figure~\ref{fig:spectral_regimes}, in the intermediate region, a crossover occurs when one of the combinations
$\mathsf k_f\pm\mathsf k_-\pm\lambda$
becomes of order one, while sufficiently large final energies eventually probe
the UV plateau at $\mathsf k_f\gg\mathsf k_-,\lambda$.

For a semiclassical initial black hole $\mathsf k_-\gg1$, the upper bound in
\eqref{eq:localized-shell-window} has no parametric overlap with the deep-edge condition $\sigma\gg\mathsf C$. The localized-shell window~\eqref{eq:localized-shell-window} instead implies
\be
{\mathsf k_-\over\mathsf C}
\ll
{1\over\sigma}
\ll
\mathsf E_-
=
{\mathsf k_-^2\over2\mathsf C}
\leq
{(\lambda+\mathsf k_-)^2\over2\mathsf C}.
\ee
Thus its source bandwidth $1/\sigma$ lies above the Schwarzian edge scale while
remaining below the scale associated with the UV plateau. The UV plateau requires
$\mathsf E_f\gg(\lambda+\mathsf k_-)^2/(2\mathsf C)$ and no longer
describes a cooling trajectory. The dominant final energies are therefore expected to lie in the intermediate spectral region. Their precise location within this region depends on
$\lambda-\mathsf k_-$ and on the final $\mathsf k_f$ selected by the
full spectral factor.

\begin{figure}[t!]
    \centering

    \begin{subfigure}[t]{0.49\linewidth}
        \centering
        \includegraphics[width=\linewidth]{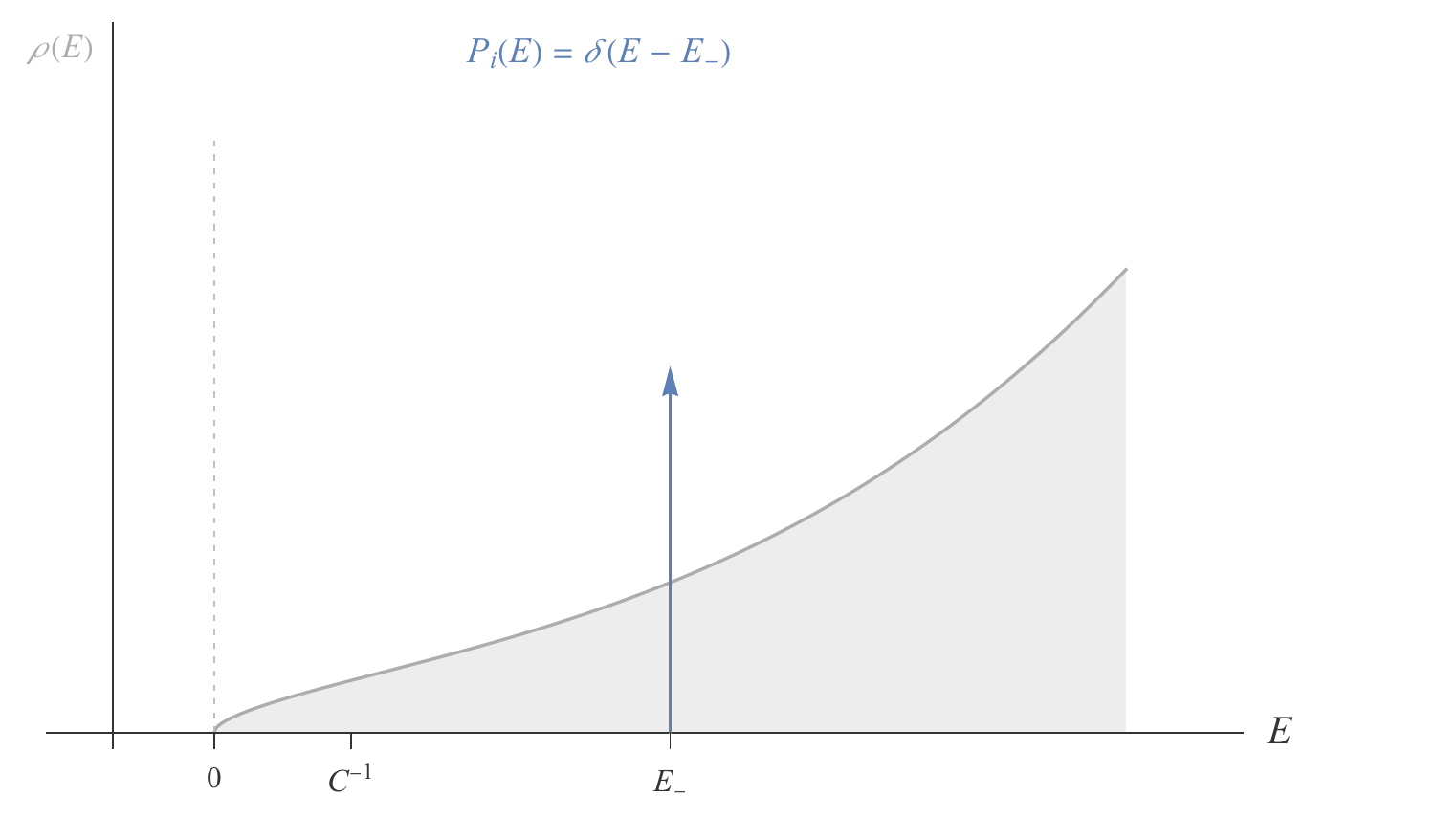}
        \caption{(a) Initial state}
        \label{fig:initial}
    \end{subfigure}
    \begin{subfigure}[t]{0.49\linewidth}
        \centering
        \includegraphics[width=\linewidth]{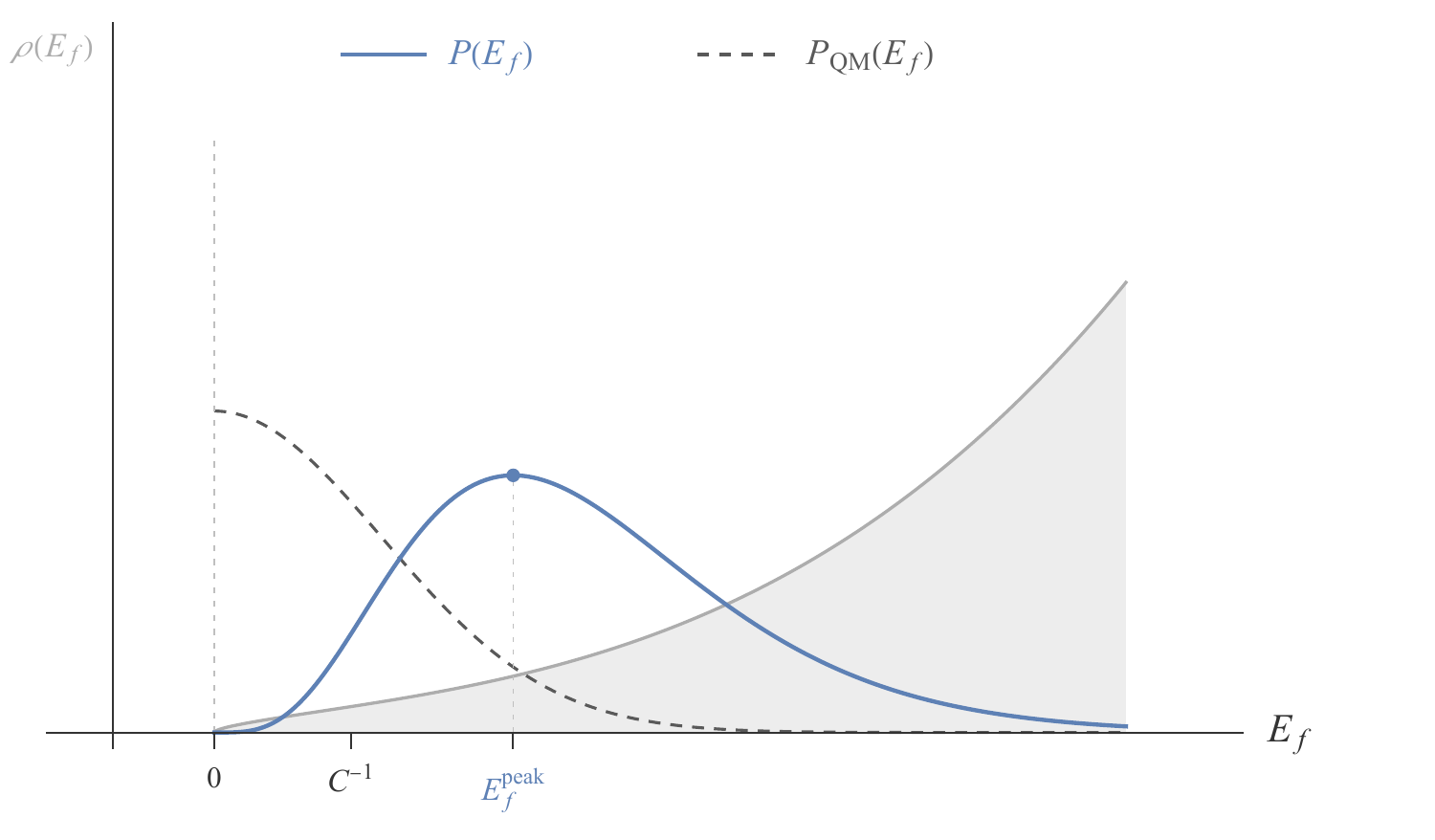}
        \caption{(b) Final state}
        \label{fig:final}
    \end{subfigure}

    \caption{Schematic illustration of the quantum transition for a third-law-violating localized shell with an exactly extremal classical target $\mathsf E_+=0$. In panel (a), the initial black hole is prepared at a fixed semiclassical
energy $\mathsf E_-$, with the blue arrow representing the delta function. The scale $\mathsf C^{-1}$ marks the crossover into the Schwarzian edge regime. In panel (b), the blue curve shows the final distribution $P(\mathsf E_f)\propto
F(\mathsf E_f)e^{-\sigma^2\mathsf E_f^2}$. The dashed curve shows the Gaussian profile
$P_{\rm QM}(\mathsf E_f)$ with the gravitational spectral factor
removed. We see that unlike $P_{\rm QM}(\mathsf E_f)$, the full distribution vanishes at the
extremal edge and peaks at $\mathsf E_f^{\rm peak}>0$. In this regime, the peak lies well above
$C^{-1}$ and the characteristic width is of order
$\sigma^{-1}\gg C^{-1}$. The grey curves with shaded areas show the corresponding Schwarzian density of states, while the vertical dashed lines mark the extremal edge.}
    \label{fig:initial_final}
\end{figure}

Finally, we comment on the genuine quantum gravitational input in these results and distinguish them from the quantum mechanical requirement of the finite frequency resolution of the source. The effect of a finite $\sigma$ is to turn the classical deterministic target into a distribution over final energies. If the gravitational spectral factor
$F({\sf E}_f)$ were removed, an exactly extremal target would instead
give
\be
P_{\rm QM}({\sf E}_f)
\propto
e^{-\sigma^2{\sf E}_f^2},
\qquad
{\sf E}_f\geq0,
\ee
whose maximum remains at ${\sf E}_f=0$. The finite width never moves the most likely final state away from extremality. The positive shift of the peak in all the regimes discussed above comes from the density of states and the transition amplitude contained in  $F({\sf E}_f)$. It is only that in the first two regimes this factor admits a semiclassical approximation, while
near the Schwarzian edge its full quantum behavior becomes essential. They are just different limits of the same quantum gravitational input. This is the sense in which the
third law is protected by quantum gravity in our setup. Here, the protection does not come from forbidding the
classical trajectory to extremality. The classical process may reach
${\sf E}_+=0$, while quantum gravity replaces this deterministic endpoint
by a distribution whose most likely outcome lies at positive energy and
whose probability density vanishes at the extremal edge. This is illustrated schematically in Figure~\ref{fig:initial_final}.

\subsection{Approaching the extremal edge}
\label{subsec:operational-third-law}

The preceding section established both the universal suppression of the extremal
endpoint and the spectral regime probed by the localized classical shell. We now ask a different operational question. What source preparations can place the dominant final distribution inside the Schwarzian edge itself? The first possibility for a source centered on the exactly extremal
classical target is to increase the source duration until $\sigma\gg\mathsf C$, as we discussed in Section~\ref{subsec:disk_prob}.  This gives a controlled edge distribution whose
statistics can be computed analytically. However, a second possibility is to tune the source past the
classical extremal target, that is, an overcooling process. This allows the conditional
distribution to approach the edge without requiring the same long source duration, yet this does not have a corresponding classical physical process.

\subsection*{Energy statistics of the Schwarzian regime}

\begin{figure}[t!]
    \centering
        \centering
        \includegraphics[width=0.55\linewidth]{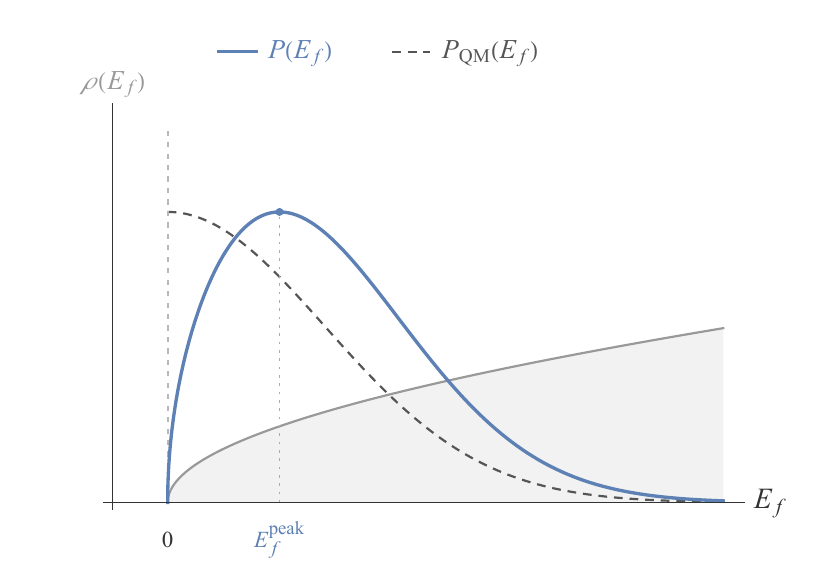}

    \caption{The distribution illustrates the long-duration $\sigma^{-1}\ll {\sf C}^{-1}$ preparation  to resolve the Schwarzian edge. The dominant final distribution lies well within the Schwarzian edge regime. Still, the gravitational spectral factor suppresses the exactly extremal endpoint and shifts the most likely final state to $\mathsf E_f^{\rm peak}>0$.}
    \label{fig:final_Schwarzian}
\end{figure}

We have shown in Section~\ref{subsec:disk_prob} that for an exactly extremal target the stationary
point of the edge approximation is
$\mathsf E_{f,\rm edge}^{\rm peak}=1/(2\sigma)$ for $\sigma\gg\mathsf C$. In this regime the dominant part of the distribution lies inside the
Schwarzian edge. For the extremal target $\mathsf E_+=0$, the unnormalized
distribution is proportional to
$\mathsf E_f^{1/2}e^{-\sigma^2\mathsf E_f^2}$. The coefficient multiplying the square-root edge is independent of
$\mathsf E_f$ and therefore cancels after normalization. Using
\be
\int_0^\infty
\mathrm{d}\mathsf E_f\,
\mathsf E_f^{1/2}
e^{-\sigma^2\mathsf E_f^2}
=
{\Gamma(3/4)\over2\sigma^{3/2}},
\ee
we obtain the normalized probability density
\be
P_{\rm edge}(\mathsf E_f)
=
{2\sigma^{3/2}\over\Gamma(3/4)}
\mathsf E_f^{1/2}
e^{-\sigma^2\mathsf E_f^2},
\qquad
\mathsf E_f\geq0.
\label{eq:normalized-edge-distribution}
\ee
The same integral gives the energy moments,
\be
\langle
\mathsf E_f^n
\rangle_{\rm edge}
=
{1\over\sigma^n}
{\Gamma\left((2n+3)/4\right)
\over
\Gamma(3/4)}.
\label{eq:edge-general-moments}
\ee
In particular,
\be
\langle
\mathsf E_f
\rangle_{\rm edge}
=
{\Gamma(5/4)\over\Gamma(3/4)}
{1\over\sigma},
\quad
\langle
\mathsf E_f^2
\rangle_{\rm edge}
=
{3\over4\sigma^2} \implies (\Delta\mathsf E_f)^2_{\rm edge}
=
\left[
{3\over4}
-
\left(
{\Gamma(5/4)\over\Gamma(3/4)}
\right)^2
\right]
{1\over\sigma^2}.
\label{eq:edge-variance}
\ee
The relative width is therefore
\be
{\Delta\mathsf E_f
\over
\left\langle\mathsf E_f\right\rangle_{\rm edge}}
\simeq
0.609.
\label{eq:edge-relative-width}
\ee
Thus increasing $\sigma$ moves the distribution toward lower energies
and reduces its absolute width, while the relative width remains of order one. The numerical value in \eqref{eq:edge-relative-width} depends on the
Gaussian source profile $e^{-\sigma^2\mathsf E_f^2}$, but the square-root suppression $\mathsf E_f^{1/2}$ at the extremal
edge is universal. The corresponding distribution is illustrated in Figure~\ref{fig:final_Schwarzian}. 

The next-order correction to the square-root edge is of relative order
$\mathsf C\mathsf E_f$. Since the edge-dominated distribution has
$\mathsf E_f\sim1/\sigma$, the corrections to the normalized
distribution and its moments are of order $\mathsf C/\sigma$.
They are therefore small when $\sigma\gg\mathsf C$.\footnote{There is also an upper limit on how large $\sigma$ can
be. The continuum edge regime requires $
\mathsf C
\ll
\sigma
\ll
e^{S_0}$. The second inequality ensures that the source does not resolve the
microscopic spectrum.  We return
to the nonperturbative edge in
Section~\ref{sec:discussion}, but before we get to these scales we have to first worry about Hawking radiation and Schwinger discharge.}

\subsection*{Tuning toward extremality by overcooling}

We next ask whether the final distribution can nevertheless be pushed closer to extremality by tuning the source. A natural possibility is to tune it past the classical extremal target by ``overcooling.''  Starting from the source with $\mathsf E_+=0$, we lower its central
frequency by an amount
$\Delta_{\rm over}>0$. The Gaussian factor in the final distribution
then becomes $\exp[
-\sigma^2
(
\mathsf E_f+\Delta_{\rm over}
)^2
]$. Here $\Delta_{\rm over}$ only specifies how far the source is
tuned past the frequency that classically produces an extremal black
hole, and does not represent a physical final black hole energy $\mathsf E_f<0$. 

In particular, a negative Schwarzian target
does not require the shell itself to have negative ADM energy since the total
energy transfer also includes $\upmu_0\mathsf q$. If we restrict to the positive-energy shell family of Section~\ref{sec:classical_thin_shell}, continuing
the parameters in this direction would encounter a
turning point such that the shell turns around. A different way of lowering the
energy of a nearly-AdS$_2$ black hole is possible if one allows an
externally controlled deformation of the boundary Hamiltonian. In the state-dependent
construction of \cite{Kourkoulou:2017zaj}, such a deformation gives an
effective negative energy contribution to the nearly-AdS$_2$ dynamics
and changes the Schwarzian trajectory so that the region which was
previously behind the horizon becomes accessible, in fact enlarging
the accessible geometry from a Rindler patch to the full Poincar\'e
patch. This is a different protocol from the positive-energy
charged shell considered in Section~\ref{sec:classical_thin_shell}.

The center of the Gaussian is now at $\mathsf E_f=-\Delta_{\rm over}<0$. The quantum calculation, however, still includes only physical final
black hole states with $\mathsf E_f\geq0$. For such a source, the final
energy distribution is
\be
P_{\rm over}(\mathsf E_f)
\propto
F(\mathsf E_f)
\exp\left[
-\sigma^2
\left(
\mathsf E_f+\Delta_{\rm over}
\right)^2
\right],
\qquad
\mathsf E_f\geq0.
\label{eq:overcooling-distribution}
\ee
As in Section~\ref{subsec:disk_prob}, the exact distribution vanishes at
$\mathsf E_f=0$ and at large $\mathsf E_f$, so its peak remains at
strictly positive energy. When the relevant final energies lie in the Schwarzian edge region,
\eqref{eq:overcooling-distribution} becomes $P_{\rm over}^{\rm edge}(\mathsf E_f)
\propto
\mathsf E_f^{1/2}
\exp\left[
-\sigma^2
\left(
\mathsf E_f+\Delta_{\rm over}
\right)^2
\right]$. The peak is
\be
\mathsf E_{f,\rm over}^{\rm peak}
=
{1\over2}
\left[
\sqrt{
\Delta_{\rm over}^2
+
{1\over\sigma^2}
}
-
\Delta_{\rm over}
\right]
>
0.
\label{eq:overcooling-edge-peak}
\ee
The source has bandwidth $1/\sigma$, while the Gaussian has a formal center at $-\Delta_{\rm over}$. Therefore, for a source tuned many bandwidths below the edge $\mathsf E_f=0$, we have $\Delta_{\rm over} \gg 1/\sigma$. Expanding
\eqref{eq:overcooling-edge-peak} in this limit gives
\be
\mathsf E_{f,\rm over}^{\rm peak}
=
{1\over
4\sigma^2\Delta_{\rm over}}
+
O\left(
{1\over
\sigma^4\Delta_{\rm over}^3}
\right).
\label{eq:overcooling-edge-peak-asymptotic}
\ee
The characteristic final energy is therefore parametrically smaller
than the original source width $\mathsf E_{f,\rm over}^{\rm peak}
\ll
{1\over\sigma}$. That is, the conditional final distribution can indeed be pushed closer to extremality. An intuitive way to describe this is that the center of the Gaussian now lies at
$\mathsf E_f=-\Delta_{\rm over}$, outside the physical spectrum
$\mathsf E_f\geq0$. The allowed final states therefore receive weight
only from the exponentially suppressed tail of the source profile. Pushing the conditional distribution
closer to the edge therefore comes with an exponentially suppressed transition weight at fixed source normalization. Increasing
$\Delta_{\rm over}$ suppresses larger positive energies more strongly
and squeezes the allowed part of the distribution toward the extremal
edge.

For the edge approximation to remain valid at this new peak, we require
$\mathsf C\mathsf E_{f,\rm over}^{\rm peak}\ll1$, which gives $\mathsf C/(\sigma^2\Delta_{\rm over})\ll1$. As anticipated in Section~\ref{subsec:disk_prob}, this is a weaker restriction than the condition
$\sigma\gg\mathsf C$ for a source centered exactly at the extremal
target. A sufficiently tuned source can make the dominant part of the final distribution probe the Schwarzian edge even when $\sigma$ is not parametrically larger than $\mathsf C$. The discussion so far applies as long as the dominant final energies remain above the scale at which the discreteness of the spectrum becomes important, as discussed in Section~\ref{sec:discussion}.

\section{Discussion and outlook}
\label{sec:discussion}

We presented a classical process that violates the third law of black hole thermodynamics, and can be understood entirely within the near-horizon AdS$_2 \times X$ throat. This provides a universal description that applies even when the throat arises from a non-spherically-symmetric black hole, such as Kerr. The classical  condition for the third law violation also implies that the matter describing the shell violates the AdS$_2$ BF bound and belongs to a principal-series representation of SL$(2, \mathbb{R})$. We constructed the corresponding fixed-incoming holographic dictionary and derived its exact Schwarzian-Maxwell-dressed energy-basis vertex.  The quantization of JT gravity allows us to formulate the process in quantum gravity
and analyze the resulting collapse. In particular, we showed that quantum-gravitational effects make the final energy probability density vanish at extremality and place its maximum at positive energy.  More strikingly, although the classical process reaches extremality in finite time, its quantum counterpart cannot probe the quantum gravity regime in a preparation time that remains finite in the semiclassical limit. Reaching that regime requires a duration that scales with the black hole entropy and therefore diverges in the classical limit. Thus quantum gravity protects the third law against the
violation seen at the classical level, and the physics is compatible with the black hole being described by an ordinary quantum system.

Before turning to open questions, we comment on dynamical effects that can
compete with the source-driven collapse. In particular, Hawking radiation
and Schwinger discharge introduce additional timescales, and we discuss
below when the closed-system treatment used in our calculation remains
appropriate.

\paragraph{Hawking radiation.}

We treated the throat as a closed quantum system, with the exterior entering only through the source that inserts the thin shell. In an asymptotically flat completion, fields can propagate out of the throat and escape to infinity, so the throat should instead be treated as an open system. In AdS, the energy exchange with the throat
depends on whether the radiation is reflected back or removed
through an absorbing boundary.   The estimates below use an asymptotically flat exterior with no incoming radiation. Including this coupling to the exterior allows us to account for Hawking radiation~\cite{Hawking:1974sw}. See \cite{Maldacena:1997ih} and, for recent applications to near-extremal black holes, \cite{Brown:2024ajk,Emparan:2025sao,Biggs:2025nzs,Kraus:2025efu,Bintanja:2026bht, Luo:2026epp}.

In the decoupling limit of the throat, for wavelengths larger than the extremal radius $r_0$, we can describe the coupling to the environment by an interaction Hamiltonian \cite{Maldacena:1997ih, Brown:2024ajk}
\beq
{\sf H} \to {\sf H} + {\sf H}_{\text{int}} ,\qquad {\sf H}_{\text{int}}  = g \, O_{\text{rad}}(u) \,\phi_{\text{ext}}(u).
\label{eq:Hawk-coupling}
\eeq
In this model, $O_{\text{rad}}(u)$ is an operator acting on the black hole Hilbert space corresponding to a 4d massless scalar $s$-wave with $\Delta_{\text{rad}}=1$. $\phi_{\text{ext}}(u)$ is the exterior quantum field evaluated at the effective location of the throat acting as a source. The coupling constant $g$ can be fixed by matching the absorption
cross-section from \eqref{eq:Hawk-coupling} with the semiclassical result.
This matching applies in the low-frequency regime
$r_+\omega_{\rm 4d}\ll1$.
In the asymptotically flat normalization of \cite{Biggs:2025nzs}, this gives
$g=2r_+$ with $r_+$ the outer horizon radius, which reduces to $g=2r_0$ at leading order in our near-extremal
expansion. We keep $g$ explicit since its value depends on the normalization
of the boundary time and fields.

We can compute the energy lost to Hawking radiation by combining Fermi's golden rule approach in \cite{Maldacena:1997ih} with the matrix elements of $O_{\text{rad}}$ extracted from the exact JT correlators of \cite{Mertens:2017mtv}. This was done in detail in \cite{Brown:2024ajk}, and we express the result in terms of $\mathsf E$. In the deep Schwarzian regime, the classical relation
${\sf E}\propto{\sf T}^2$ no longer holds. A canonical description requires
averaging over the quantum density of states and gives a different
temperature dependence \cite{Brown:2024ajk}. For the neutral scalar considered here, the leading energy dependence is 
\be
\mathcal{F}({\sf E})
\equiv
-\frac{\d {\sf E}}{\d u}
\propto
\begin{cases}
g^2{\sf E}^2, & {\sf C}{\sf E}\gg 1,\\
g^2{\sf E}^{7/2}, & {\sf C}{\sf E}\ll 1.
\end{cases}
\ee
Here we display only the dependence on ${\sf E}$ at fixed ${\sf C}$ and
keep the remaining normalization-dependent prefactors implicit. We define the corresponding evaporation timescale by
\beq
\tau_{\rm H} = \frac{{\sf E}}{\mathcal{F}({\sf E})}.
\eeq
At fixed $g$ and ${\sf C}$, this gives $\tau_{\rm H}\propto{\sf E}^{-1}$ in the semiclassical regime and $\tau_{\rm H}\propto{\sf E}^{-5/2}$ in the deep Schwarzian regime.

For the closed-system treatment to remain appropriate, the timescale of the
collapse should be shorter than $\tau_{\rm H}$. Classically the crossing proper time and advanced time are finite, as we have shown in Section~\ref{sec:classical_thin_shell}, while the shell has zero thickness. Quantum mechanically the source is smeared over a time
$\sigma$ in Section~\ref{sec:disk-level-protection}. To turn this into a useful bound we also need to decide at which energy we evaluate the rate of evaporation. To be conservative, we evaluate $\tau_{\rm H}$ at the initial energy ${\sf E}_-$. This gives the shortest evaporation timescale along
the cooling process. Taking $\sigma$ to characterize the temporal width of the source, the
closed-system treatment is therefore controlled when
$\sigma\ll\tau_{\rm H}({\sf E}_-)$. If we further require the change in the black hole energy caused by Hawking
radiation while the source is active to be smaller than the energy resolution
$1/\sigma$ of the source, we obtain the stronger condition
$\mathcal{F}({\sf E}_-)\sigma\ll1/\sigma$. 

Let us make these conditions explicit for the
asymptotically flat massless scalar example. Restoring the normalization
of the semiclassical flux in~\cite{Brown:2024ajk}, in throat units one
finds
\be
\mathcal{F}({\sf E}_-)
\simeq
\frac{{\sf E}_-^2}{30\pi{\sf C}^2},
\qquad
{\sf C}{\sf E}_-\gg1.
\ee
The corresponding
evaporation timescale is $\tau_{\rm H}({\sf E}_-)
\simeq
\frac{15}{\pi}
\frac{{\sf C}}{{\sf T}_-^2}$. We see that even the source duration required to begin probing the
Schwarzian regime is compatible with the ordinary evaporation time criterion for neglecting Hawking radiation. At
$\sigma\sim{\sf C}$, the ratio of the source duration to the evaporation
timescale is
\be
\frac{\sigma}{\tau_{\rm H}({\sf E}_-)}
\sim
{\sf T}_-^2\ll1.
\ee
Thus, for a near-extremal initial black hole, the ordinary evaporation timescale is parametrically longer than the source duration.  At $\sigma\sim{\sf C}$, the stronger condition requires ${\sf T}_-\ll{\sf C}^{-1/2}$. Since a semiclassical initial state only
requires ${\sf T}_-\gg{\sf C}^{-1}$, there is a parametrically controlled
window
${\sf C}^{-1}\ll{\sf T}_-\ll{\sf C}^{-1/2}$, in which the initial black hole is semiclassical while Hawking radiation
can consistently be neglected at the energy resolution relevant for the Schwarzian crossover.

It would be interesting to study the collapse when Hawking radiation indeed
competes with the source-driven transition discussed here. A natural extension would be
to evolve the full energy probability distribution using the corresponding probability
evolution equation, along the lines of \cite{Biggs:2025nzs} for
Schwarzian-corrected Hawking evaporation. We leave this for future work.

\paragraph{Schwinger discharge.}

We have checked that third law violation from a semiclassical initial
black hole requires
${\sf q}^2-{\sf m}^2>(\delta S/2\pi)^2\gg1$.
In particular, this places the matter field above the AdS$_2$
pair-production threshold
${\sf q}^2>{\sf m}^2+1/4$. The relevance of this threshold
for quantum protection of the third law has recently been emphasized in
\cite{Hod:2026hql}. This is a curved-space analog of Schwinger pair production first studied in the black hole context in 
\cite{Damour:1975,Gibbons:1975kk,Page:1977um}. Unlike Hawking radiation, Schwinger pair production is a vacuum instability of the charged field already visible within the throat. The AdS$_2$ evaluation of the production rate is discussed in
\cite{Pioline:2005pf,Kim:2008xv} and revisited in \cite{Brown:2024ajk,Rakic:2025svg}. Here we assume that the effective scalar describing the shell is a
dynamical species that can be pair-produced. For a genuinely composite shell, the relevant production rate should also depend on its microscopic charged constituents. Furthermore, it is worth emphasizing that whether a locally produced pair actually discharges the black hole
also depends on the finite throat and on whether the charges separate
so that one is absorbed while the other escapes into the exterior, as emphasized in~\cite{Brown:2024ajk}. See also~\cite{Coviello:2025ryn} for discharge effects in more realistic environments.

Since the channel is above the pair production threshold, the Schwinger channel is open, but we have not determined if the production rate is large or exponentially suppressed. It is physically sensible to ignore the Schwinger pair production if such events are rare while the
source is active. Denoting the inverse effective event rate by
$\tau_{\text{S}}$, we require
\be
\sigma\ll\tau_{\text{S}},
\qquad
\tau_{\text{S}}\propto e^{S_{\rm inst}},
\ee
where, at leading semiclassical order,
\be
S_{\rm inst}
=
2\pi\left(
|{\sf q}|-\sqrt{{\sf q}^2-{\sf m}^2}
\right)
\simeq
\frac{\pi{\sf m}^2}{|{\sf q}|}.
\ee
The exponent is the on-shell action of the Euclidean
instanton consisting
of the charged particle loop around the horizon~\cite{Pioline:2005pf}. The last expression applies in the
additional strong-field regime ${\sf q}^2\gg{\sf m}^2\gg1$, which is the familiar flat-space Schwinger exponent~\cite{Pioline:2005pf,Kim:2008xv}. We display only the
leading exponential dependence, since the prefactor depends on the
finite-throat matching and boundary conditions.
Together with the Hawking condition, for generic parameters the condition on $\sigma$ that we need is
\be
\sigma
\ll
\min\left[
\tau_{\rm H}(\mathsf E_-),
\tau_{\text{S}}
\right].
\label{eq:combined-Hawking-Schwinger-bound}
\ee
Both rapidly discharging and long-lived channels, with the latter requiring $S_{\rm inst}\gg1$, are compatible with the condition for third law violation. There is a parametrically large overlap between third law
violation and $S_{\rm inst}\gg1$, including but not limited to the
strong-field regime discussed above. For example, taking ${\sf m}$ and
$\sqrt{{\sf q}^2-{\sf m}^2}$ to be of the same parametric order as
$\delta S$ gives $S_{\rm inst}=O(\delta S)\gg1$. When this hierarchy is not satisfied, Schwinger discharge can compete with
the source-driven collapse and the probability distribution must instead be
evolved jointly in energy and charge. We leave this problem for future work.

\paragraph{Open questions.} There are two clear generalizations of these results that are worth pursuing. First, we mostly considered collapse due to a thin shell, and it would be interesting to consider the fate in quantum gravity of fully smooth processes, including the charged scalar and Einstein-Maxwell-Vlasov constructions of~\cite{Kehle:2022,Kehle:2024vyt}. An intermediate step in this direction
could be to study a continuous charged matter profile, such as the null fluid
construction of~\cite{Ori:1991}. The relation between the thin shell, null fluid, and smooth Vlasov descriptions
is discussed in Appendix~\ref{app:Ori_discussion}. At the quantum level, a related question is whether such a smooth or composite configuration can still be described by the single effective matter channel used here. Smearing the classical shell does not necessarily require higher orders in the source expansion, but a
description in terms of independent quantum constituents can involve a many-particle state and multiple insertions. 

Second, it would be interesting to learn how to incorporate quantum gravity in the construction when the initial black hole is either very far from extremality, or when the charge of the shell is comparable to that of the black hole as in panel (b) of Figure~\ref{fig:two_panel_throats}.  In either case the process cannot be studied within a single throat either because there is no throat to begin with, or because the initial shell is very far from the initial throat. We expect the conclusions in this paper regarding the properties of the final quantum state to be the same at a qualitative level. 

It would also be interesting to realize analogs of classical third-law violations in SYK models. The analog of the classical gravity approximation is to solve the large-$N$ saddle-point equations for the bilocal fields $G$ and $\Sigma$~\cite{Maldacena:2016hyu}, neglecting fluctuations even in regimes where they become important. One could therefore search for driven, nonequilibrium solutions whose correlators evolve from finite-temperature to zero-temperature behavior in finite time. Comparing these solutions with finite-$N$ time evolution obtained by exact diagonalization would provide a microscopic test of how quantum effects modify the mean-field prediction. Reproducing our mechanism also requires a matter channel with a complex scaling dimension $\Delta=1/2+\i s$, with real $s\neq0$, as explained in Section~\ref{subsec:near-horizon-jt}. Such dimensions occur for fermion bilinears in coupled SYK models, where they signal an instability of the conformal saddle \cite{Kim:2019upg}. Constructing a suitable protocol and determining its quantum fate remain interesting directions for future work.

Our results also suggest a connection with classical critical collapse~\cite{Choptuik:1992jv}. Classical extremal black holes can arise as codimension-one threshold solutions in gravitational collapse~\cite{Kehle:2024vyt,East:2025nfb,Angelopoulos:2026bez, Mittal:2026ryg}.  In the Vlasov construction, the extremal solution separates
dispersion from subextremal black hole formation
\cite{Kehle:2024vyt}. Relatedly, codimension-one nonlinear asymptotic stability of
extremal RN has been established in spherical symmetry
\cite{Angelopoulos:2024yev}. This fits into a broader conjectural phase
portrait of near-extremal black hole dynamics~\cite{Dafermos:2025int}. Recent work has studied dynamical extremal thresholds,
analytically in the near-horizon JT description and numerically in
nonlinear charged scalar evolution~\cite{Porfyriadis:2025pov, Gelles:2026bix}. Our calculation gives a quantum version of approaching this threshold. Tuning the collapse toward the critical surface takes the final temperature and the energy above extremality to zero, but the deterministic classical endpoint is replaced by a probability distribution over final energies. The classical critical surface is therefore a useful organizing limit, yet cannot be reached as a sharply defined final state. This suggests that the quantum counterpart of the threshold
is no longer represented by a single extremal geometry. There is a close parallel with quantum effects in Choptuik collapse
\cite{Tomasevic:2025clf,Tomasevic:2025kqy,Wu:2026xbr}. There the classical critical solution is also reached by fine tuning, while vacuum polarization introduces a quantum growing mode that shifts the threshold and produces a finite mass gap. These examples motivate the conjecture that this is a universal feature: the approach to a classically fine-tuned critical endpoint enters a regime in which quantum effects qualitatively change the collapse.

We also comment on the nonperturbative effects associated with higher topologies~\cite{Saad:2019lba, Stanford:2019vob, Maxfield:2020ale}. The nonperturbative completion of the 4d black hole need not coincide with that of pure JT gravity, but it is nevertheless interesting to ask what the JT result would suggest. In this case, the sum over topologies gives a genus
expansion controlled by $e^{-S_0}$, whose matrix integral completion is
sensitive to the discreteness of the spectrum. This can become important on timescales exponential in the black hole entropy. Although this regime lies beyond the validity of our analysis, it is interesting to ask how nonperturbative corrections might modify collapse near threshold. Suppose that, for a chosen nonperturbative completion, the source and matrix element remain smooth, so that $P({\sf E}_f) \propto \rho({\sf E}_f)$ continues to hold and that $\rho({\sf E}_f)$ has the same nonperturbative behavior as the density of states in pure JT gravity. By the forbidden region, we mean ${\sf E}_f<0$, below the perturbative
extremal edge, where the density of states vanishes to all orders in the
topological expansion. In the near-edge forbidden region, away from the immediate vicinity of the spectral edge, it would give~\cite{Saad:2019lba}
\be
P({\sf E}_f)
\sim
\frac{1}{8\pi |{\sf E}_f|}
\exp\left(
-2\pi\i
\int_0^{{\sf E}_f}
\d{\sf E}\,\rho({\sf E}+\i\epsilon)
\right),
\qquad
{\sf E}_f<0.
\ee
Here $\rho({\sf E})$ is the disk density in \eqref{eq:density},
including its overall factor $e^{S_0}$, and $\epsilon$ is positive and
infinitesimal. The exponent is real and negative in this region.  The effect is doubly nonperturbative, of order
$\exp[-O(e^{S_0})]$, and is therefore invisible to every order in the
$e^{-S_0}$ topological expansion. Under these assumptions, the nonperturbative tail would give nonzero transition weight at ${\sf E}_f<0$, relative to the
perturbative extremal reference. This should not be interpreted as preparing a classical superextremal geometry, and the density alone does not determine the probability of preparing the exact ground state.  This differs from the near-BPS case in supergravity~\cite{Heydeman:2020hhw, Boruch:2022tno}, where supersymmetry imposes an exact lower bound on the
spectrum and protects states at the BPS threshold. It would be interesting to study collapse in this regime in more detail. 

Finally, the approach to extremality can also enhance corrections that are not included in the JT description. Higher-derivative effects can produce large tidal forces near extremal rotating black holes~\cite{Horowitz:2023xyl,Horowitz:2024dch}. This effect is absent for the 4d RN black hole limit considered in our main analysis. But it can become relevant for extending the discussion to near-extremal Kerr and Kerr-Newman horizons. These corrections can
become important before the Schwarzian regime is reached, and understanding
the interplay between these effects is an interesting open problem.

\paragraph{Acknowledgements} 
We thank Netta Engelhardt, Gary Horowitz, Luca Iliesiu, Guanda Lin, and Mukund Rangamani for useful discussions. GJT would like to thank Roberto Emparan, Harvey Reall, and the participants of the GGI school ``Pathways to Quantum Black Holes'' for multiple discussions on these topics. This work was supported by the DOE Early Career Award DE-SC0026287. We used an internal Google research agent powered by Gemini Deep Think and the OpenAI Pro academic subscription. 

\begin{appendix}

\section{The massless shell limit and bouncing charged matter}
\label{app:Ori_discussion}

In this Appendix, we explain how the massless limit of the distributional timelike shell separating $\mathcal{M}_-$ and $\mathcal{M}_+$ in Section~\ref{sec:classical_thin_shell} is related to the charged null dust construction given by Ori~\cite{Ori:1991}. The relation is direct for a single shell, although the matter models are not the same. In particular, some new assumptions enter when many shells are assembled together.

It is instructive to consider a charged massless thin shell collapsing, and null junction conditions can be formulated from the beginning~\cite{Barrabes:1991ng}. Here it is simpler to take a controlled $m \to 0$ limit of the timelike equation~\eqref{eq:radial-potential}. This involves proper time which is obviously ill-defined for massless shells. Since $k^\mu = m\, \d x^\mu/\d \tau$, it is natural to define a worldline parameter $ s = \tau / m$ such that $k^\mu=\d x^\mu/\d s$. For nonzero $m$, this is only a change of parametrization. Its advantage is that $k^\mu$ has the normalization of the shell momentum and remains meaningful as $m\to0$, with $k^\mu k_\mu=-m^2\to0$.

We consider an ultrarelativistic limit by holding $\Delta M$ and $\Delta (Q^2)$ fixed while taking $m \to 0$. The result is
\be
\frac{\d R}{\d s}= \pm  \left(\Delta M - \frac{\Delta (Q^2)}{2R} \right).
\label{eq:RdotMassless}
\ee
At first sight, either sign merely gives one of the two radial null directions. This might suggest that the trajectory of the shell is not affected by the charge, but this is not the case. We see that the charge still enters through the effective energy in parentheses, and it vanishes at a bounce radius
\be
R=R_b = \frac{\Delta(Q^2)}{2\Delta M}.
\label{eq:finite-Ori-radius}
\ee
At this point $\d x^\mu/\d s=0$ and the shell momentum vanishes $k^\mu \to 0$. This happens at infinite worldline parameter since near $R_b$ the equation becomes $ \d R/ \d s \sim (R-R_b)$ which implies $s \sim \log |R-R_b|$.

The limiting equation does not choose a continuation through this endpoint, and at $R=R_b$ we have two choices to continue the trajectory.  Keeping the shell in the ingoing null congruence makes its momentum past directed after the zero and gives the familiar negative energy continuation. Gluing it to the outgoing congruence keeps the momentum future directed. The latter is selected by the $m>0$ case studied in Section~\ref{subsec:classical-thin-shell}. Indeed, the physical turning point is the zero of the potential function $U(R)$ and approaches $R_b$ as $m \to 0$. The Lorentz force has the effect of reversing the direction of the shell even in the massless case. A small angular momentum gives another regulator and selects the same outgoing continuation~\cite{Ori:1991}.

A direct connection with Ori's charged null dust appears when the finite jump considered in this paper is resolved into infinitesimal layers. For one layer, we write $\Delta M \to \d M$, $\Delta (Q^2) \to 2Q \d Q$. If the layers are labeled by an advanced coordinate $v$, this gives
\be
R_b(v)=Q \frac{\dot{Q}}{\dot{M}},
\ee
precisely the vanishing radius in charged ingoing Vaidya. Here dots denote derivatives with respect to $v$. The matter contribution to the ingoing flux vanishes at $R_b(v)$ and changes sign in the formal ingoing continuation. Ori replaces that continuation with the future-directed outgoing flow. 

The vanishing points of the different layers form the bounce hypersurface $\mathcal{B}$. The usual gluing assumes that $\mathcal{B}$ is spacelike~\cite{Ori:1991, Chatterjee:2015cyv}. This is an important assumption since then an outgoing Vaidya region is attached without an overlap of ingoing and outgoing dust.\footnote{The spacelike assumption about $\mathcal{B}$ should be distinguished from the mechanism described here. If $\mathcal{B}$ is timelike, particles that have turned overlap with particles that are still ingoing. One must then solve an interacting two-stream dust system. It has been found recently that the system partly decouples~\cite{Bick:2026naa}, with $\partial_u \partial_v Q=0$ in double-null coordinates, and the equations for $r$ and the conformal factor can close before the number currents are reconstructed. This permits a broad class of timelike bounce surfaces, and the parameters can even remain far away from extremality.} It is worth mentioning that this bounce is not necessarily a local minimum of the areal radius. It may even occur between the inner and outer horizons of a subextremal geometry~\cite{Chatterjee:2015cyv}, where $f<0$ and $r$ is a time coordinate. Both future-directed null congruences then move toward smaller $r$. In our case with an extremal final geometry, $f_+(R)>0$ away from $R=r_e$, and $R_b<r_e$. The outgoing continuation has increasing $R$, so here the limiting shell makes a literal radial turn.

Now we compare with the construction in Kehle and Unger~\cite{Kehle:2024vyt}, where they first construct an Ori null dust spacetime and then desingularize it with a smooth beam in the Einstein-Maxwell-Vlasov system. The beam has a small spread in phase space and small nonzero angular momentum. Its characteristics turn smoothly under electromagnetic repulsion for both $m=0$ and $m>0$. In this case, the smoothing is not coming from the rest mass alone. Our finite-mass shell regulates the individual trajectory in a simpler way, but the matter configuration remains distributional.

There is one final distinction worth pointing out. A distributional shell specifies only two endpoint geometries $\mathcal{M}_-$ and $\mathcal{M}_+$ with no intermediate mass and charge profile. But if we choose a smooth profile, there is a simple local criterion. Let the final point $v_f$ lie on the charged AdS extremal curve $M_{\rm ext}(Q)$, with extremal radius $r_e$. At this point,
\be
M(v_f)=M_{\rm ext}(Q(v_f)),
\qquad
\left.
\frac{\d M_{\rm ext}}{\d Q}
\right|_{v_f}
=
\frac{Q(v_f)}{r_e}.
\ee
Expanding around $v_f$ gives
\be
M(v)-M_{\rm ext}(Q(v))
\approx {}
\dot M(v_f)
\left(
1-\frac{R_b(v_f)}{r_e}
\right)
(v-v_f),
\label{eq:superextremal-interlude}
\ee
where $R_b(v)=Q(v)\dot Q(v)/\dot M(v)$ is the bounce radius of an infinitesimal layer. If $\dot M(v_f)>0$ and $R_b(v_f)<r_e$, the coefficient of $v-v_f$ is positive. Since $v-v_f<0$ immediately before the endpoint, we find
\be
M(v)<M_{\rm ext}(Q(v)),
\qquad
v<v_f,
\ee
for $v$ sufficiently close to $v_f$. The profile therefore lies briefly on the superextremal side of the stationary parameter space. For positive charge in flat space this reads $Q(v)>M(v)$. This behavior is explicit in the Ori model used by Kehle and Unger \cite{Kehle:2024vyt}. However, this is local and does not describe a globally naked superextremal black hole. Superextremality is also not required for the bounce itself, since charged null matter can turn in a subextremal geometry \cite{Chatterjee:2015cyv,Bick:2026naa}.

\section{Effective action for the charged scalar field}
\label{app:fixed_incoming_kernel}

The goal of this Appendix is to derive the effective action after integrating out a charged matter field in rigid AdS$_2$. We begin working in Euclidean signature and then continue to Lorentzian.

The Euclidean throat metric and gauge field are
\be
\mathrm{d}\mathsf s^2
=
\mathsf{r}^2\mathrm{d}\mathsf t_{\rm E}^2+{\mathrm{d}\mathsf{r}^2\over\mathsf{r}^2},
\qquad
\mathsf A =- \i \mathsf{r} \d \mathsf t_{\rm E}.
\label{eq:Euclidean-fixed-AdS2}
\ee
For a Fourier mode
$
\psi(\mathsf t_{\rm E},\mathsf{r})
=
e^{-\mathrm{i}\omega\mathsf t_{\rm E}}\psi_\omega(\mathsf{r}),
$
the equation of motion is
\be
\left[
\partial_{\mathsf{r}}\mathsf{r}^2\partial_{\mathsf{r}}
-
{(\omega-\mathrm{i}\mathsf q\mathsf{r})^2\over\mathsf{r}^2}
-
\mathsf m^2
\right]
\psi_\omega(\mathsf{r})
=0.
\label{eq:Euclidean-radial-eq-rho}
\ee
Using $\lambda^2
=
\mathsf q^2-\mathsf m^2-{1\over4}$, the radial equation can be put in Whittaker form, and the solution regular in
the Euclidean interior is
\be
\psi_\omega(\mathsf{r})
=
C_\omega\,
W_{\mathrm{i} \,\text{sgn}(\omega) \mathsf q,\,\mathrm{i}\lambda}
\left(
{2|\omega|\over\mathsf{r}}
\right),
\label{eq:Whittaker-regular-solution}
\ee
with $C_{\omega}$ an overall normalization which we now determine in terms of the source $J_{\text{in}}$. To do this, expand the solution near the boundary with ${\sf r} \to\infty$ as
\be
W_{\mathrm{i}\, \text{sgn}(\omega)\mathsf q,\,\mathrm{i}\lambda}
\left(
{2|\omega|\over\mathsf{r}}
\right)
\sim
B_W(\omega)\mathsf{r}^{-1/2-\mathrm{i}\lambda}
+
A_W(\omega)\mathsf{r}^{-1/2+\mathrm{i}\lambda},
\label{eq:Whittaker-asymptotic}
\ee
where
\be
B_W(\omega)
=
{\Gamma(-2\mathrm{i}\lambda)\over
\Gamma\left({1\over2}-\mathrm{i}\lambda-\mathrm{i}\, \text{sgn}(\omega) \mathsf q\right)}
(2|\omega|)^{{1\over2}+\mathrm{i}\lambda},
\label{eq:BW-def}
\ee
\be
A_W(\omega)
=
{\Gamma(2\mathrm{i}\lambda)\over
\Gamma\left({1\over2}+\mathrm{i}\lambda-\mathrm{i} \, \text{sgn}(\omega) \mathsf q\right)}
(2|\omega|)^{{1\over2}-\mathrm{i}\lambda}.
\label{eq:AW-def}
\ee
Fixing the incoming source means fixing the coefficient of
$\mathsf{r}^{-1/2-\mathrm{i}\lambda}$.  Thus, if
\be
\psi_\omega(\mathsf{r})
\sim
J_{\rm in}(\omega)\mathsf{r}^{-1/2-\mathrm{i}\lambda}
+
R_{\text{out}}(\omega)\mathsf{r}^{-1/2+\mathrm{i}\lambda},
\label{eq:incoming-response-expansion-frequency}
\ee
then regularity in the Euclidean interior gives
\be
R_{\text{out}}(\omega)
=
R_E(\omega)J_{\rm in}(\omega),
\label{eq:AE-response}
\ee
with
\be
R_E(\omega)
=
{A_W(\omega)\over B_W(\omega)}
=
{\Gamma(2\mathrm{i}\lambda)\over\Gamma(-2\mathrm{i}\lambda)}
{\Gamma\left({1\over2}-\mathrm{i}\lambda-\mathrm{i} \, \text{sgn}(\omega)\mathsf q\right)
\over
\Gamma\left({1\over2}+\mathrm{i}\lambda-\mathrm{i}  \, \text{sgn}(\omega)\mathsf q\right)}
(2|\omega|)^{-2\mathrm{i}\lambda}.
\label{eq:Euclidean-response-function}
\ee

The renormalized nonlocal on-shell action in the fixed-incoming ensemble is
bilinear in the fixed-incoming sources,
\be
S_{\rm in}
=
\int {\mathrm{d}\omega\over2\pi}\,
\widehat{\widetilde J}_{\rm in}(-\omega)\,
K_E(\omega)\,
\widehat J_{\rm in}(\omega).
\label{eq:Win-frequency}
\ee
Here we use hats to denote sources written in the fixed AdS$_2$ time $\mathsf t$.  The kernel
is determined by the response coefficient conjugate to the fixed-incoming
source.  In the principal-series polarization, this conjugate response is
$2\mathrm{i}\lambda R_{\rm out}(\omega)$, with the factor $2\mathrm{i}\lambda$ the radial symplectic normalization.  Since Euclidean
regularity gives $R_{\rm out}(\omega)=R_E(\omega)\widehat J_{\rm in}(\omega)$, the nonlocal kernel is
\be
K_E(\omega)
=
2\mathrm{i}\lambda\,R_E(\omega).
\label{eq:KE-response}
\ee
At finite cutoff there is also a local term proportional to
$\mathsf{r}_c^{-2\mathrm{i}\lambda}\widehat{\widetilde J}_{\rm in}\widehat J_{\rm in}$.
This term is local in boundary time and will be subtracted into the definition of
the renormalized source.  The nonlocal part is therefore determined by
\eqref{eq:Euclidean-response-function}.

Fourier transforming gives the fixed-frame time-domain kernel.  Let
\be
C_+
\equiv
{\Gamma\left({1\over2}-\mathrm{i}\lambda-\mathrm{i}\mathsf q\right)
\over
\Gamma\left({1\over2}+\mathrm{i}\lambda-\mathrm{i}\mathsf q\right)},
\qquad
C_-
\equiv
{\Gamma\left({1\over2}-\mathrm{i}\lambda+\mathrm{i}\mathsf q\right)
\over
\Gamma\left({1\over2}+\mathrm{i}\lambda+\mathrm{i}\mathsf q\right)} .
\label{eq:Cpm-def}
\ee
The frequency-space kernel is
\be
K_E(\omega)=2 \mathrm{i} \lambda \frac{\Gamma(2 \mathrm{i} \lambda)}{\Gamma(-2 \mathrm{i} \lambda)} (2|\omega|)^{-2 \mathrm{i} \lambda}\left[C_+ \theta(\omega)+C_- \theta(-\omega)\right],
\ee
with $\theta(\omega)$ the step function. Then, away from coincident points $\mathsf t_1\neq\mathsf t_2$, or equivalently $\mathsf t_{\rm E} \neq 0$,
\be
K_E(\mathsf t_{\rm E})
=
{\cal N}_+
(\mathsf t_{\rm E}-\mathrm{i}0)^{-2\Delta_{\rm in}}
+
{\cal N}_-
(\mathsf t_{\rm E}+\mathrm{i}0)^{-2\Delta_{\rm in}},
\qquad
\Delta_{\rm in}={1\over2}-\mathrm{i}\lambda,
\label{eq:time-domain-kernel-two-branches}
\ee
with
\be
{\cal N}_+
=
{2\lambda^2\over\pi}\,
2^{-2\mathrm{i}\lambda}\Gamma(2\mathrm{i}\lambda)\,
C_+\,
e^{-{\mathrm{i}\pi\over2}(1-2\mathrm{i}\lambda)},
\label{eq:Nplus-def}
\ee
\be
{\cal N}_-
=
{2\lambda^2\over\pi}\,
2^{-2\mathrm{i}\lambda}\Gamma(2\mathrm{i}\lambda)\,
C_-\,
e^{+{\mathrm{i}\pi\over2}(1-2\mathrm{i}\lambda)}.
\label{eq:Nminus-def}
\ee
The two coefficients differ because the Euclidean gauge field is imaginary and
therefore distinguishes positive and negative Euclidean frequencies. Indeed, one can write this kernel in the time domain and get 
\beq
S_{\text{matter}}
=
{\cal N}_{\Delta_{\text{in}},{\sf q}}
\int \d\tau_2\,\d\tau_1\,
\widetilde J(\tau_2)\,
J(\tau_1)\,
\frac{
e^{+\pi{\sf q}}\,
\theta(\tau_2-\tau_1)
+
e^{-\pi{\sf q}}\,
\theta(\tau_1-\tau_2)
}{
|\tau_2-\tau_1|^{2\Delta_{\text{in}}}
},
\label{eq:matter_kernel_functional}
\eeq
with the Euclidean time as $\tau=\mathsf t_{\rm E}$. The common normalization for the two branches is 
\be
    {\cal N}_{\Delta,{\sf q}}
    =
    -
    \frac{
        2^{2\Delta-2}(2\Delta-1)
    }{
        \pi\,\Gamma(2\Delta-1)
    }
    \Gamma(\Delta+\i{\sf q})
    \Gamma(\Delta-\i{\sf q}) .
    \label{eq:charged-kernel-normalization}
\ee
For real $\Delta$, the Gamma function product reduces to
$|\Gamma(\Delta+\i{\sf q})|^2$. Here it is generally complex for $\Delta=\Delta_{\rm in}$, as expected for the fixed-incoming problem.

We see the relative weight of the two Euclidean orderings is $e^{2 \pi \sf q}$. It should be clear that this is not due to a complex $\Delta_{\rm in}$ since $\Delta_{\rm in} $ only depends on $\sf q^2$. We therefore cannot use a single normalization to absorb both $e^{\pm \pi \sf q}$. However, for a calculation which keeps
the ordering $\tau_2>\tau_1$, the factor $e^{+\pi{\sf q}}$ may be
included in a normalization ${\cal N}_{\rm in}$.  The reverse ordering then retains
the relative factor $e^{-2\pi{\sf q}}$.

At finite temperature, $\beta=1/\mathsf T$, we define the oriented Euclidean separation as
\be
    \tau_{21}
    \equiv
    (\tau_2-\tau_1)\bmod\beta,
    \qquad
    0<\tau_{21}<\beta .
    \label{eq:thermal-oriented-separation}
\ee
The analogous result obtained at finite temperature is given by
\beq
S_{\text{matter}}
=
{\cal N}_{\Delta_{\text{in}},{\sf q}}
\int_0^\beta \d\tau_2
\int_0^\beta \d\tau_1\,
\widetilde J(\tau_2)\,
J(\tau_1)\,
e^{\pi{\sf q}
(
1-\frac{2\tau_{21}}{\beta}
)}
\left[
\frac{\pi/\beta}{
\sin\!\left(
\pi\tau_{21}/\beta
\right)
}
\right]^{2\Delta_{\text{in}}},
\eeq
Since $\sin(\pi\tau_{21}/\beta)>0$, the Euclidean power is well-defined even for complex $\Delta_{\rm in}$. For $\tau_2>\tau_1$ or $\tau_1>\tau_2$, the $\beta \to \infty$ limit gives the two pieces in~\eqref{eq:matter_kernel_functional}, respectively. The same charge-dependent factor appears in~\cite{Gu:2019jub}. Upon continuation to Lorentzian signature with $\tau_{21} = \epsilon + \i {\sf t}_{21}$ and $\epsilon>0$, one gets at finite temperature
\beq
\begin{aligned}
S_{\text{matter}}
={}&
{\cal N}_{\Delta_{\text{in}},{\sf q}}\,
e^{\pi{\sf q}}e^{-\i\pi\Delta_{\text{in}}}
\int_{-\infty}^{\infty} \d \mathsf t_2
\int_{-\infty}^{\infty} \d \mathsf t_1\,
\widetilde J(\mathsf t_2)\,J(\mathsf t_1)
\\
&\qquad \times
e^{-\frac{2\pi \i{\sf q}}{\beta}(\mathsf t_2-\mathsf t_1)}
\left[
\frac{\pi/\beta}{
\sinh\!\left[
\frac{\pi}{\beta}
\left(
\mathsf t_2-\mathsf t_1-\i\epsilon
\right)
\right]
}
\right]^{2\Delta_{\text{in}}}.
\end{aligned}
\eeq

\end{appendix}
\addcontentsline{toc}{section}{References}
\bibliographystyle{JHEP}
\bibliography{bibliography}

\end{document}